\documentclass[10pt,oneside]{article}
\usepackage{amsmath, amsfonts, amsthm, amssymb, amscd}
\usepackage[mathcal]{euscript} 
\usepackage{dsfont, pxfonts, verbatim}
\usepackage{mathrsfs}  
\newtheorem{theorem}{Theorem}[section]
\newtheorem{proposition}[theorem]{Proposition} 
\newtheorem{lemma}[theorem]{Lemma}
 
\newtheorem{definition}[theorem]{Definition} 
\newtheorem{remark}[theorem]{Remark}  
\numberwithin{equation}{section} 
\newcommand \gt {{\widetilde g}}
\newcommand \la \langle
\newcommand \ra \rangle 
\newcommand \thetab {\overline \theta}
\newcommand \gh {\widehat g}  
\newcommand \Trace   {\text{Tr}}  
\newcommand \del	     \partial  
\newcommand \auth 	{\textsc}   
\newcommand \Mcal 	{\mathcal M}  
\newcommand \Bcal 	{\mathcal B}  
  
\newcommand \Rcal 	{\mathcal R}  
\newcommand \Lcal 	{\mathcal L}  
\newcommand \Ecal 	{\mathcal E}

\newcommand \Gb 		{\overline G}
\newcommand \Hb 		{\overline H}
\newcommand \Gbz 	{\displaystyle{\mathop{\overline G}_{0} }}
\newcommand \Hbz 	{\displaystyle{\mathop{\overline H}_{0} }}
\newcommand \Ubzero {\overline U_0}
\newcommand \Abzero {\overline A_0}
\newcommand \Ubone  {\overline U_1}
\newcommand \Abone  {\overline A_1}
\newcommand \Xb   {\overline X}
\newcommand \Yb   {\overline Y}

\newcommand \etabzero  {\overline \eta_0} 
\newcommand \etabone  {\overline \eta_1} 
\newcommand \abar  {\overline a} 
\newcommand \Ubz  	{\displaystyle{\mathop{\overline U}_{0}}}
\newcommand \Abz  	{\displaystyle{\mathop{\overline A}_{0}}}
\newcommand \Pbz  	{\displaystyle{\mathop{\overline P}_{0}}}
\newcommand \nubz  	{\displaystyle{\mathop{\overline \nu}_{0}}}

\newcommand \Rbz  	{\displaystyle{\mathop{\overline R}_{0}}}

\newcommand \habz       {\displaystyle{ \mathop{h_{ab}}_0}}
\newcommand \hbdz       {\displaystyle{ \mathop{h_{bd}}_0}}
\newcommand \alephz       {\displaystyle{ \mathop{\aleph_{ab}}_0}}
\newcommand \alephcdz       {\displaystyle{ \mathop{\aleph_{cd}}_0}}
\newcommand \alephadz       {\displaystyle{ \mathop{\aleph_{ad}}_0}}
\newcommand \Rb 		{\overline R}

\newcommand \nub 	{\overline \nu}
\newcommand \etab 	{\overline \eta}
\newcommand \Ub 		{\overline U}
\newcommand \Pb 		{\overline P}
\newcommand \Ab 		{\overline A} 

\newcommand \ab {\overline a} 
\newcommand \loc   	{\text{loc}} 
\newcommand \RR 		{\mathbb R}   
\newcommand \eps 	\epsilon  
\newcommand \be 		{\begin{equation}}
\newcommand \ee 		{\end{equation}} 
\let\oldmarginpar\marginpar
\renewcommand\marginpar[1]{\-\oldmarginpar[\raggedleft\footnotesize #1]%
{\raggedright\footnotesize #1}}

\usepackage{hyperref} 
\begin{document}

\title{Weakly regular $T^2$--symmetric spacetimes. 
\\
The global geometry of future developments}  
\author{ 
%    Spelling:   LeFloch  or LeFLOCH
Philippe G. LeFloch\footnote{Laboratoire Jacques-Louis Lions \& Centre National de la Recherche Scientifique, 
Universit\'e Pierre et Marie Curie (Paris 6), 4 Place Jussieu, 75252 Paris Cedex, France. 
\newline 
Blog: {\sl philippelefloch.wordpress.com.} E-mail : {\sl pgLeFloch@gmail.com.} 
}
\, and Jacques Smulevici\footnote{
Max-Planck-Institut f\"ur Gravitationsphysik, 
Albert-Einstein Institut, Am M\"uhlenberg 1, 14476 Potsdam, Germany. 
E-mail: {\sl Jacques.Smulevici@aei.mpg.de.} \, 
%\newline
% 2000\textit{\ AMS Subject Classification.} ... 
}
}
\date{April 2011}
\maketitle

\begin{abstract}  Under {\sl weak regularity} assumptions, only,  
we develop a fully geometric theory of vacuum Einstein spacetimes with $T^2$--symmetry,
establish the {\sl global well-posedness} of the initial value problem for 
Einstein's field equations, 
and investigate the global causal structure of the constructed spacetimes. 
Our weak regularity assumptions are the minimal ones 
allowing to give a meaning to the Einstein equations 
under the assumed symmetry and to solve the initial value problem. 
First of all, we introduce a frame adapted to the symmetry in which each 
Christoffel symbol can be checked to belong to $L^p$ for $p=1,2,$ or $\infty$. 
We identify certain cancellation properties taking place in the 
expression of the Riemann and Ricci curvatures, 
and this leads us to a reformulation of 
the initial value problem for the Einstein field equations when the initial data set has weak regularity. 
Second, we investigate the future development of a weakly regular initial data set. We check that 
the area $R$ of the orbits of symmetry must grow to infinity in the future timelike directions, and    
we establish the existence of a global foliation by the level sets of the function $R$.  
Our weak regularity assumptions only require that $R$ is Lipschitz continuous
while the metric coefficients describing the initial geometry of the orbits of symmetry 
are in the Sobolev space $H^1$ and the remaining coefficients have even weaker regularity. 
We develop here the compactness arguments required to cover the natural level of regularity associated 
with the {\sl energy} of the system of partial differential equations determined from Einstein's field equations. 
\end{abstract}

%===========================================================================

\section{Introduction}
\label{IN}

This is the first of a series of papers \cite{LeFlochSmulevici2,LeFlochSmulevici3} 
devoted to {\sl weakly regular} spacetimes of general relativity satisfying Einstein's vacuum field equations
under certain symmetry assumptions.  
One of the main difficulties we overcome here is precisely to determine the natural weak regularity conditions 
that are required to handle the Einstein equations under the assumed symmetry. 
In this framework, 
for any initial data set with weak regularity, 
we determine the {\sl global geometric structure} of the associated Cauchy development. 
Specifically, we impose that the initial data are defined on a manifold diffeomorphic to the $3$-dimensional torus $T^3$ 
and are invariant under the action of the Lie group $T^2$. 
This requirement characterizes the so-called {\sl $T^2$--symmetric spacetimes on $T^3$}
with possibly non-vanishing twist constants.  This symmetry assumptions allows to study 
the propagation and dispersion of gravitational waves.

A large literature is available on such spacetimes when {\sl sufficiently high regularity}
on the initial data is assumed. 
Let us especially refer to Moncrief \cite{Moncrief}, Chru\'sciel \cite{Chrusciel}, 
Berger, Chru\'sciel, Isenberg, and Moncrief \cite{BergerChruscielIsenbergMoncrief}, 
and Isenberg and Weaver \cite{IsenbergWeaver}. (For further references, cf.~\cite{Rendall2,Smulevici2}.)  
The present paper is also motivated by the earlier work 
by LeFloch and Stewart \cite{LeFlochStewart,LeFlochStewart2} (see also \cite{BLSS}) and LeFloch and Rendall \cite{LeFlochRendall}, 
who treated a special case of $T^2$-symmetric spacetimes, namely Gowdy--symmetric spacetimes, coupled to matter 
and satisfying the Einstein-Euler equations. 
There\-in, it was recognized that, due to the formation of shock waves in the fluid and by virtue of the 
Einstein equations, only weak regularity on the geometry can be allowed and, therefore, these papers provide a 
motivation for the present work. 
In addition, we recall that Christodoulou's proof \cite{Christodoulou} 
of Penrose's cosmic censorship conjecture for spherically symmetric spacetimes
also relied on the introduction of a class of spacetimes with weak regularity. 

In the present paper, we introduce a {\sl 
fully geometric\footnote{Only coordinate-independent results 
are of interest in general relativity, although certain gauge choices must be made for the Einstein equations to be amenable to
techniques of partial differential equations.}
well-posedness theory} covering a large class of weakly regular spacetimes. 
We determine the {\sl optimal regularity conditions} of weak regularity that allow one to 
establish the existence of future developments of weakly regular initial data sets. 
We also thoroughfully describe the {\sl global geometry} of the constructed spacetimes. 
To this end, we provide {\sl novel a priori estimates} for the Einstein equations under the assumed symmetry,
which only assume our weak regularity  assumptions and, in turn, require us to develop a new compactness
strategy leading to the existence, uniqueness, and stability of solutions to the Einstein
equations under the assumed symmetry.  For additional results, 
we refer to the companion papers \cite{LeFlochSmulevici2,LeFlochSmulevici3}
in which, especially, 
we provide a complete description of the {\sl long-time asymptotics} of $T^2$--symmetric spacetimes. 
These results were first announced in \cite{LeFlochSmulevici0}.

Let us recall briefly the formulation of the initial value problem in general relativity (in the vacuum case). 
An initial data set for the vacuum Einstein equations is a triple $(\Sigma,h,K)$ such that 
$(\Sigma,h)$ is a $3$-dimensional Riemannian manifold,
$K$ is a symmetric $2$-tensor field defined on $\Sigma$, 
and satisfying the so-called Einstein's constraint equations 
\begin{eqnarray}
R^{(3)} - |K|^2 + (tr K)^2 = 0,  &
\label{ee:constham} 
\\
\nabla^{(3)j}K_{ij}-\nabla_i^{(3)} tr K = 0, & 
\label{ee:constmom}
\end{eqnarray} 
in which the covariant derivative $\nabla^{(3)}$ and
the scalar curvature $R^{(3)}$ are computed from the Riemannian metric $h$.   
Then, a solution to the initial value problem associated with the initial data set $(\Sigma,h,K)$, by definition,  
is a $(3+1)$-dimensional Lorentzian manifold $(\Mcal,g)$ satisfying the vacuum Einstein equations
\be
\label{Eins1}
R_{\mu \nu}=0,
\ee
together with an embedding $\phi: \Sigma \to \Mcal$ such that $\phi(\Sigma)$ is a Cauchy surface of $(\Mcal,g)$ 
and the pull-back of its first and second fundamental forms coincides with $h$ and $K$, respectively. 
By convention, for instance in \eqref{ee:constham}--\eqref{Eins1},  
Greek indices describe $0,\ldots, 3$ while Latin indices describe $1,2,3$. 

Recall that the existence of a unique (up to diffeomorphism) maximal globally hyperbolic 
solution $(\Mcal,g)$, or {\sl maximal Cauchy development},  
was established in pioneering work by Choquet-Bruhat \cite{ChoquetBruhat} and 
Choquet-Bruhat and Geroch \cite{ChoquetBruhatGeroch}.
The local existence theorem given by Hughes, Kato, and Marsden \cite{HughesKatoMarsden}
requires that the initial data $(h,K)$ belong to the Sobolev space 
$H_{loc}^{s}(\Sigma) \times H_{loc}^{s-1}(\Sigma)$ for some $s>5/2$. 
The current state of the art is provided by Klainerman and Rodnianski \cite{KlainermanRodnianski} (see also \cite{SmithTataru}) 
and requires asymptotically flat initial data with $s>2$, only.  

We are now in a position to sketch the main results established in the present paper about
the existence and the global structure of the development of weakly regular initial data sets.  For the precise definitions
and concepts used now we refer to Section~\ref{mainR}, below. 
While we restrict attention to the class of $T^2$--symmetric spacetimes, 
our regularity assumptions are far below those covered in all previous works (with or without symmetry). 
As far as the initial data set is concerned, our weak regularity conditions can be summarized as follows. 
First of all, 
we assume that the area $R$ of the orbits of $T^2$--symmetry is Lipschitz continuous, 
and we observe that additional regularity on the function $R$ (i.e.~it admits integrable second-order derivatives)
is implied by Einstein's constraint equations. 
The remaining components of the data set prescribed on the initial slice $\Sigma$ 
either represent the geometry of the $T^2$--orbits and are assumed to belong to
the Sobolev space $H^1(\Sigma)$, or represent its orthogonal complement and have even lower regularity. 
A weak regularity property is also imposed on the second fundamental form. 
(See again Section~\ref{mainR}, below.) 

The first result established in this paper is as follows. 

\begin{theorem}[Weak formulation of the Einstein equations for weakly regular spacetimes]
\label{theorem61} 
If $(\Sigma, h, K)$ is a weakly regular $T^2$-symmetric triple, 
then Einstein's constraint equations \eqref{ee:constham}-\eqref{ee:constmom} 
can be reformulated in a weak sense. Similarly, 
if $(\Mcal,g)$ is a weakly regular 
$T^2$-symmetric Lorentzian manifold, then Einstein's (constraint and evolution) field equations 
\eqref{Eins1}
can be reformulated in a weak sense. 
\end{theorem}

Section~\ref{mainR}, below, is devoted to proving this theorem and therein, in particular, we define 
our terminology. Importantly, we prove that 
{\sl not all} Christoffel symbols and curvature components make sense as distributions, but 
only those that turn out next to be relevant to the weak formulation of the Einstein equations. 
To do otherwise, additional regularity would be necessary, as recognized by LeFloch and Mardare~\cite{LeFlochMardare}. 
For the proof of Theorem~\ref{theorem61}, we introduce a {\sl frame suitably adapted to the symmetry}
and we 
uncover certain {\sl cancellation properties} within the standard expressions of the Riemann and Ricci 
curvatures, which allow us to suppress certain (otherwise ill-defined) terms.  

Our second main result  
establishes the {\sl existence of a weakly regular, future Cauchy development} of any given initial data set,
and provides detailed information about the {\sl global geometric structure} of the constructed spacetime.  
In particular, we establish that these weakly regular developments may be covered 
by a global foliation whose spacelike leaves coincide with the level sets of the area function $R$, as stated now. 

\begin{theorem}[Well-posedness theory for the Einstein equations of weakly regular spacetimes]
\label{theorem62} 
Given any weakly regular $T^2$-symmetric initial data set $(\Sigma,h,K)$  with topology $T^3$
whose orbits of symmetry have initially constant area denoted by $R_0>0$, 
there exists a  
weakly regular, vacuum spacetime with $T^2$--symmetry on $T^3$, say $(\Mcal,g)$, 
which is a future development of $(\Sigma, h, K)$, is maximal among all $T^2$-symmetric developments, 
and admits a unique global foliation by the level sets of $R \in [R_0,\infty)$. 
\end{theorem}

The restriction that the initial slice has constant area is not an essential assumption and is made only for convenience in the 
presentation. The proof of this theorem will rely on the material developed in Sections~\ref{sec2} to \ref{sec5}
 and be finally provided at the end of Section~\ref{sec56}. 

Importantly, the new compactness arguments developed in the present paper for the Einstein equations with 
weakly regular initial data 
are completely different from those known for regular initial data. Indeed, many key estimates derived 
in \cite{Moncrief,BergerChruscielIsenbergMoncrief} no longer hold for 
our larger class of spacetimes, 
especially the estimates involving second-order derivatives of the metric coefficients. 

Interestingly, our analysis relies on two different coordinate charts. 
One difficulty arises from the fact that these coordinates systems
must be constructed together with the solution to the Einstein equations and, therefore, also enjoy weak regularity.  
The passage from one coordinate chart to another and the preservation of weak regularity is rigorously 
established in this paper (cf.~Section~\ref{sec52}, below). 
On one hand, the required compactness property is established by identifying a certain 
null structure of the Einstein equations in the so-called conformal gauge under consideration; 
the conformal coordinates are better suited to establish compactness properties of solutions since the equations
become semi-linear, whereas the constraints and certain nonlinear terms take a more involved form.
On the other hand, in order to analyze the global structure of the spacetime, 
areal coordinates are necessary and allows us to control the long-time behavior of solutions; they 
have the advantage that the constraints degenerate to first-order equations and 
the evolution equations admit a monotone energy-like functional. 
  
To establish an existence result for the Einstein equations, we need to investigate the constraints
imposed on the initial data by the Einstein equations. We propose here a new {\sl regularization scheme} that allows us to 
approximate any weakly regular initial data set by a sequence of smooth initial data sets,
while preserving the Einstein constraints. In addition, we establish the {\sl existence of weakly regular initial data 
sets,} in which each metric coefficient has the assumed regularity, only.  

An outline of this paper follows. 
In Section~\ref{mainR}, we define the class of weakly regular initial data and spacetimes of interest, 
and we provide a fully geometric reformulation of the Einstein constraint and evolution equations
and arrive at a proof of Theorem~1.1, above.
Then, in Section~\ref{sec2}, we reformulate the assumed regularity in certain (admissible, conformal, areal) coordinates
adapted to the symmetry. In Section~\ref{sec4}, 
we express the weak form of the Einstein equations as a system of partial differential equations
whose (generalized) solutions are understood in the sense of distributions. 
Section~\ref{sec45} contains some preliminary results and, in particular, includes a discussion of the regularization 
of initial data sets. 
Section~\ref{sec5} is concerned with local existence and compactness arguments, and 
takes advantage of the structure of the Einstein equations under the assumed symmetry. 
Finally, in Section~\ref{sec56} we analyze the global geometry of the constructed spacetime, 
and complete the proof of Theorem~\ref{theorem62}, above. 
 
%===============================================================================

\section{Geometric formulation}
\label{mainR}

\subsection{Weakly regular $T^2$--symmetric Riemannian manifolds}
\label{sec210}

All topological manifolds\footnote{All manifolds in this paper are assumed to be Hausdorff, orientable, connected, and paracompact.}
under consideration are of class $C^\infty$, 
that is, are defined by local charts such that the overlap maps are of class $C^\infty$. 
On the other hand, metric structures under consideration have {\sl low regularity,}  
specified in the course of our analysis. 
The Lie derivative $\Lcal_Z h$ of a measurable and locally integrable $2$-tensor 
$h$ (on a differentiable manifold) is defined in the weak sense, for any $C^1$ vector fields $X,T,Z$, by  
\be
\label{284}
(\Lcal_X h) (T,Z) := X(h(T,Z)) - h(\Lcal_X T,Z) - h(T, \Lcal_X Z),  
\ee
in which the last two terms are (classically defined as) locally integrable functions (that is, in $L^1_\loc$),
but the first one is defined in a weak sense, only, as is now explained. 
Namely, given a measurable and locally integrable function $f$, its differential $df$ is defined 
in the distributional sense by using local coordinate charts, 
then $df$ is extended to apply in a weak sense to $C^1$ vector fields 
and, finally, one sets $X(f) := df(X)$.  
 
Throughout, we use a standard notation for Lebesgue and Sobolev spaces, 
and we begin several definitions. 

\begin{definition} 
\label{def21}
A {\bf weakly regular $T^2$--symmetric Riemannian manifold} $(\Sigma, h)$
is a compact, $C^\infty$ differentiable
$3$-manifold endowed with a tensor field, enjoying the following properties:

\begin{enumerate}

\item {\bf Riemannian structure.} The field $h$ is a Riemannian metric in $L^\infty$,   
whose associated volume form has its coefficient bounded below (in any given smooth frame of $\Sigma$).

\item {\bf Symmetry property.} 
The Riemannian manifold $(\Sigma,h)$ is invariant under the action of the Lie group $T^2$ generated by 
two (smooth, linearly independent, commuting) Killing fields $X,Y$ 
(with closed orbits) satisfying, therefore, in particular   
\be
\label{ass:t2sym} 
\Lcal_X h =0, \qquad \Lcal_Y h = 0, 
\ee 
understood in the weak sense \eqref{284}. 

\item {\bf Regularity on the orbits.} The functions $h(X,X)$, $h(X,Y)$, and $h(Y,Y)$ 
belong to the Sobolev space $H^{1}(\Sigma;h)$, and the area  $\Rb$ of the orbits of symmetry 
defined by 
\be
\label{ass:rconstant}
\Rb^2 := h(X,X) \, h(Y,Y) - h(X,Y)^2, 
\ee  
which then lies in $W^{1,1}(\Sigma;h)$, actually belongs to $W^{1,\infty}(\Sigma)$.

\item {\bf Regularity on the orthogonal of the orbits.}
There exists a {\rm (smooth)} vector field $\Theta$ defined on $\Sigma$ such that $(X,Y,\Theta)$ 
forms 
a frame of commuting vector fields ($\Lcal_X \Theta = \Lcal_Y \Theta = 0$)
for which, by introducing
 the (non-smooth!) vector field 
\be
\label{395}
Z := \Theta + a \, X + b \, Y, \qquad Z \in \big\{ X,Y \big\}^\perp,  
\ee
for some real functions $a,b$, 
the regularity $h(Z,Z) \in W^{1,1}(\Sigma;h)$ holds with\footnote{One can easily check  
that, in fact, \eqref{minn} is a consequence of the assumed regularity and symmetry.} 
\be
\label{minn}
\inf_\Sigma h(Z,Z)>0.
\ee
\end{enumerate} 
\end{definition}

Observe that the existence of the (non-vanishing) 
commuting vector fields $(X,Y,\Theta)$ implies that $\Sigma$ is diffeomeorphic to 
the $3$-torus $T^3$. (See \cite{Chrusciel}.) 
In the context of Definition~\ref{def21},
we refer to the triple $(X,Y,Z)$ as an {\bf adapted frame} on $\Sigma$. 
The above definition is fully geometric, 
as it is easily checked (using \eqref{combi}, below) that it does not depend 
 on the specific choice of Killing fields within the generators of the $T^2$-symmetry. 
We emphasize that {\sl no regularity is required} on the derivatives of the ``cross-terms'' $h(X,\Theta)$ and $h(Y,\Theta)$. 
On the other hand, since $h$ is a Riemannian metric the definition \eqref{ass:rconstant} yields a positive function
 $\Rb^2$
and,  
since $\Rb$ is a continuous function defined on a compact set,  
\be
\label{minR}
\min_\Sigma \Rb > 0. 
\ee 
The strict positivity conditions \eqref{minn} and \eqref{minR}
ensure that the isomorphism~$^\sharp$ (and the isomorphism~$^\flat$, respectively)
which transforms covectors into vectors (and vice-versa, resp.) 
corresponds to a multiplicative operator with $L^\infty$ coefficients.  

To fully describe the class of initial data sets of interest,  we need to consider Riemannian manifolds 
endowed with a $2$-covariant tensor field
which, later, will stand for the second fundamental form describing  
the extrinsic geometry of the initial slice. 

\begin{definition} 
\label{def:t2tri}
A {\bf weakly regular $T^2$--symmetric triple} $(\Sigma,h,K)$
is a weakly regular $T^2$--symmetric Riemannian manifold $(\Sigma,h)$, with adapted frame $(X,Y,Z)$,  
satisfying the following conditions: 
\begin{enumerate}

\item {\bf Regularity property.} 
The field $K$ is a symmetric $2$-tensor field on $\Sigma$ enjoying the regularity
\be
K(Z,Z) \in L^1(\Sigma;h)
\ee
and, for all pair $(U,V) \neq (Z,Z)$ with $U,V \in \big\{X,Y,Z\big\}$,  
\be
K(U,V) \in L^2(\Sigma;h).  
\ee

\item {\bf Symmetry property.} The field $K$ is invariant under the action of the Lie group $T^2$ generated by $(X,Y)$: 
\be
\label{assum:t2symk} 
\Lcal_X K = \Lcal_Y K = 0, 
\ee
understood in the weak sense \eqref{284}. 

\item {\bf Additional regularity.} The following trace of $K$  on the orbuts of symmetry 
\be 
\label{assum:R}
\aligned
\Trace^{(2)} (K) 
:&= h^{-1}(X,X) \, K(X,X) + 2 h^{-1}(X,Y) \, K(X,Y) + h^{-1}(Y,Y) \, K(Y,Y) 
\\
& \in L^\infty(\Sigma), 
\endaligned 
\ee   
in which each product involves an $H^1$ function and an $L^2$ function, 
and $h^{-1}$ denotes the inverse metric. 
\end{enumerate}
\end{definition}
 
As far as solutions to the Einstein equations (which are not assumed yet) are concerned, 
the additional regularity in \eqref{assum:R} corresponds to a Lipschitz continuous bound
on the time derivative of the area $\Rb$ of the orbits of symmetry, and therefore is a natural 
regularity condition in view of the assumption $\Rb \in W^{1,\infty}$ made in Definition~1.1. 
Importantly, the weak regularity described in the above two definitions is precisely the one suitable to 
deal with the constraints associated with Einstein's field equations, 
as will be shown in Section~\ref{defcurv}, below.

%----------------------------------------------------------------------------------------------------

\subsection{Weakly regular $T^2$--symmetric Lorentzian manifolds}
\label{2222}

We now introduce a class of spacetimes endowed with a time function.  

\begin{definition} 
\label{def:wrt2rm000}
An {\bf $L^\infty$ Lorentzian structure} is a $(3+1)$-dimensional manifold $\Mcal$ (possibly with boundary) 
endowed with a Lorentzian metric $g$ in $L_\loc^\infty(\Mcal)$
whose volume form is $L^\infty_\loc$ bounded below. 
\end{definition} 

\begin{definition} 
\label{def:wrt2rm00}
On an $L^\infty$ Lorentzian structure $(\Mcal, g)$ with topology $\Mcal \simeq I \times \Sigma$ 
($I$ being a non-empty interval and $\Sigma$ a compact $3$-manifold), 
an {\bf $L^\infty$ normal foliation compatible with the metric $g$}  
is determined by  a function $t \in C^\infty(\Mcal)$ which 
defines a foliation by hypersurfaces $\Sigma_t$ of constant time $t$, $C^\infty$ diffeomorphic to $\Sigma$, 
and satisfies the following conditions  uniformly within any compact subset of $I$: 
\begin{enumerate}

\item The normal vector field $T:=\nabla t$ is timelike,  

\item there exists a family of Riemannian metrics $h(t)$ defined on $\Sigma_t$ and 
belonging to $L^\infty$, together with their inverse $h^{-1}(t)$, and 

\item the metric reads  
\be
\label{922}
g =g(t) = - n(t)^2 \, dt^2 + h(t), \qquad h(t) = (h_{ij}(t)),  
\ee 
with $n(t) := (-g(T,T))^{1/2}>0$ is referred to as the lapse function and 
is assumed to be bounded below in  
$L^\infty(\Sigma)$.
\end{enumerate} 
\end{definition}
 
\begin{definition} 
\label{def:wrt2rm22}
Let $\Mcal$ be a $C^\infty$ differentiable 
manifold endowed with 
an effective action $G: \Mcal \times T^2 \to \Mcal$ of the Lie group $T^2$, 
and 
a $C^\infty$ function $t: \Mcal \to \RR$ whose level surfaces are smooth embedded 
hypersurfaces $\Sigma_t$.  
This $t$-foliation is said to be {\bf compatible with the $T^2$ symmetry} if 
$G$ induces an action on each leave of the foliation, that is,  
the restriction of $G$ to $\Sigma_t\times T^2$ takes its image in $\Sigma_t$ and defines an action of $T^2$ on $\Sigma_t$. 
\end{definition}

\begin{definition} 
\label{def:wrt2rm}
A {\bf weakly regular $T^2$--symmetric Lorentzian manifold} with spatial topology $\Sigma$   
is a $4$-manifold  $(\Mcal,g)$ 
endowed with an $L^\infty$ Lorentzian structure
and an effective action of the Lie group $T^2$ satisfying the following property. 
For any normal foliation $\Sigma_t$ ($t \in I$) compatible with $g$ and determined by a $T^2$--symmetric compatible, 
time-function $t$, the following conditions hold:  
\begin{enumerate}

\item {\bf Symmetry property for the lapse.} 
For any two vector fields $X,Y$ generating the $T^2$--action  
\be
\label{prope:taue}
\Lcal_X ( n^2 ) = \Lcal_Y (n^2 ) = 0 
\ee
in the weak sense, where $n$ is the lapse function associated with $t$. 

\item {\bf Timelike regularity.} 
There exists a {\rm smooth} vector field $\Theta$ defined 
on $\Mcal$ such that $(X,Y,\Theta)$ is orthogonal to $T$ and 
forms a frame of commuting vector fields (tangent to the slices). 
Introducing the vector field 
\be
\label{395-2}
Z := \Theta + a \, X + b \, Y, \qquad Z \in \big\{ T, X,Y \big\}^\perp, 
\ee  
one imposes that  the components of the field $\Lcal_T(h(t))$ in the frame $(X,Y,Z)$ belong to
$L^1(\Sigma; h(t))$,  uniformly in $t$ in any compact subset of $I$.   

\item {\bf Spacelike regularity.}
Finally, for each $t$ (and uniformly within any compact subset of $I$), the triple $(\Sigma_t, g(t),K(t))$, 
with 
\be
\label{92} 
K(t)(U,V) := - {1 \over 2n} \, (\Lcal_T g) (U,V), \qquad U,V \in \big\{X,Y,Z\big\} 
\ee
is a weakly regular $T^2$--symmetric triple in the sense of Definition~\ref{def:t2tri}
such that the restriction of $(X,Y,Z)$ to $\Sigma_t$ is an adapted frame.
The implied constants are assumed to be bounded on each compact subset of $I$, and one can define 
the corresponding area function 
\be
\label{ass:rconstant-time}
R^2 := g(X,X) \, g(Y,Y) - g(X,Y)^2.  
\ee  
 
\item{\bf Conformal metric regularity.}  Finally, one also imposes that   
\be
\label{reg-rho}
\rho^2(t) := {h(Z,Z) \over n(t)^2} = - {g(Z,Z) \over g(T,T)} \in W^{1,\infty}(\Sigma).
\ee
\end{enumerate}
\end{definition} 

Spacetimes satisfying the above definition will indeed be constructed in the present work 
by solving the initial value problem for the Einstein equations
from initial data sets satisfying Definitions~1.1 and 2.2. 
We will first construct one specific foliation along which the regularity conditions in the definition are satisfied
and, second,  deduce the same regularity along general foliations. 
Note also that the function $\rho^2= - g(Z,Z)/g(T,T)$ in \eqref{reg-rho}
determines the conformal quotient metric and the wave operator relevant later in this paper to deal with the evolution part of 
the Einstein equations. 
Observe that the definition \eqref{92} together with (2.1)
allow us to write, for all $e_i,e_j\in \big\{X,Y\big\}$, 
$$
\aligned
K(t)(e_i,e_j)  & = -{1 \over 2 n} \, T\big( g(t)(e_i,e_j)\big),
\qquad 
K(t)(Z,Z) = -{1 \over 2 n} \, T\big( g(t)(Z,Z)\big),
\\
K(t)(Z,e_i) & = K(t)(e_i,Z)= \frac{1}{2n} g(e_i, \mathcal{L}_Z T),
\endaligned
$$
which, by our definition, are $L^1$ or $L^2$ functions on each slice. 

%--------------------------------------------------------------------------------------------
 
Let us now discuss the condition \eqref{assum:R} which, as we claimed, 
is equivalent to a sup-norm bound on the first-order time-derivative of the function $R$, 
provided we specify the timelike direction in which to compute the derivative. 
Precisely, recall that when the data are {\sl sufficiently regular,} at least, 
and whenever $\Sigma$ is identified with a hypersurface with unit timelike normal $N$, then 
$\Trace^{(2)} (K)$ is given by 
\be
\label{299}
\Trace^{(2)} (K) = N (\ln R ). 
\ee
In Lemma~\ref{ABCD}, below, we sate this formula within the low regularity class of interest. 
Let us begin with an analogous result first and observe that the identity \eqref{5678}, below, 
holds almost everywhere, only, unless additional regularity is imposed. 

\begin{lemma}[Normal derivative of the volume element] 
\label{933} 
Let $(\Mcal,g)$ be a weakly regular $T^2$-symmetric Lorentzian manifold with time function $t$, and 
let $K$ be the second fundamental form associated with a slice of constant time $t$ which is
defined within an adapted frame as presented in Definition \ref{def:wrt2rm}. 
Then, 
the trace $\Trace (K) := h^{ij} \, K_{ij}$ of the second fundamental form 
is the time derivative of (a nonlinear function of) the determinant $h:=\det h_{ij}$ and, more precisely, 
\be
\label{5678} 
\Trace (K) := N \big( \ln \sqrt{h}\big) \qquad \text{ almost everywhere in } \Mcal. 
\ee 
\end{lemma}

\begin{proof} For completeness, we provide a proof of this result. 
Without loss of generality, assume that $N= {\del \over \del t} $ and
that $e_i$ and $e_j$ are commuting vector fields tangent to $\Mcal$. On one hand, we have
$$
\Trace (K)=h^{ij}K_{ij}=-h^{ij}g(N,\nabla_{e_i}e_j)
=-h^{ij}g_{\alpha \beta} N^\alpha \Gamma^\beta_{ij}
=\frac{1}{2} \, h^{ij}h_{ij,t},
$$
where $, t$ denotes a partial derivative. 
On the other hand, we can compute $\det h_{ij}$ by developing around the line with index $i$, that is, 
$\det h_{ij}=H^{(i)j}h_{(i)j}$, 
where $(i)$ is a fixed index (no sum over $i$) and $H^{ij}$ denote the cofactor of $h_{ij}$. In particular, 
one sees from the last expression that $h$ seen as a function of the $h_{ij}$ satisfies
$\frac{\del h }{\del h_{ij}}=H^{ij}=h h^{ji}$, 
and we find 
\[
N(\ln \sqrt{h})=\frac{1}{2h}\frac{\del h }{\del h_{ij}}h_{ij,t}=\frac{1}{2} \, h^{ij}h_{ij,t}. 
\qedhere
\]
\end{proof}

We are now in a position to introduce the second fundamental form 
associated with the orbits of symmetry (of any $T^2$-symmetric weakly regular Riemannian manifold). 
Observe that, again due to our low regularity assumptions, 
a {\sl family} of second fundamental forms {\sl defined almost everywhere,} only, can be introduced here. 

\begin{definition}
\label{wsec}
Let $(\Mcal,g)$ be a weakly regular $T^2$--symmetric Lorentzian manifold 
with adapted frame $(X,Y,Z)$. 
Then, the {\bf weak version of the second fundamental form in the $Z$-direction} 
associated with the orbits of symmetry is defined (almost everywhere, only) as the tensor field  
$$
\chi_{ij} := -\frac{1}{2} \, \big( h(t)(Z,Z) \big)^{-1/2} \gamma^a_i \, \gamma^b_j \,  Z\big( h(t)_{ab} \big)
\qquad \text{ almost everywhere in } \Mcal,  
$$ 
where $\gamma^a_i$ is the projector on the space generated by $X, Y$,    
with $i,j=X,Y,Z$ and $a,b=X,Y$.
Similarly, the {\bf weak version of the second fundamental form in the $T$-direction} 
associated with the orbits of symmetry  is defined (almost everywhere, only) as the tensor field  
$$
\kappa_{ij} := -  {1 \over 2} \, \big( h(t)(T,T)\big)^{-1/2} \gamma^a_i \, \gamma^b_j \, T \big( h(t)_{ab} \big) 
\qquad \text{ almost everywhere in } \Mcal.  
$$
\end{definition}

Recall that, by definition, $\gamma^a_i = h^a_i - h(Z,Z)^{-1} \, Z^a Z_i$. 
Thus, in view of Definition~\ref{def21},
 the components of $\gamma^a_i$ in the frame $(X,Y,Z)$
 are in $L^\infty(\Sigma, h(t))$, 
and it follows from the $H^1$ regularity of $h_{ab}$ 
that $\chi$ belongs to $L^2(\Sigma; h(t))$, uniformly in the time variable on any compact time interval. 
The same is true for the components of $\kappa$, so
\be
\label{205}
\chi_{ij}, \, \kappa_{ij} \in L^2(\Sigma; h(t)) \qquad \text{ locally uniformly in $t$.}  
\ee

Similarly to Lemma \ref{933}, we can prove the following statement whose proof is omitted.

\begin{lemma}[Normal derivative of the area element]
\label{ABCD}
If $(\Mcal,g)$ is a weakly regular Lorentzian manifold with adapted frame $(X,Y,Z)$, 
then $\Trace^{(2)} (\chi)$ is determined by the (normalized) $Z$-derivative
of the area element:  
\be
\label{305A}
\Trace^{(2)} (\chi) := h^{ab} \, \chi_{ab} = \big( h(t)(Z,Z) \big)^{-1/2} Z  \big( \ln R\big) \qquad \text{ almost everywhere in } \Mcal. 
\ee
Similarly, $\Trace^{(2)} (\kappa)$ is determined by the (normalized) $T$-derivative
of the area element: 
\be
\label{305B}
\Trace^{(2)} (\kappa) := h^{ab} \, \kappa_{ab} = -\frac{1}{n} T  \big( \ln R \big) 
\qquad \text{ almost everywhere in } \Mcal.  
\ee 
\end{lemma}

%-----------------------------------------------------------------------------------------------------------

\subsection{Weak version of Einstein's constraint equations}  
\label{defcurv}

\subsubsection*{Christoffel symbols}

The standard definition of the Christoffel symbols involves certain nonlinear terms that {\sl cannot be defined}
even as distributions, under the weak regularity conditions 
introduced in Sections~\ref{sec210} and \ref{2222}, above.
A fortiori, it is unclear whether any component of the curvature could be well-defined. 
In fact, for general manifolds, the minimal regularity assumption for the curvature to make sense as a tensor distribution
is known to be $H^1 \cap L^\infty$, 
as discussed in \cite{LeFlochMardare}. In the present paper, we assume a {\sl weaker regularity}
for {\sl certain components} of the metric and
take advantage of the symmetry of the spacetimes under consideration. 
We are then able to {\sl re-formulate} the Einstein's constraint and evolution equations 
so that, for weakly regular $T^2$--symmetric spacetimes, 
all of the geometric objects of interest are well-defined in a suitably weak sense, 
and our definitions reduce to the classical ones when sufficient regularity is assumed. 

First of all, we emphasize that a fully geometric standpoint 
based on an adapted frame, as we propose in this work, is required.  
Indeed, under the conditions stated in Proposition \ref{prop:regchri1}, below, 
one {\sl cannot define} the Christoffel symbol $\Gamma_{\theta\theta}^\theta$, as 
this would involve products of the form $h^{i\theta} \, \del_\theta h_{\theta b}$ (with $b=x,y$),
which cannot be defined in the weak sense when $h_{\theta b} \in L^\infty(\Sigma)$. 
This is why we introduce a (non-smooth) adapted frame $(X,Y,Z)$ (as defined in \eqref{395} or \eqref{395-2})
in which the problematic terms {\rm vanish} by construction since $Z$ is orthogonal to $X,Y$. 

A preliminary remark is in order. The vector field $Z$ introduced 
is {\sl not smooth} so that it does not apply to {\sl general} functions of class $L^1$, 
but yet can be applied to {\sl $T^2$--symmetric} functions,  by defining 
$$
Z(f) := \Theta(f) \qquad \text{ for $T^2$--symmetric $f \in L^1(\Sigma)$,} 
$$
in which the right-hand side involves the $C^\infty$ vector field $\Theta$ determining 
the frame $(X,Y,\Theta)$, as in Definition~\ref{def21}.  
In the following, this observation will be used without further notice.

\begin{proposition}[Regularity of the Christoffel symbols in an adapted frame] 
\label{prop:regchri1}
Let $(\Sigma,h)$ be a weakly regular $T^2$--symmetric Riemannian manifold with adapted frame $(X,Y,Z)$
and, for all $i=X,Y,Z$ and $a,b=X,Y$, consider the formal expressions $\Gamma^i_{jk}$ defined by  
$$
\aligned
& \Gamma_{ab}^c :=0, \qquad \Gamma^Z_{ab} := - \frac{1}{2} \, h^{ZZ} \, Z(h_{ab}), 
\qquad 
\Gamma^Z_{aZ} = \Gamma^Z_{Za} := 0, 
\\
& \Gamma^b_{aZ} = \Gamma^b_{Za} := \frac{1}{2} \, \big( h^{bX} \, Z(h_{aX}) + h^{bY} \, Z(h_{aY}) \big), 
\\
& \Gamma^X_{ZZ} = \Gamma^Y_{ZZ} : = 0,
\qquad \Gamma^Z_{ZZ} := \frac{1}{2} \, h^{ZZ} \, Z(h_{ZZ}). 
\endaligned
$$
Then, for $(j,k) \neq (Z,Z)$, the symbols $\Gamma^i_{jk}$ are well-defined as functions in $L^2(\Sigma; h)$, 
while $\Gamma^Z_{ZZ}$ is well-defined as a function in $L^1(\Sigma; h)$
and, in addition,
\be
\label{930} 
\Gamma_{aZ}^a \in L^\infty(\Sigma). 
\ee
Moreover, if  $(\Sigma,h)$ is sufficiently regular, then these functions 
$\Gamma^i_{jk}$ coincide with the standard Christoffel symbols (in the frame $X,Y, Z$) associated with the metric $h$.
\end{proposition}

\begin{proof} We observe that the given expressions do make sense and have the claimed regularity, as 
follows immediately from Definition~\ref{def21}.  The additional regularity on $\Gamma_{aZ}^a$ is a
direct consequence of Lemma~\ref{ABCD}. 
On the other hand, when the data are sufficiently regular and since $X,Y,Z$ commute, 
the Christoffel symbols can be computed in a standard way.
For $i=X,Y,Z$ and $a,b=X,Y$ and by using the symmetry properties of $h$ 
and the orthogonality condition $Z \in \big\{ X,Y \big\}^\perp$, we find 
\be
\aligned 
\Gamma^i_{ab} = & \frac{1}{2} h^{ij} 
\left( h_{aj,b}+h_{jb,a}-h_{ab,j} \right) 
=-\frac{1}{2} h^{iZ}Z(h_{ab}), 
\\
\Gamma^{Z}_{aZ} = &\frac{1}{2} h^{Zj} 
\left( h_{aj,Z}+h_{jZ,a}-h_{aZ,j} \right)=0, 
\\
\Gamma^b_{aZ} = & \frac{1}{2} h^{bj} \left( h_{aj,Z}+h_{jZ,a}-h_{aZ,j} \right)
=\frac{1}{2} \, \big( h^{bX} \, Z(h_{aX}) + h^{bY} \, Z(h_{aY}) \big),
\\
\Gamma^a_{ZZ} = & \frac{1}{2}h^{aj}\left(2h_{jZ,Z}-h_{ZZ,j} \right)=0,
\\
\Gamma^{Z}_{ZZ} = & \frac{1}{2}h^{Zj}\left( 2h_{Zj,Z}-h_{ZZ,j} \right)=\frac{1}{2} \, h^{ZZ} \, Z(h_{ZZ}).
\endaligned 
\ee 
Hence,  when the data are sufficiently regular,
$\Gamma^i _{jk}$  do coincide with the standard Christoffel symbols. 
\end{proof}

%--------------------------------------------------------------------------------------------------

\subsubsection*{Weak version of the second fundamental forms of the orbits}

It follows from Proposition~\ref{prop:regchri1} that, for weakly regular $T^2$-symmetric Riemannian manifolds,
the main obstacle in defining the curvature tensor in the sense of distributions (in the frame $X,Y,Z$) comes from 
the component $\Gamma^Z_{ZZ}$ which is only in $L^1(\Sigma;h)$ and, therefore, can not be multiplied 
by Christoffel coefficients ---which are solely in $L^2(\Sigma;h)$ or $L^1(\Sigma;h)$. 
Fortunately, as we check it below, for {\sl sufficiently regular} $T^2$--symmetric spacetimes and within the 
expression of the curvature, the formal products involving such coefficients {\sl cancel out.} 
This suggests to {\sl redefine} the Ricci scalar by a {\sl new formula}  taking into account this cancellation, 
as we now explain.

We now rely on Definition~\ref{wsec} above, especially on the tensor $\chi$. 
Indeed, to arrive at a weak version of the curvature, we now take advantage of a {\sl $(2+1)$--decomposition} into, on one hand,
intrinsic and extrinsic contributions of the orbits of symmetry and, on  the other hand, a remainder containing 
the contributions of the orthogonal to the orbits. 
The orbits of symmetry being flat by assumption, their intrinsic geometry is trivial, while 
their extrinsic geometry is characterized by their second fundamental form, as introduced in Definition \ref{wsec} above. 
Once again, we take into account the symmetry assumptions in order to go {\sl below} 
the standard regularity assumption in \cite{LeFlochMardare}. 

%--------------------------------------------------------------------------------------------------

\subsubsection*{Weak version of the Hamiltonian constraint}

To write down a weak form of the Ricci scalar, denoted by $R^{(3)}$, we need first to {\sl redefine the component}
$R_{Z Z}^{(3)}$ of the Ricci tensor, as follows. 
The following definition will be fully justified below in the proof of Proposition~\ref{Equiv}, 
where terms of the form $\pm \Gamma_{ZZ}^Z \, \Gamma_{ZZ}^Z$ will be checked to cancel out
and, for that reason, do not arise in the following definition. 

\begin{definition} 
Let $(\Sigma,h)$ be a weakly regular $T^2$--symmetric Riemannian manifold 
with adapted frame $(X,Y,Z)$.   
The {\bf weak version of the Ricci curvature in the direction $(Z,Z)$} is defined as 
\be
\label{eq:wvriczz}
R^{(3)}_{ZZ} := -Z( \Gamma^a_{\phantom{a}aZ})
                +\Gamma^a_{\phantom{a}aZ}\Gamma^Z_{ZZ} 
                -\Gamma^a_{bZ} \, \Gamma^b_{aZ}, 
\ee
where the first term of the right-hand side is defined in the weak sense, only, 
and the other terms are products of the type $L^\infty L^1$ or $L^2 L^2$.
\end{definition}

Based on the above definition, we can now formulate the Hamiltonian constraint in the weak sense.
First, we observe that, in viex of Gauss equation 
and the flatness of the orbits of symmetry,
\be 
\label{Gauss44} 
R^{(3)}=2 h(Z,Z)^{-1}R^{(3)}_{ZZ} + |\chi|^2 - \big(\Trace^{(2)} ( \chi) \big)^2 \qquad \text{\sl for sufficiently regular metrics} 
\ee
(with $\chi$ given in Definition~\ref{wsec}). 
However, 
in our setting, the Ricci curvature term $R^{(3)}_{Z Z}$ is defined in the weak sense \eqref{eq:wvriczz}, only, 
and it {\sl does not make sense} to multiply it by the factor $h(Z,Z)^{-1}$.  
This motivates us to introduce the following {\sl normalized} version. 

\begin{definition}
\label{def:whe}
Let $(\Sigma,h)$ be a weakly regular $T^2$--symmetric Riemannian manifold with adapted frame $(X,Y,Z)$. 
Then,  
the {\bf weak version of the normalized scalar curvature} of $(\Sigma,h)$ is defined as 
$$
R_\text{norm}^{(3)} := 2 \, R^{(3)}_{ZZ} + h(Z,Z) \, \big( |\chi|^2 - (\Trace^{(2)}(\chi))^2 \big),
$$
in which $\chi$ is the weak version of the second fundamental form in the $Z$-direction. 
In addition, a weakly regular $T^2$--symmetric triple $(\Sigma,h,K)$ is said to satisfy  
the {\bf weak version of the Hamiltonian constraint} 
if 
\be 
\label{weak-Hamilton}
R_\text{norm}^{(3)} + h(Z,Z) \, \left( \big( \Trace ( K) \big)^2 - |K|^2  \right) = 0.
\ee 
\end{definition} 

\begin{remark}
The weak form of the Hamiltonian constraint 
is independent of the specific choice of Killing vectors $X,Y$. Indeed, the orthogonal complement 
to the orbits is uniquely determined and, thus, the vector field $Z$ is uniquely determined
up to multiplication by a $C^\infty$ function, which does not change the set of solutions to \eqref{weak-Hamilton}.
\end{remark} 

\begin{proposition}[Equivalence with the classical definition] 
\label{Equiv}
Let $(\Sigma,h,K)$ be a weakly regular $T^2$--symmetric triple.
If $h,K$ are sufficiently regular, then $(\Sigma,h,K)$ satisfies the weak version
of the Hamiltonian constraint equation 
(in the sense
of Definition~\ref{def:whe}) 
if and only if it satisfies the constraint equation \eqref{ee:constham}
in the classical sense. 
\end{proposition}

\begin{proof} In view of \eqref{Gauss44}, 
the result follows if, assuming now {\sl sufficient regularity,}
we can prove that the classical definition for 
$R^{(3)}_{ZZ}$ coincides with the one adopted in Definition \ref{wsec}. 
Namely, computing $R_{ZZ}^{(3)}$ in the classical sense from the trace of the Riemann curvature, 
we find 
$R^{(3)}_{ZZ} 
= \Omega_1 + \Omega_2$ 
with (a coma indicating differentiation) 
$$
\Omega_1 := \Gamma^i_{ZZ,i}-\Gamma^i_{iZ,Z}, 
\qquad 
\Omega_2 := \Gamma^j_{ji}\Gamma^i_{ZZ}-\Gamma^j_{Zi}\Gamma^i_{jZ}. 
$$
On one hand, since $\Gamma^a_{ZZ,a}=0$ we have 
$$
\Omega_1=\Gamma^a_{ZZ,a}+\Gamma^Z_{ZZ,Z}-\Gamma^Z_{ZZ,Z}-\Gamma^a_{aZ,Z}
=-Z(\Gamma^a_{aZ}). 
$$
in which we have cancelled out the  terms $\pm \Gamma^Z_{ZZ,Z}$.
On the other hand, we have 
$$ 
\aligned
\Omega_2
= & \, \Gamma^{Z}_{ZZ}\Gamma^Z_{ZZ}+\Gamma^{a}_{aZ}\Gamma^Z_{ZZ}
+\Gamma^Z_{Za}\Gamma^a_{ZZ}+\Gamma^a_{ab}\Gamma^b_{ZZ} 
\\
& \, -\Gamma^{Z}_{ZZ}\Gamma^Z_{ZZ}-\Gamma^Z_{Zb}\Gamma^b_{ZZ}
-\Gamma^a_{ZZ}\Gamma^Z_{aZ}-\Gamma^a_{Zb}\Gamma^b_{aZ} 
\\ 
= & \, \Gamma^a_{aZ}\Gamma^Z_{ZZ}-\Gamma^a_{Zb}\Gamma^b_{aZ},
\endaligned
$$
where we have used $\Gamma^Z_{aZ}=\Gamma^a_{ZZ} = 0$ and cancelled out the products  
$\pm \Gamma^{Z}_{ZZ}\Gamma^Z_{ZZ}$.
This leads us to \eqref{eq:wvriczz}, as claimed.
\end{proof}

%-------------------------------------------------------------------------------------------------------------

\subsubsection*{Weak version of the momentum constraints}

Next, we introduce the following definition. 

\begin{definition} 
\label{def:wme}
A weakly regular $T^2$--symmetric triple  $(\Sigma,h,K)$ is said to satisfy  
the {\bf weak version of the momentum constraints} if 
the equations   
\be
\label{weak-momentum} 
\aligned 
Z( \Trace^{(2)} K) - \, h(Z,Z)^{1/2} \, \Trace^{(2)} ( \chi)  \, K_Z^Z - \Gamma^a_{Z b}K_a^b
& = 0, 
\\
Z\big( h(Z,Z)^{1/2}K^Z_a \big) - \Gamma^b_{bZ}K^Z_a & = 0, 
\qquad a = X,Y, 
\endaligned 
\ee
hold in the weak sense, with  $\Trace^{(2)} (K )=  \Trace( K) - K_{Z}^Z$.   
\end{definition}

Observe that the second set of equations in \eqref{weak-momentum}
has been {\sl weighted} by the scalar 
$h(Z,Z)^{-1}$ ---in order for it to be well-defined in a weak sense, while 
the first equation has a different homogeneity in $Z$. 

\begin{proposition}[Equivalence with the classical definition] 
\label{Equiv-2}
Let $(\Sigma,h,K)$ be a weakly regular $T^2$-symmetric triple.
If $h,K$ are sufficiently regular, then $(\Sigma,h,K)$ satisfies the weak version
of the momentum constraint equations 
(in the sense
of Definition~\ref{def:wme}) 
if and only if it satisfies the constraint equations \eqref{ee:constmom}
in the classical sense. 
\end{proposition}

\begin{proof} Assuming sufficient regularity and
 that the momentum constraint equations hold in the classical sense, i.e. 
$$
\nabla^{(3)j}K_{ij}-\nabla_i^{(3)}tr K=0, 
$$
we begin by computing $\nabla^{(3)j}K_{Zj}$ in an adapted frame:  
$$
\aligned 
\nabla^{(3)j}K_{Zj}
= & \, K^Z_{Z,Z}+K^a_{Z,a}-\Gamma^i_{jZ}K_i^j+\Gamma^j_{ji}K_Z^i 
\\
= & \, K^Z_{Z,Z}-\Gamma^Z_{ZZ}K^Z_Z-\Gamma^a_{ZZ}K_a^Z-\Gamma^Z_{aZ}K^a_Z-\Gamma^a_{bZ}K^b_a 
\\
 & \, + \Gamma^{Z}_{ZZ}K^Z_Z+\Gamma^Z_{Za}K^a_{Z}+\Gamma^b_{bZ}K_Z^Z+\Gamma^b_{ba}K_Z^a
\\
= & \, K^Z_{Z,Z}-\Gamma^a_{bZ}K_{a}^b+\Gamma^b_{bZ}K_Z^Z,
\endaligned
$$ 
where we used $\Gamma^Z_{aZ}=\Gamma^a_{ZZ}=0$ 
and {\sl cancelled} (potentially problematic)  terms $\pm \Gamma^Z_{ZZ}$. 
Since $h(Z,Z)^{1/2}tr\chi=-\Gamma^b_{bZ}$, 
the momentum constraint equation in the $Z$-direction  
is equivalent  
to the first equation in \eqref{weak-momentum}.
For the remaining two momentum constraint equations, we write 
$$
\aligned 
\nabla^{(3)}_j K_a^j
& = K^j_{a,j}-\Gamma^i_{aj} K_i^j +\Gamma^j_{ji} K_a^i 
\\
& = Z(K_a^Z)-\Gamma^Z_{aZ}K_Z^Z - \sum_{(i,j) \neq (Z,Z) } \Gamma^i_{aj}K_i^j 
+\Gamma^Z_{ZZ}K_a^Z+\sum_{(i,j) \neq (Z,Z)}\Gamma^j_{ji}K^i_a 
\\
& = Z(K_a^Z)+\Gamma^Z_{ZZ}K_a^Z + \sum_{(i,j) \neq (Z,Z)} \Big( \Gamma^j_{ji}K^i_a - \Gamma^i_{aj}K_i^j \Big). 
\endaligned
$$ 

Recalling that the term involving $\Gamma^Z_{ZZ}$ is not well-defined
for weakly regular spacetimes, 
we multiply the above equations by $h(Z,Z)^{1/2}$ 
and expand the Christoffel symbol of the second term, in order to get 
$$
\aligned
& h(Z,Z)^{1/2}\nabla^{(3)}_j K_a^j
\\
& = h(Z,Z)^{1/2} \, \Big(  Z(K_a^Z)+\frac{1}{2} h(Z,Z)^{-1} Z(h(Z,Z))K_a^Z
      + h(Z,Z)^{1/2} \hskip-.3cm  \sum_{(i,j) \neq (Z,Z)} \Big( \Gamma^j_{ji}K^i_a - \Gamma^i_{aj}K_i^j \Big).
\endaligned
$$
Combining the first two terms on the right-hand side yields the second set of equations 
in \eqref{weak-momentum}, as expected. We conclude that \eqref{ee:constmom}
and \eqref{weak-momentum} 
are equivalent for sufficiently regular data.  
\end{proof}

%-----------------------------------------------------------------------------------------------------

\subsection{Weak version of Einstein's evolution equations}
\label{209} 

We are now in a position to discuss the Einstein equations. 
As before, we need first to examine the regularity of the Christoffel symbols,  now 
associated with a spacetime metric.

\begin{proposition}[Regularity of the Christoffel symbols in an adapted frame]
Let $(\Mcal,g)$ be a weakly regular $T^2$-symmetric Lorentzian manifold
with adapted frame $(T,X,Y,Z)$, time-function $t$, spacelike slices $\Sigma_t$,
and second fundamental form $K$,  
as introduced in Definition~\ref{def:wrt2rm}, and  
for all $a,b=X,Y$ define  
$$
\aligned 
& \Gamma^T_{ab} := -\frac{1}{n} \, K_{ab}, 
\qquad 
&& \Gamma^T_{ZZ} := -\frac{1}{n} \, K(Z,Z), 
\qquad 
&&&& \Gamma^T_{TZ} = \Gamma^T_{ZT} := \frac{1}{n}Z(n),
\\
& \Gamma^T_{Za} = \Gamma^T_{aZ} := -\frac{1}{n} \, K(\cdot, Z)_a, 
\qquad 
&& \Gamma^T_{Ta} = \Gamma^T_{aT} := 0,
\qquad 
&&&&\Gamma^{Z}_{TZ} = \Gamma^{Z}_{ZT} := -g^{ZZ} \, n \, K(Z,Z), 
\\
& \Gamma^Z_{aT} = \Gamma^Z_{Ta} := - n g^{ZZ} \, K(\cdot, Z)_a, 
\qquad 
&&\Gamma^{a}_{Tb} = \Gamma^{a}_{bT} := -g^{ac} \, n \, K_{bc}, 
\qquad 
&&&& \Gamma^a_{TZ} = \Gamma^a_{ZT} := n K(\cdot, Z)^a, 
\\
&\Gamma^a_{TT}: = 0,
\qquad 
&& \Gamma^T_{TT} := T(n)n^{-1}, 
\qquad 
&&&&\Gamma^Z_{TT} :=  n g^{ZZ} \, Z(n). 
\endaligned
$$
In addition, define $\Gamma^{i}_{kj}$ for $i,j,k=X,Y, Z$ 
as in Proposition~\ref{prop:regchri1}, but with $h$ replaced by $h(t)$.
Then, these functions are well-defined, have the regularity
\be
\Gamma^T_{ZZ}, \, \Gamma^{Z}_{TT}, \, \Gamma^T_{TT} \in L^1(\Sigma_t),
\qquad \quad 
 \Gamma^T_{ab} \in L^2(\Sigma_t),
\quad \qquad 
\Gamma^T_{Zi}, \, \Gamma^Z_{iT}, \, \Gamma^i_{ZT} \in L^\infty(\Sigma_t),
\ee
uniformly in the time variable.  Furthermore, the following 
linear combinations of Christoffel symbols are better behaved: 
\be
\label{more5} 
\Gamma^a_{Ta} = \Gamma^a_{aT} \in L^\infty(\Sigma_t),
\ee
\be
\label{more5b} 
\Gamma^T_{TT} - \Gamma^Z_{ZT} \in L^\infty(\Sigma_t), 
\qquad 
\Gamma^Z_{ZZ}-\Gamma^T_{ZT} \in L^\infty(\Sigma_t). 
\ee
Finally, if $(\Mcal,g)$ is sufficiently regular, then the above definition coincides 
with the standard definition of the Christoffel symbols.
\end{proposition}

\begin{proof} In the frame $(T, X,Y,Z) = (e_0, e_1, e_2,e_3)$ (which is not induced by coordinates)
and provided the data are sufficiently regular, 
we have the classical definition: 
$$ 
\label{eq:chrisni}
\aligned
& \Gamma^\alpha_{\beta\gamma} = \frac{1}{2} \, g^{\alpha \delta} \left( g_{\beta \delta,\gamma} + g_{\gamma \delta,\beta}
- g_{\beta \gamma,\delta} + c_{\delta \beta \gamma} + c_{\delta \gamma \beta} + c_{\beta \gamma \delta} 
\right), 
\\
& c_{\beta \gamma\delta}:= [e_\beta,e_\gamma]_\delta = g_{\delta \rho} \, [e_\beta, e_\gamma]^\rho. 
\endaligned
$$
Except for \eqref{more5} and \eqref{more5b}, the regularity properties stated in the proposition follow immediately 
from our definitions. Then, 
\eqref{more5} is an immediate consequence of Lemma \ref{933} and of the assumptions on $tr^{(2)} K$. 
To derive \eqref{more5b}, we use the regularity assumption
made on $\rho= n^{-2} \, g_{ZZ}$ and obtain 
$$
\aligned 
\Gamma^T_{TT}-\Gamma^Z_{ZT}
= & {1 \over 2n^2} \, \left( T(n^2)-T(g_{ZZ})n^2g^{ZZ} \right)
 =  {1 \over 2n^2} \, g_{ZZ} T(\rho^{-2})  
\endaligned
$$
and
$$
\aligned 
\Gamma^Z_{ZZ}-\Gamma^T_{ZT}
= & \frac{1}{2}g^{ZZ} Z(g_{ZZ}) - \frac{1}{n}Z(n^2)
\\
= &\frac{n^2}{2}g^{ZZ} Z\left(\frac{g_{ZZ}}{n^2}\right)
=\frac{n^2}{2}g^{ZZ}Z(\rho^2).
\endaligned
$$ 

It remain to check that the above definition agrees with the standard one
when the spacetime is sufficiently regular, which we now assume.  We 
recall that $[e_1,e_2]=[e_2,e_3]=0$ and from the definition of $T^2$-symmetric spacetime,
we also have $[e_0,e_2]=[e_0,e_3]=0$. Indeed, we have 
$$
g([e_0,e_a],e_0)=2e_a \left(g(e_0,e_0)\right)=0 
$$
by using the symmetry assumptions, and
$$
g([e_0,e_a],e_i)=-g(e_0,[e_a,e_i])=0 
$$
by using the orthogonality of $e_0, e_1$, and
the commutation property of $e_a, e_i$.

Moreover, it follows from the definition of $Z$ that $[Z,T]= f X+ gY$ 
for some functions $f$ and $g$ and in particular, 
$[Z,T]$ is orthogonal to both $Z$ and $T$. We now compute the Christoffel symbols as follows. 
For $i,j=1,2,3$, we obtain 
$$
g(T,\nabla_i e_j )= g_{TT} \Gamma^T_{ij}=g(nN,\nabla_i e_j )=n K_{ij},
$$
where $N$ is the timelike unit normal to $\Sigma_t$, thus  
$\Gamma^T_{ij}=-\frac{1}{n} K_{ij}$. 

For $\Gamma^T_{TZ}$, $\Gamma^T{TT}$, $\Gamma^T_{ta}$ and $\Gamma^Z_{TT}$, 
one derives the desired formulas directly from \eqref{eq:chrisni}, 
for instance 
$$
\Gamma^T_{TZ} = \frac{1}{2}g^{TT}\left(g_{TZ,T} + g_{TT,Z} - g_{TZ,T}
+c_{TTZ}+c_{TZT}+c_{ZTT}\right)=\frac{Z(n)}{n}.
$$
On the other hand, for $i=1,2,3$ we find
$$
\aligned 
g(e_1, \nabla_T e_i )
= & \, g_{ZZ}\Gamma^Z_{Ti}
\\
= & \, g(e_1,\nabla_i T) =-g(\nabla_i e_1,T)=-n g(\nabla_i e_1, N) = -n K_{iZ}.
\endaligned
$$
For $\Gamma^a_{Tb}$, we get 
$$
\aligned 
g(e_c,\nabla_T e_b)
= & \, g_{ca} \Gamma^a_{Tb}
\\
= & \, g(e_c, \nabla_b T )= - g(\nabla_b e_c , T )=-n K_{bc}, 
\endaligned
$$
while, for $\Gamma^a_{TZ}$,
$$
\aligned 
g(e_c,\nabla_t Z)
= & \, g_{ca} \Gamma^a_{TZ} = -g( \nabla_T e_c, Z) 
\\
= & \, -g(\nabla_c T, Z ) = g(T, \nabla_c Z )=n K_{cZ}
\endaligned
$$
and finally, for $\Gamma^a_{ZT}$, 
$$
\aligned 
g(e_c,\nabla_Z T)
= & \, g_{ca} \Gamma^a_{ZT}
\\
= & \, -g( \nabla_Z e_c, T)= - n K_{Zc}.
\endaligned
$$
\end{proof}

Finally, we introduce a weak version of Einstein's evolution equations. 
Given a $(3+1)$-splitting of the Einstein equations and provided the constraint equations 
are satisfied on each slice, the evolution equations are equivalent to
$R_{ij}=0$ (cf.~\cite[Sec.~VI-3.1]{Choquet-Bruhat2}.) 
Hence, since we have already derived the constraint equations in a weak form in the previous section, 
we can restrict attention now to the components $R_{ij}$ of the Ricci curvature
and recover all the remaining components from the (tensorial) Gauss equations.  

\begin{definition}
\label{def-wek-rzz}
When $(\Mcal,g)$ is a weakly regular $T^2$--symmetric Lorentzian manifold, 
the {\bf weak version of the component $R_{ZZ}$} of the Ricci tensor
is defined as 
\be
\aligned
R_{ZZ} : = \, 
& T(\Gamma^T_{ZZ}) - Z(\Gamma^T_{TZ}) - Z(\Gamma^a_{aZ}) 
\\
&  - \Gamma^{a}_{Zb} \, \Gamma^{b}_{aZ}
+ \Gamma^{a}_{aT} \, \Gamma^T_{ZZ} + \Gamma^{a}_{aZ} \, \Gamma^{Z}_{ZZ} 
\\
&+2\Gamma^T_{Za}\Gamma^a_{ZT}
+ \Gamma^T_{ZZ}\left( \Gamma^T_{TT}-\Gamma^{Z}_{ZT} \right) 
+ \Gamma^T_{tZ}\left( \Gamma^Z_{ZZ}-\Gamma^T_{ZT} \right),
\endaligned
\ee 
in which the first three terms are derivatives of $L^p$ functions on each slice, while the remaining terms
belong to $L^1$ on each slice. (Observe that the last two terms make sense, thanks to \eqref{more5}.)  
\end{definition}

\begin{definition}\label{def-wek-rzd}
When $(\Mcal,g)$ is a weakly regular $T^2$--symmetric Lorentzian manifold,
the {\bf weak version of the components $R_{Zd}$} of the Ricci tensor is defined as  
$$
R_{Zd} := T(\Gamma^T_{dZ}) +\Gamma^T_{dZ} \left(\Gamma^T_{TT}- \Gamma^Z_{ZT}\right) 
                +\Gamma^a_{aT}\Gamma^T_{dZ}. 
$$ 
\end{definition}

Finally, the components $R_{cd}$, $c,d=X,Y$ need to be suitably weighted by the norm of the vector field $Z$, 
in order to be well-defined as distributions.  
 
\begin{definition}\label{def-wek-rcd}
When $(\Mcal,g)$ is a weakly regular $T^2$--symmetric Lorentzian manifold, 
the {\bf weak version of the normalized components $R^{\text{norm}}_{cd}$} of the Ricci tensor is defined as
$$
\aligned
R^{\text{norm}}_{cd}
:= 
& T \left( n g(Z,Z)^{1/2} \Gamma^T_{dc} \right)
+ Z \left( n g(Z,Z)^{1/2} \Gamma^Z_{dc} \right)
\\
& + n g(Z,Z)^{1/2} \bigg(\Gamma^a_{aT} \Gamma^T_{dc}
+ \Gamma^a_{aZ}\Gamma^Z_{dc}
   -\Gamma^T_{dZ}\Gamma^Z_{Tc} - \Gamma^Z_{dT}\Gamma^T_{Zc}
   \\
   & \hskip2.2cm
- \Gamma^T_{da}\Gamma^a_{Tc}-\Gamma^a_{dT}\Gamma^T_{ac} 
- \Gamma^Z_{da}\Gamma^a_{Zc}-\Gamma^a_{dZ}\Gamma^Z_{ac} \bigg).
\endaligned
$$
\end{definition}

\begin{definition}
\label{def-wek-Einstein}
A weakly regular $T^2$--symmetric Lorentzian manifold $(\Mcal, g)$ 
is said to satisfy the {\bf weak version of Einstein's evolution equations} if 
\be
\label{weak-Einstein}
R_{ZZ}=0,\quad  R_{Zd}=0, \quad R^{norm}_{cd} = 0, \qquad c,d = X,Y, 
\ee 
in the weak sense introduced in Definitions~\ref{def-wek-rzz} to \ref{def-wek-rcd}. 
\end{definition} 

Again, we have an equivalence result establishing the link with the classical definition. 

\begin{proposition}[Equivalence with the classical definition] 
\label{eins-9} 
For any sufficiently regular $T^2$-symmetric spacetime $(\Mcal,g)$, 
the weak version of the Einstein equations \eqref{weak-Einstein}
is satisfied if and only if the Ricci flat condition
$$
Ric(e_i,e_j) = 0, \qquad e_i,e_j \in \big\{X,Y,Z\big\}
$$
holds, where $Ric$ denotes the Ricci tensor of $g$ defined in the classical sense.
\end{proposition}
 
Before we provide a proof of this result, we summarize our conclusions in this section, as follows. 

\begin{theorem}[Weak formulation of the Einstein equations] 
If $(\Sigma, h, K)$ is a weakly regular $T^2$--symmetric triple, then
Einstein's constraint equations \eqref{ee:constham}-\eqref{ee:constmom} 
make sense in the weak form 
\eqref{weak-Hamilton}-\eqref{weak-momentum}. 
Similarly, if $(\Mcal, g)$ is a weakly regular $T^2$--symmetric Lorentzian manifold,  
Einstein's evolution equations \eqref{Eins1} make sense in the weak form \eqref{weak-Einstein}. 
Furthermore, these new geometric objects coincide with the classical ones when sufficient regularity is assumed. 
\end{theorem}

We thus have re-stated and established Theorem~\ref{theorem61} (stated earlier in the introduction). 
We refer to a weakly regular $T^2$--symmetric triple satisfying
the weak version of the Hamiltonian and momentum constraint equations 
as a {\bf weakly regular $T^2$--symmetric initial data set.} 
Analogously, we refer to a weakly regular $T^2$--symmetric Lorentzian manifold
{\sl satisfying} the weak version of the Einstein constraint and evolution equations 
as a {\bf weakly regular $T^2$--symmetric vacuum spacetime.}

\begin{proof}[Proof of Proposition~\ref{eins-9}]  
We will show that the distributions defined by $Ric(e_i,e_j)$, for $e_i,e_j=X,Y,Z$, 
agree with the ones introduced in \eqref{def-wek-rzz}, \eqref{def-wek-rzd}, and \eqref{def-wek-rcd} 
(where an additional weight must be introduced for $Ric(e_c,e_d)$).
Abusing notation, we denote the Christoffel symbols 
defined in the classical sense by $\Gamma^\alpha_{\beta \delta}$. 
With $c^f_{aZ}=[e_c,e_d]^f$, we find 
\be
\label{3404}
Ric(Z,Z)=R^\alpha_{Z\alpha Z}=\Gamma^\alpha_{ZZ,\alpha}-\Gamma^\alpha_{\alpha Z,Z}
      +\Gamma^\alpha_{\alpha \beta}\Gamma^\beta_{ZZ}-\Gamma^\alpha_{Z\beta}\Gamma^\beta_{YZ}
      -c^\beta_{\alpha Z}\Gamma^\alpha_{Z\beta},
\ee

We expand the right-hand side of \eqref{3404} by focusing our attention on the terms which 
for $T^2$-symmetric solution having only weak regularity are a priori not well-defined:
$$
\aligned
Ric(Z,Z)
= \, & Z( \Gamma^Z_{ZZ})+T(\Gamma^T_{ZZ})
   - \Big( Z(\Gamma^Z_{ZZ}) + Z(\Gamma^T_{tZ}) + Z(\Gamma^a_{aZ}) \Big) 
\\
& +\Gamma^T_{TT}\Gamma^T_{ZZ}+\Gamma^Z_{ZZ}\Gamma^Z_{ZZ} 
  +\Gamma^T_{tZ}\Gamma^Z_{ZZ}+\Gamma^Z_{Zt}\Gamma^T_{ZZ} 
\\
& +\Gamma^T_{ta}\Gamma^a_{ZZ}+\Gamma^a_{at}\Gamma^T_{ZZ} 
 +\Gamma^Z_{Za}\Gamma^a_{ZZ}+\Gamma^a_{aZ}\Gamma^Z_{ZZ} 
\\
& +\Gamma^a_{ab}\Gamma^b_{ZZ} 
  - \Big( \Gamma^T_{Zt}\Gamma^T_{tZ} + \Gamma^Z_{ZZ}\Gamma^Z_{ZZ}  
             +\Gamma^T_{ZZ}\Gamma^Z_{tZ} +\Gamma^Z_{Zt}\Gamma^T_{ZZ} \Big) 
\\
& - \Big( \Gamma^T_{Za}\Gamma^a_{tZ} + \Gamma^a_{Zt}\Gamma^T_{aZ} 
           +\Gamma^Z_{Za}\Gamma^a_{ZZ} + \Gamma^a_{ZZ}\Gamma^Z_{aZ} \Big) 
      -\Gamma^a_{Zb}\Gamma^b_{aZ} - c^b_{tZ}\Gamma^T_{Zb}. 
\endaligned
$$
To handle the latter term, we observe that the only non-vanishing commutator is $[T,Z]$ 
and that this commutator is orthogonal to both $Z, T$.
Note that this last term can be rewritten in term of the connection coefficients since
$$
[T,Z]^b=\Gamma^b_{tZ}-\Gamma^b_{Zt}.
$$
Taking into account the cancelations in 
$Z( \Gamma^Z_{ZZ})$, $\Gamma^Z_{ZZ}\Gamma^Z_{ZZ}$ and
$\Gamma^Z_{Zt}\Gamma^T_{ZZ}$, as well as the antisymmetry of $\Gamma^a_{Zt}$ 
and the fact that $\Gamma^T_{ta}=\Gamma^Z_{Za}=0$, we obtain
$$
\aligned
Ric(Z,Z)
= \, &
T(\Gamma^T_{ZZ})-Z(\Gamma^T_{tZ})-Z(\Gamma^a_{aZ}) 
+ \Big( \Gamma^T_{TT}\Gamma^T_{ZZ}+\Gamma^T_{tZ}\Gamma^Z_{ZZ} \Big) 
\\
& + \Big( \Gamma^a_{at}\Gamma^T_{ZZ}  +\Gamma^a_{aZ}\Gamma^Z_{ZZ} \Big) 
  +\Gamma^a_{ab}\Gamma^b_{ZZ}
\\
& - \Big( \Gamma^T_{Zt}\Gamma^T_{tZ} + \Gamma^T_{ZZ}\Gamma^Z_{tZ}\Big) 
 -\Gamma^a_{Zb}\Gamma^b_{aZ} 
 +2 \Gamma^T_{Za}\Gamma^a_{Zt} 
\endaligned
$$
and the expression for $Ric(T,T)$ then follows by factorizing out $\Gamma^Z_{Zt}$ and $\Gamma^Z_{TT}$.

For $Ric(Z,e_d)$, we proceed similarly and obtain 
$$
\aligned
Ric(Z,e_d)
= \, & 
\Gamma^{\alpha}_{dZ,\alpha}-\Gamma^\alpha_{\alpha Z,d} 
+\Gamma^{\alpha}_{\alpha \beta} \Gamma^\beta_{dZ}
   -\Gamma^\alpha_{d\beta}\Gamma^\beta_{\alpha Z}-c^\beta_{\alpha d}\Gamma^\alpha_{Z\beta}
\\
= \, & T ( \Gamma^T_{dZ})+ Z( \Gamma^Z_{Zd} ) 
+ \Big(\Gamma^T_{TT}\Gamma^T_{dZ}+\Gamma^Z_{ZZ}\Gamma^Z_{dZ}
+\Gamma^{t}_{tZ}\Gamma^Z_{dZ}+\Gamma^Z_{Zt}\Gamma^T_{dZ} \Big) 
\\
& +\Gamma^T_{ta}\Gamma^a_{dZ}+\Gamma^a_{at}\Gamma^T_{dZ}
  +\Gamma^Z_{Za}\Gamma^a_{dZ}+\Gamma^a_{aZ}\Gamma^Z_{dZ}
\\
& +\Gamma^a_{ab}\Gamma^b_{dZ}
  -\Big( \Gamma^T_{dt}\Gamma^T_{tZ} + \Gamma^Z_{dZ}\Gamma^Z_{ZZ} 
             + \Gamma^T_{dZ}\Gamma^Z_{tZ} + \Gamma^Z_{dt}\Gamma^T_{ZZ}\Big) 
\\
& -\Big( \Gamma^T_{da}\Gamma^a_{tZ} + \Gamma^a_{dt}\Gamma^T_{aZ}
                +\Gamma^Z_{da}\Gamma^a_{ZZ} + \Gamma^a_{dZ}\Gamma^Z_{aZ}\Big) 
    -\Gamma^a_{db}\Gamma^b_{aZ}.
\endaligned
$$
Next, using $\Gamma^a_{ZZ}=\Gamma^Z_{Za}=\Gamma^T_{ta}=\Gamma^a_{bc}=0$ 
and the fact that $X,Y$ commutes with $Z,T$, we obtain
$$
\aligned
Ric(Z,e_d)
= \, & T(\Gamma^T_{dZ}) +\Gamma^T_{TT}\Gamma^T_{dZ} +\Gamma^Z_{Zt}\Gamma^T_{dZ} +\Gamma^a_{aT}\Gamma^T_{dZ} 
\\
& - \Big( \Gamma^T_{dZ}\Gamma^Z_{TZ} + \Gamma^Z_{dT}\Gamma^T_{ZZ} \Big)
   - \Big( \Gamma^T_{da}\Gamma^a_{TZ} + \Gamma^a_{dT}\Gamma^T_{aZ} \Big).
\endaligned
$$
We also note the cancellation in $\Gamma^{Z}_{ZT}\Gamma^T_{dZ}$ 
as well as the identities 
$$
\Gamma^T_{TT}\Gamma^T_{dZ}-\Gamma^z_{dT}\Gamma^T_{ZZ}
=\Gamma^T_{Zd} 
\left(\Gamma^T_{TT}-\Gamma^Z_{ZT} \right), 
\qquad 
\quad 
-\Gamma^T_{da}\Gamma^a_{TZ}-\Gamma^a_{dT}\Gamma^T_{aZ}=0,
$$
and arrive at the desired formula 
$$
Ric(Z,e_d) 
= T(\Gamma^T_{dZ}) +\Gamma^T_{Zd} \left(\Gamma^T_{TT}- \Gamma^Z_{ZT}\right) 
+\Gamma^a_{aT}\Gamma^T_{dZ}. 
$$

Next, for $Ric(e_c,e_d)$, we have 
$$
\aligned
Ric(e_c,e_d)
= \, & \Gamma^\alpha_{dc,\alpha}-\Gamma^\alpha_{\alpha c,d} 
        +\Gamma^\alpha_{\alpha \beta}\Gamma^\beta_{dc}
       -\Gamma^{\alpha}_{d\beta}\Gamma^\beta_{\alpha c}-c^\beta_{\alpha d}\Gamma^\alpha_{c \beta} 
 \\
= \, & T(\Gamma^T_{dc})+Z(\Gamma^z_{dc})  
     + \Big( \Gamma^T_{TT}\Gamma^T_{dc}+\Gamma^Z_{ZZ}\Gamma^Z_{dc}\Big) 
+ \Big( \Gamma^T_{TZ}\Gamma^Z_{dc}+\Gamma^Z_{ZT}\Gamma^T_{dc} \Big) 
\\
&
 + \Big( \Gamma^T_{Ta}\Gamma^a_{dc}+\Gamma^Z_{Za}\Gamma^a_{dc} \Big) 
+ \Big( \Gamma^a_{aT}\Gamma^T_{dc}+\Gamma^a_{aZ}\Gamma^Z_{dc} \Big) 
  +\Gamma^a_{ab}\Gamma^b_{dc}
 - \Big( \Gamma^T_{dT}\Gamma^T_{Tc} + \Gamma^z_{dz}\Gamma^Z_{Zc}\Big) 
\\
& - \Big( \Gamma^T_{dZ}\Gamma^z_{Tc} + \Gamma^Z_{dT}\Gamma^T_{Zc}\Big) 
   - \Big( \Gamma^T_{da}\Gamma^a_{Tc} + \Gamma^a_{dT}\Gamma^T_{ac}\Big) 
   - \Big( \Gamma^Z_{da}\Gamma^a_{Zc} + \Gamma^a_{dZ}\Gamma^Z_{ac}\Big) 
   -\Gamma^b_{da}\Gamma^a_{bc}
\endaligned
$$
and, using $\Gamma^Z_{Za}=\Gamma^T_{Ta}=\Gamma^a_{bc}=0$, 
$$
\aligned
Ric(e_c,e_d)
= \, &  
   T(\Gamma^T_{dc})+Z(\Gamma^z_{dc})  
   + \Big( \Gamma^T_{TT}\Gamma^T_{dc} + \Gamma^Z_{ZZ}\Gamma^Z_{dc}\Big) 
 + \Big( \Gamma^T_{TZ}\Gamma^Z_{dc}+\Gamma^Z_{ZT}\Gamma^T_{dc}\Big)
 \\
& + \Big( \Gamma^a_{at}\Gamma^T_{dc} + \Gamma^a_{aZ}\Gamma^Z_{dc}\Big)
   - \Big( \Gamma^T_{dz}\Gamma^Z_{Tc} + \Gamma^z_{dT}\Gamma^T_{Zc}\Big)
 \\
& - \Big( \Gamma^T_{da}\Gamma^a_{Tc} + \Gamma^a_{dT}\Gamma^T_{ac}\Big) 
 - \Big( \Gamma^Z_{da}\Gamma^a_{Zc} + \Gamma^a_{dZ}\Gamma^Z_{ac} \Big).
\endaligned
$$
The first six terms of the right-hand side above can be rewritten as 
$$
\aligned
T(\Gamma^T_{dc})+\Gamma^T_{TT}\Gamma^T_{dc}+\Gamma^T_{dc}\Gamma^Z_{ZT}
= & T(\Gamma^T_{dc}) + \Gamma^T_{dc}\, \frac{T(n)}{n} + \Gamma^T_{dc} \, (g(Z,Z))^{-1/2} \, T\big((g(Z,Z))^{1/2}\big) 
\\
= & n^{-1}( g(Z,Z) )^{-1/2} T \left( n g(Z,Z)^{1/2} \Gamma^T_{dc} \right) 
\endaligned
$$
and, similarly,
$$
\aligned
Z(\Gamma^Z_{dc})+\Gamma^Z_{ZZ}\Gamma^Z_{dc}+\Gamma^Z_{dc}\Gamma^T_{TZ}
= \, & Z(\Gamma^z_{dc})
      + \Gamma^Z_{dc} \, \frac{Z(n)}{n} +\Gamma^Z_{dc} \, (g(Z,Z))^{-1/2} \, Z\big((g(Z,Z))^{1/2}\big)
\\
= \, & n^{-1}( g(Z,Z) )^{-1/2} Z \left( n g(Z,Z)^{1/2} \Gamma^Z_{dc} \right).  
\endaligned
$$
This suggests to introduce a weight in $Ric(e_c, e_d)$, that is, $n g(Z,Z)^{1/2}$, which leads us to 
 the desired expression: 
\[
\aligned
n g(Z,Z)^{1/2} Ric(e_c, e_d)
= \, & T \left( n g(Z,Z)^{1/2} \Gamma^T_{dc} \right)
+ Z \left( n g(Z,Z)^{1/2} \Gamma^Z_{dc} \right) 
\\
& + n g(Z,Z)^{1/2} \Big(\Gamma^a_{aT}\Gamma^T_{dc}
  +\Gamma^a_{aZ}\Gamma^Z_{dc}
 - \Big( \Gamma^T_{dZ}\Gamma^Z_{Tc} + \Gamma^Z_{dT}\Gamma^T_{Zc}\Big) 
\\
& \qquad \qquad \qquad - \Big( \Gamma^T_{da}\Gamma^a_{Tc} + \Gamma^a_{dT}\Gamma^T_{ac}\Big) 
   - \Big( \Gamma^Z_{da}\Gamma^a_{Zc} + \Gamma^a_{dZ}\Gamma^Z_{ac} \Big) 
\Big). \hskip3.cm \qedhere
\endaligned
\]
\end{proof}

%---------------------------------------------------------------------------------------------------- 

\subsection{Twist coefficients} 

We end this section with an important property of the {\bf twist coefficients} associated with two Killing fields $X,Y$
and defined by  
$$
C_X := \Ecal_{\alpha\beta\gamma\delta} X^\alpha Y^\beta \nabla^\gamma X^\delta, 
\qquad 
C_Y := \Ecal_{\alpha\beta\gamma\delta} Y^\alpha Y^\beta \nabla^\gamma X^\delta, 
$$
where $\Ecal_{\alpha\beta\gamma\delta}$ is the volume form of $(\Mcal,g)$. 
We recall that the twists vanish if and only if the family of $2$-planes 
orthogonal to $X$ and $Y$ is integrable. 
For all {\sl sufficiently regular} spacetimes, it is also well-known that the vacuum Einstein equations
imply that the twists are constant \cite{Chrusciel}. We show now that this latter property is preserved at our level of 
(weak) regularity. 

\begin{proposition}[Constant twist property] 
\label{constant-twist}
The twist coefficients of any weakly regular $T^2$--symmetric spacetime are constants.
Furthermore, one can always choose the Killing fields $X,Y$ in such a way that, one of them vanishes identically. 
\end{proposition}

\begin{proof} It follows from the anti-symmetry of the volume form that 
$$
C_X=\Ecal_{XYTZ}g^{TT} \Gamma^Z_{TX}+\Ecal_{XYZT} \, g^{ZZ}\Gamma^T_{ZX},
$$
which contains products of $L^\infty$ functions by $L^2$ functions, only, 
and is therefore well-defined. 
Moreover, in view of the relation $\Gamma^Z_{TX}=n^2 \, g^{ZZ} \Gamma^T_{ZX}$, we have 
$$
C_X=2\Ecal_{XYZT}g^{ZZ}\Gamma^T_{ZX}=-2 \frac{R}{\rho}\Gamma^T_{ZX}. 
$$
It follows immediately from one of the Hamiltonian constraint equations 
and the evolution equation $R_{ZX}=0$ that $C_X$ is a constant. The same holds for $C_Y$
and, moreover,  
one of the twists can be made to vanish 
by introducing a suitable linear combination of the Killing vectors, say
\be
\label{combi}
X' = a \, X + b \, Y, \qquad Y' = c \, X + d \, Y, \qquad ad - bc = 1, 
\ee
where the latter restriction on the coefficients $a,b,c,d$ ensures that the transformation 
preserves the periodicity property.
Then, the conclusion follows easily from 
\[  
C_{X'} = \Ecal_{\alpha\beta\gamma\delta} {X'}^\alpha {Y'}^\beta \nabla^\gamma {X'}^\delta 
= (ad-bc) \Big( a \, C_X + b \, C_Y \Big).  
\qedhere
\]
\end{proof}  

%==========================================================================

\section{Weakly regular metrics in admissible coordinates}
\label{sec2}

\subsection{Weakly regular Riemannian manifolds in admissible coordinates}

In this section, we introduce several choices of coordinates ---in which we will express later (cf.~Section~\ref{sec4}) 
the weak version of the Einstein equations in a form amenable to 
 techniques of analysis for nonlinear partial differential equations.  
We determine here the regularity of the metric coefficients that is implied by the 
geometric regularity assumptions made in the previous section. 
From now on, functions invariant by the action of the Killing fields 
are identified with functions defined on the circle $S^1$ (and, later in this section, also depending on a time variable).  
We refer to the expression \eqref{eq:inducedmetric} introduced now as the 
{\bf spatial metric in admissible coordinates.}

\begin{lemma}[Weakly regular $T^2$-symmetric metrics in admissible coordinates]   
\label{206}
Let $(\Sigma,h)$ be a weakly regular $T^2$--symmetric Riemannian manifold 
and $(x,y,\theta)$ be coordinates associated with an adapted frame. Then, the metric $h$ takes the form
\be 
\label{eq:inducedmetric}
h = {e^{2\nub-2\Pb} \over \Rb} d\theta^2 +  e^{2\Pb}\Rb \big(
dx + \Ab \, dy + \big(\Gb + \Ab \, \Hb\big) \, d\theta \big)^2
+ e^{-2\Pb} \, \Rb \, \big( dy + \Hb \, d\theta \big)^2, 
\ee
in which the coefficients $\Rb$, $\Pb, \Ab, \nub, \Gb, \Hb$ depend on the variable $\theta \in S^1$, only, 
and satisfy 
$$
\Pb, \Ab \in H^1(S^1), \qquad \nub \in W^{1,1}(S^1), \qquad \Gb, \Hb \in L^\infty(S^1), 
$$
while the area function 
$\Rb$ (already defined in \eqref{ass:rconstant}) satisfies
$
\Rb \in W^{1,\infty}(S^1)$
and is bounded above and below by positive constants. 
\end{lemma}

\begin{proof} We rely here on the conditions introduced in Definition~\ref{def21}. 
Clearly, any metric can be expressed in the form \eqref{eq:inducedmetric}, 
provided one defines $\nub, \Pb, \Ab, \Gb$ and $\Hb$ by $h(X,X)=:e^{2\Pb}\Rb$, 
$h(X, Y)=:\Rb e^{2\Pb}\Ab$,\ldots  
Since the metric is $T^2$--symmetric, 
all coefficients are independent of the variables $(x,y)$.
By our assumption \eqref{ass:rconstant}, we have $\Rb \in W^{1,\infty}(\Sigma)$ and, after
identifying $\Rb$ with a function on $S^1$, it follows that $\Rb \in W^{1,\infty}(S^1)$.
Since $e^{2\Pb}\Rb = h(X,X) \in H^1(\Sigma)$, we obtain $e^{2\Pb}\in H^1(\Sigma)$, and 
after identifying $e^{2\Pb}$ with a function of $\theta \in S^1$, 
it follows that $e^{2\Pb} \in H^1(S^1)$. In particular, $\Pb$ (defined almost everywhere)
admits an H\"older continuous representative. Since $e^{2\Pb} \in C^0(S^1)$ is positive and defined on the compact set $S^1$, its inverse 
$e^{-2\Pb}$ also belongs to the space $L^\infty(S^1)$. From this, 
it also follows that $\Pb_\theta= 1/2e^{-2\Pb} \, \big( e^{2\Pb} \big)_\theta$ belongs to $L^2(S^1)$, 
and we conclude that  $\Pb \in H^1(S^1)$.
A completely similar argument applies to $\Ab$ and shows that $\Ab \in H^1(S^1)$. 

For $\Gb$ and $\Hb$, we have $h(X, Z) = e^{2\Pb}\Rb(\Gb+\Ab \, \Hb) \in L^\infty(S^1)$ and thus
$\Pb \in C^0(S^1)$ and $\Rb \in C^0(S^1)$. So, we find 
\be
(\Gb+\Ab \, \Hb) \in L^\infty(S^1). 
\label{aghl2} 
\ee
On the other hand, from the assumptions on $h(Y,Z)$, we also know that  
$$
h(Y,Z)
=
\Rb e^{2\Pb}\Ab(\Gb+\Ab \, \Hb) +e^{-2\Pb} \, \Rb  \, \Hb \in L^\infty(S^1), 
$$
in which the first term is in $L^\infty(S^1)$ by (\ref{aghl2}), and so 
we have 
$e^{-2\Pb} \Rb \, \Hb \in L^\infty(S^1)$. Moreover, using the lower bound on $\Rb$, 
the function 
$e^{2\Pb}/ \Rb$ belongs to $L^\infty(S^1)$ and thus $\Hb \in L^\infty(S^1)$. 
From \eqref{aghl2}, it then follows that $\Gb \in L^\infty$.
Finally, by observing that $h(Z,Z)$ provides a control of $e^{2\nub}$, similar arguments show that $\nub$ belongs to $W^{1,1}(S^1)$. 
\end{proof}

Relying on Definition~\ref{def:t2tri}, we now introduce a decomposition of the tensor field $K$ 
and specify the regularity of each component. The proof of the following statement is omitted.  

\begin{lemma}[Decomposition of weakly regular tensor fields $K$]
\label{lem:decompK}  
Let $(\Sigma,h,K)$ be a weakly regular $T^2$-symmetric triple in admissible coordinates \eqref{eq:inducedmetric}. 
Then, there exist functions $\Pbz$, $\Abz$, $\Gbz$, $\Hbz$, $\Rbz$, $\nubz$ and a symmetric $2$ tensor $\habz$ 
so that, in an adapted frame $(X,Y,Z)$, the components of $K$ read   
$$
\aligned
K_{ab} & =\frac{1}{2} \, \big( h(Z,Z) \big)^{-1/2} \, \habz, 
\qquad \qquad \qquad \qquad 
K(X,Z) = \frac{1}{2} \, e^{-\nub+3\Pb}\left( \Gbz+\Ab \, \Hbz \right),
\\
K(Y,Z) & = \frac{1}{2} \, e^{-\nu-\Pb} \, \Rb^2 \, \Hbz + \Ab \, K(X,Z)
= \frac{1}{2} \, e^{-\nub+\Pb}\left( \Rb^2 e^{-2\Pb} \, \Hbz
          +\Ab \, e^{2\Pb}\left( \Gbz +\Ab \, \Hbz \right) \right), 
\\
tr^{(2)} K & = e^{-\nub+\Pb} \, \Rbz \, \big( \Rb \big)^{-1} = e^{-\nub+\Pb} \, \big(\Rb\big)^{-1/2} \, \Rbz, 
\quad \qquad \qquad \qquad 
K_{ZZ} = e^{\nub - \Pb} \, \big( \Rb\big)^{-1/2} \big( \nubz - \Pbz - \Rbz \, (2\Rb)^{-1}\big), 
\endaligned
$$
with 
$$
\aligned
\habz 
= & \, \, (\Rb)^{-1} \Rbz \, h_{ab} 
+ \Rb \,  \Big( e^{2P} 2 \Pbz (dx + \Ab \, dy)^2 -2 \Pbz e^{-2\Pb} dy^2 \Big) 
     \\&   + \Rb e^{2 \Pb} \big( 2 \Abz \, dxdy + 2 \Ab \, \Abz dy^2 \big), 
\endaligned
$$
and the following regularity properties hold:
$$
\Pbz, \, \Abz, \, \Gbz, \, \Hbz, \, \habz \in L^2(S^1), 
\qquad 
\Rbz \in W^{1,\infty}(S^1), \quad \nubz \in L^1(S^1).  
$$
\end{lemma}

%---------------------------------------------------------------------------------------------------------

\subsection{Weakly regular Lorentzian manifolds in admissible coordinates}
 
In the context of Definition~\ref{def:wrt2rm}, Lemma~\ref{206} applies to each slice of the foliation and, again,
 we refer to the expression \eqref{eq:metric3} below as the {\bf  metric in admissible coordinates.}

\begin{lemma}[Weakly regular $(3+1)$-metrics in admissible coordinates]  
\label{206-2}
Let $(\Mcal,g)$ be a weakly regular $T^2$--symmetric spacetime 
and $(t, x,y,\theta)$ be admissible coordinates adapted to the symmetry and a given foliation. 
Then, the spacetime metric $g$ takes the form
\be 
\label{eq:metric3}
g = - n^2 \, dt^2 + {e^{2\nu - 2\, P} \over R} d\theta^2 +  e^{2P} R \big(
   dx + A \, dy + \big(G + A \, H\big) \, d\theta \big)^2
   + e^{-2P} R \big( dy + H \, d\theta \big)^2 
\ee 
with coefficients $P, A, \nu, G, H$ depending only on $t \in I$ and $\theta \in S^1$, 
satisfying 
$$
P, A \in L^\infty_\loc(I, H^1(S^1)), 
\quad
\nu \in L^\infty_\loc(I, W^{1,1}(S^1)),
\quad
G, H \in L^\infty_\loc(I, L^\infty(S^1)).
$$ 
and such that the area function $R$ (defined in \eqref{ass:rconstant}) satisfies
$
R \in W^{1,\infty}(S^1)
$
and is bounded above and below by positive constants. 
\end{lemma}

From now on, a subscript (like $t$ and $\theta$) denotes a partial derivative, possibly 
understood in the weak sense. The regularity assumed on the second fundamental form 
implies some regularity on the time-derivative of the metric coefficients.

\begin{lemma}[Timelike regularity in admissible coordinates]   
\label{207} 
Let $(\Mcal,g)$ be a weakly regular $T^2$--symmetric spacetime 
and $(t, x,y,\theta)$ be admissible coordinates adapted to the symmetry and a given foliation in the variable $t \in I$. 
Then, the metric coefficients in \eqref{eq:metric3} enjoy the following regularity in time: 
$$
\aligned
& P_t, A_t \in L^\infty_\loc(I, L^2(S^1)), \qquad R_t \in  L^\infty_\loc(I, L^\infty(S^1)),
\\
& \nu_t \in  L^\infty_\loc(I, L^1(S^1)), \qquad G_t, H_t \in  L^\infty_\loc(I, L^2(S^1)). 
\endaligned
$$ 
\end{lemma}

\begin{proof} In view of Definition~\ref{def:wrt2rm}, the components of $K$ satisfy 
$$
L^1(S^1) \ni 2 n \, K(Z, Z) 
= \left(e^{-2P}H^2 R + R e^{2P}(G+AH)^2+e^{2\nu-2P} R^{-1} \right)_t, 
$$
while all other components belong to $L^2(S^1)$
$$
\aligned
&2 n \, K(Z, X)   
= \left( e^{2P} R (G+AH) \right)_t,
%\\
%L^2(S^1) \ni 
\qquad
\qquad 
&& 2 n \, K(Z, Y) 
= \left( e^{2P} R A(G+AH)+e^{-2P}R \, H \right)_t,
\\
%L^2(S^1) \ni 
& 2 n \, K(X, X) 
= \big(e^{2P} R\big)_t,
\qquad \qquad 
%L^2(S^1) \ni 
&&2 n \, K(X, Y) 
= (e^{2P} R A)_t,
\qquad \qquad 
%L^2(S^1)  \ni 
\\
& 2 n \, K(Y,Y) 
= \left( e^{2P} R A^2 + R e^{-2P} \right)_t
\endaligned
$$
with, moreover,
$$
\aligned
L^\infty(S^1) 
\ni \,\Trace^{(2)} (K) 
& = e^{2 P} \, R^{-1} K(Y,Y) - 2 \, A e^{2P} \, R^{-1} K(X,Y)
 + \big( e^{2P} A^2 \, R^{-1} + 1\big) \, K(X,X).
\endaligned
$$  
We first use the conditions $\big(e^{2P} R \big)_t  \in L^2(S^1)$ and $(e^{2P} R A)_t  \in L^2(S^1)$
and deduce that $ P_t, A_t \in L^2(S^1)$. Then, the condition on $\Trace^{(2)}(K)$ implies that 
$ R_t \in L^\infty(S^1)$. We then deduce a control on the functions $G_t, H_t$, and finally 
the condition $2n \, K(Z, Z) \in L^1(S^1)$ yields $ \nu_t \in L^1(S^1)$. 
\end{proof}

%---------------------------------------------------------------------------------------------------------

\subsection{Conformal coordinates for weakly regular metrics}

A well-known problem in general relativity and, more generally, in geometric analysis is to exploit the gauge freedom 
at our disposal to simplify the analysis. This typically means choosing a coordinate system or a frame well-adapted to the problem.
Here, it will turn out that we need to make 
two different gauges, specifically the so-called conformal and areal gauges.
We begin by proving the existence of conformal coordinates, as follows. 

\begin{lemma}[Existence of conformal coordinates]  
\label{lem:exiconf}
Let $(\Mcal,g)$ be a weakly regular $T^2$--symmetric spacetime 
and $(t, x,y,\theta)$ be admissible coordinates adapted to the symmetry and a given foliation, 
with $t \in I$ and $x,y,\theta \in S^1$.  
Then, there exist functions $\tau,\xi: \Mcal \to \RR$ such that:
\begin{enumerate}

\item In the coordinates $(t,x,y,\theta)$, the functions $\tau, \xi$ depend on $(t,\theta)$, only,
and belong to $W^{1,\infty}_{loc}(I \times S^1)$.  

\item The functions $\tau, \xi, x, y$ determine a global chart on $\Mcal$ and, hence,
defines a smooth differential structure (not necessarily $C^\infty$ compatible with that defined by $(t, \theta,x,y)$) 
but at least $W^{1,\infty}$ compatible).

\item In the coordinate system $(\tau,\xi,x,y)$, the metric takes the form 
\be 
\label{eq:metric}
\aligned
g = \, & {e^{2\nu-2P} \over R} \big( - d\tau^2 + d\xi^2 \big) 
      +  e^{2P} R \, \big( dx + A \, dy + \big(G + A \, H\big) \, d\xi \big)^2
  + e^{-2P} R \, \big( dy + H \, d\xi \big)^2,  
\endaligned
\ee
where the coefficients $\nu, P, A, R, G, H$
depend on $\tau \in J$ and $\xi \in S^1$, only, where $J$ is an interval. 

\item The hypersurface $t=t_0$ coincides with a level set of $\tau$.
\end{enumerate}
\end{lemma}
In fact, the coefficients of the metric will also enjoy the same regularity properties as those presented 
in Lemmas~\ref{206-2} and \ref{207}, provided the weak version of the Einstein equations hold true. 
This fact will be checked later in Section \ref{sec45}, below.

\begin{proof} We restrict attention to the two-dimensional quotient metric
$
\gh = {e^{2\nu-2P} \over R} \left( -\rho^2 \, dt^2 + d\theta^2 \right), 
$
and establish the existence of functions $\tau, \xi$ such that 
$$
\gh = {e^{2\hat{\nu}-2P} \over R} \big( - d\tau^2 + d\xi^2 \big). 
$$
We are going to construct null coordinates $u,v: \Mcal \to \RR$ enjoying the following properties:
\begin{enumerate}

\item The functions $u, v$ depend $(t,\theta)$, only, and belong to $W^{1,\infty}(I \times S^1)$.  

\item The following equations hold:  
\be
\label{syst-uv}
\aligned
u_t + \rho \, u_\theta = 0,
\qquad
v_t - \rho \, v_\theta  = 0.
\endaligned
\ee

\item The following periodicity conditions hold: 
\be
\label{peri}
u(t,\theta+2\pi) = u(t,\theta) -2\pi, \qquad 
v(t,\theta+2\pi)=v(t,\theta)+2\pi.
\ee 

\item The map $(t,\theta) \in I \times S^1 \mapsto (u,v)$ is a $W^{1,\infty}$ diffeomorphism on its image.

\end{enumerate}
\noindent Once this is established, one easily checks that the functions  
$$
\xi := \frac{v-u}{2}, \qquad \tau := \frac{v+u}{2} 
$$
satisfy the desired requirements. 

The equations \eqref{syst-uv} are linear transport equations, and are easily solved by the method of characteristics. 
First of all, setting $I=[t_1, t_2]$, 
initial data $u_1, v_1$ for the functions $u,v$ are chosen at the time $t_1$ to be periodic, as stated in \eqref{peri}: 
$$
u(t_1, \cdot) : = u_1, \qquad  v(t_1, \cdot) : = v_1. 
$$
Then, we consider the characteristic equations 
$$
{d\thetab \over dt} = \pm \rho\big(t,\thetab(t)\big),
$$
with initial condition $\thetab(t_1, \theta) = \pm \theta$, 
and we denote by $\thetab_\pm = \thetab_\pm(t,\theta)$ the corresponding solutions. 
Since $\rho \in W^{1,\infty}(I \times S^1)$, from a standard theorem on 
ordinary differential equations it follows that $\thetab_\pm \in W^{1,\infty}(I \times S^1)$,
 and  
$$
\aligned
\thetab_{\pm, \theta}(t,\theta)
& = \exp \Bigg( \int^t_{t_1} { \rho_\theta \over \rho} (t',\thetab_\pm(t',\theta)) \, dt' \Bigg)
\, \,  \in L^\infty_\loc
\endaligned
$$ 
never vanishes. Thus, the maps $(t,\theta) \in I \times \RR \mapsto (t, \theta_\pm) \in I \times \RR$ 
are $W^{1,\infty}$-diffeomorphism.
Since solutions are uniqueness and the data are periodic, we obtain 
$\thetab_\pm(t,\theta+2\pi) = \thetab_\pm(t,\theta) \pm 2\pi$. 
Finally, we arrive at the desired conclusion by defining the functions $u,v$ by   
\[
u(t, \theta) := u_1(t, \theta_+(t, \theta)), \qquad v(t, \theta) := v_1(t, \theta_-(t, \theta)). 
\qedhere
\]
\end{proof}

%---------------------------------------------------------------------------------------------------------

\subsection{Areal coordinates for weakly regular metrics} 
 
We will also use a time function coinciding with the area of the orbits of symmetry. In such coordinates, 
the area function is obviously of class $C^\infty$ and, instead, the metric coefficient $a$ introduced below
in \eqref{metric:areal} has weak regularity. 

Later we will justify this choice and show (cf.~Proposition~\ref{prop:arr}, below) that the gradient of the area function $R$ 
is timelike so that the area can be used as a time coordinate. In 
the so-called areal coordinates, the metric takes the form  
\be
\label{metric:areal} 
\aligned  
g = \, & e^{2(\eta-U)} \big( - dR^2 + a^{-2} \, d\theta^2 \big) 
  +e^{2U} \big(
dx + A \, dy + \big(G + AH\big) \, d\theta \big)^2
+ e^{-2U} R^2 \big( dy + H \, d\theta \big)^2,  
\endaligned 
\ee
where $U,A,\eta, a, \displaystyle G, H$ are functions of $t$ and $\theta \in S^1$. 
The variable $R$ describes some interval $[R_0, R_1)$ and the variables $x,y,\theta$ describe $S^1$. 
As in the conformal case, we will prove later in Section \ref{sec45}, that areal coordinates are admissible 
if the weak version of the Einstein equations holds and that, in particular, the regularity 
in Lemmas~\ref{206-2} and \ref{207} holds in areal coordinates, as now stated. 

\begin{lemma}[Weak regularity in areal coordinates]
\label{weakdefinitionT2}
Let $(\Mcal,g)$ be a weakly regular $T^2$--symmetric spacetime 
and suppose that the area function has a timelike gradient $\nabla R$. 
Consider areal admissible coordinates $(R,x,y,\theta)$,  
in which the metric takes the form \eqref{metric:areal} 
where all functions depend only on $R,\theta$ with $R \in I \subset (0, +\infty)$ (an interval) 
Then, the following regularity properties hold:  
\be
\label{regl}
\aligned 
& U_R, A_R, U_\theta, A_\theta \in L_\loc^\infty(I, L^2(S^1)),
\qquad 
&& \eta_R, \eta_\theta, \displaystyle G, H \in L_\loc^\infty(I, L^1(S^1)), 
\\
& a \in L_\loc^\infty(I, W^{1,\infty}(S^1)).  
\endaligned 
\ee
\end{lemma} 

%============================================================

\section{Field equations in admissible coordinates} 
 \label{sec4}
  
\subsection{Constraint equations in admissible coordinates}

In this section, we derive the Einstein equations in admissible coordinates from the geometric formulation 
of the equations presented in the previous sections. We then deduce that 
areal coordinates exists and twists are constant. To begin with, we consider the constraint equations. 
 
\begin{lemma}[Weak version of the constraint equations in admissible coordinates]
 \label{prop:constlc}
Let $(\Sigma,h,K)$ be a weakly regular $T^2$--symmetric triple and consider the metric in admissible coordinates
as described in Lemmas~\ref{206} and \ref{lem:decompK}. 
Then, the weak version of the constraint equations defined in \eqref{weak-Hamilton}-\eqref{weak-momentum}
is equivalent to the following five equations:  
\begin{eqnarray} 
&&  \Rb_{\theta\theta} +\frac{1}{4 \Rb}\left( \Rb_\theta^2+ \Rbz^2 \right)
- \Rb_\theta \left( \nub_\theta-\Pb_\theta \right)  
  -  \Rbz \, \left(\nubz-\Pbz \right) \nonumber
\\
&&\hbox{} +\Rb \left( \Pb_\theta^2+ \Pbz^2 \right) +\frac{1}{4}\Rb \left(\Ab_\theta^2 +\Abz^2 \right)e^{4\Pb}
+\frac{1}{4} e^{-2\nub+4\Pb} \Rb^2 \left( \Gbz+\Ab \, \Hbz \right)^2+\frac{1}{4} e^{-2\nub} \, \Rb^3 \Hbz^2 = 0,  \label{201} \\
&&  (\Rbz)_\theta - (\nub_\theta-\Pb_\theta)\Rbz - (\nubz-\Pbz)\Rb_\theta +\frac{1}{2\Rb}\Rb_\theta \Rbz
+ \Rb \left( 2\Pbz \Pb_\theta+(1/2) \Abz \, \Ab_\theta e^{4\Pb}\right)= 0,
 \label{202} \\ 
\label{203}
&&\left(\Rb e^{4\Ub-2\nub} \left( \Gbz+\Ab \Hbz \right) \right)_\theta = 0, 
\qquad \qquad
\left(\Rb^3 e^{-2\nub} \, \Hbz +\Ab \, \Rb e^{4\Ub-2\nub} \left( \Gbz+\Ab \, \Hbz \right)\right)_\theta = 0, 
\end{eqnarray}
 in which $\Gbz, \, \Hbz, \, \Abz, \, \Ubz, \, \nubz$, and $\Rbz$ were introduced in Lemma~\ref{lem:decompK}, and 
 the equations above hold in the weak sense. 
\end{lemma}
 
Since the term $\Rb_{\theta\theta}$ is the only one containing second-order derivatives, 
if one evaluates the constraint equations above on a hypersurface of {\sl constant} area $R$, 
then no second-order derivative of the metric arises in the constraints;  
doing so suppresses the elliptic nature of these equations and is the key reason why 
the analysis of $T^2$--symmetric spacetimes is natural in areal coordinates. 

\begin{proof} 
We consider first the Hamiltonian equation \eqref{weak-Hamilton} and compute the normalized scalar curvature
$R_\text{norm}^{(3)}$ in terms of
the metric coefficients (introduced in Lemmas \ref{206} and \ref{lem:decompK}):
$$
\aligned
R_\text{norm}^{(3)}
& = 2 R_{ZZ}^{(3)}+h(Z,Z)\left( |\chi|^2-tr (\chi)^2 \right) 
\\
& = - 2Z(\Gamma^a_{aZ})+2\Gamma^a_{aZ}\Gamma^Z_{ZZ}-2 \Gamma^{a}_{bZ}\Gamma^b_{aZ} \nonumber
+h(Z,Z) \left(  |\chi|^2-tr (\chi)^2 \right). 
\endaligned
$$
Observe then that 
$$
\chi_{ab}= g \left(h(Z,Z)^{-1/2} Z , \nabla_{e_a}e_b \right) 
= h(Z,Z)^{1/2} \Gamma^Z_{ab}= - \frac{1}{2} \left(h^{ZZ}\right)^{1/2} Z(h_{ab}), 
$$
thus
$$
\aligned 
R_\text{norm}^{(3)}
= \, & - 2Z(\Gamma^a_{aZ})+2\Gamma^a_{aZ}\Gamma^Z_{ZZ}
-2 \frac{1}{2} h^{ac}Z(h_{cb}) \frac{1}{2} h^{bd}Z(h_{da}) 
\\
& + h(Z,Z) \Bigg(    \frac{1}{2} \left(h^{ZZ}\right)^{1/2} Z(h_{ab})  \frac{1}{2} \left(h^{ZZ}\right)^{1/2} Z(h_{cd})h^{ac} h^{bd} 
 - \left(\frac{1}{2} \left(h^{ZZ}\right)^{1/2} Z(h_{ab})h^{ab} \right)^2  \Bigg). 
\endaligned
$$
Hence, we obtain 
$$
R_\text{norm}^{(3)}=- 2Z(\Gamma^a_{aZ})+2\Gamma^a_{aZ}\Gamma^Z_{ZZ} 
- \frac{1}{4}Z(h_{cb}) Z(h_{da}) h^{ac}h^{bd}
-\frac{1}{4}\left( Z(h_{ab})h^{ab} \right)^2.  
$$
Using the identity 
$$
\frac{1}{2}h^{ab}Z(h_{ab})= Z( \ln R )= \Gamma^a_{aZ}=-h(Z,Z)^{1/2} tr \chi, 
$$
where $R^2= det (h_{ab})$, we find 
\be
\aligned
R_\text{norm}^{(3)}
= \, & - 2 Z\left( Z(\ln R) \right) +2 Z(\ln R)( -\frac{R_\theta}{2 R}+\nu_{\theta}-P_\theta)- \left( Z(\ln R) \right)^2
\\
& - \frac{1}{4}Z(h_{cb}) Z(h_{ad}) h^{ac}h^{bd} 
\\
= \, & -2 Z\left( \frac{R_\theta}{R} \right) 
+ 2 \frac{R_\theta}{R} \left( \nu_{\theta}-P_{\theta} \right)
-2 \left(\frac{R_\theta}{R}\right)^2-\frac{1}{4}Z(h_{cb}) Z(h_{da}) h^{ac}h^{bd}.  
\endaligned
\label{eq:r3normcor}
\ee
Thus, we need to evaluate $\frac{1}{4}Z(h_{cb}) Z(h_{da}) h^{ac}h^{bd}$. 

To this end, by decomposing $h_{ab}$ in the form 
$
h_{ab}= R \aleph_{ab}$ with
$det (\aleph_{ab}) = 1$, 
we obtain 
$$
\aligned
-\frac{1}{4}Z(h_{cb}) Z(h_{da}) h^{ac}h^{bd} 
& = -\frac{1}{4}Z(R\aleph ) (R \aleph)^{-1}Z(R\aleph)( R\aleph)^{-1} 
\\
& = -\frac{1}{4}Z(\aleph)\aleph^{-1}Z(\aleph)  \aleph^{-1} 
    -\frac{1}{2} Z(\aleph) \frac{R_{\theta}}{R} \aleph^{-1}
    -\frac{1}{4} 2 \left(\frac{R_\theta}{R}\right)^2 
\\
& = -\frac{1}{4}Z(\aleph) \aleph^{-1}Z(\aleph) \aleph^{-1}- \frac{1}{2}\left(\frac{R_\theta}{R}\right)^2, 
\endaligned
$$
where we used $tr Z(\aleph) \aleph^{-1}=0$ (since $\aleph$ has constant determinant). 
Therefore, we have
$$
\aleph= \left(\begin{array}{cc}
e^{2P} & Ae^{2P} \\
Ae^{2P} & A^2 e^{2P}+e^{-2P} 
\end{array}
\right), 
$$
and a straightforward computation gives
$$
-\frac{1}{4}Z(\aleph)\aleph^{-1}Z(\aleph) \aleph^{-1}= -2 P_{\theta}^2-\frac{1}{2}A_{\theta}^2 e^{4P},
$$
from which it follows that
\be
\label{0304}
R^{(3)}_{norm}=-2 Z\left( \frac{R_\theta}{R} \right) 
+ 2 \frac{R_\theta}{R} \left( \nu_{\theta}-P_{\theta} \right)
- \frac{5}{2} \left(\frac{R_\theta}{R}\right)^2 -2 P_{\theta}^2-\frac{1}{2}A_{\theta}^2 e^{4P}.
\ee
To complete the derivation of the Hamiltonian constraint equations in admissible coordinates, it remains 
to determine the contribution of the tensor $K$.

Note that
\be
\label{eq:contribKwH}
\aligned
h(Z,Z) \left( (tr K)^2-|K|^2 \right)
= 
& \frac{R_0^2}{R^2}+\left(K_{ZZ}\right)^2h^{ZZ}+2K_{ZZ} (h^{ZZ})^{1/2} \frac{1}{\Rb} \, \Rbz
\\
& -\left(K_{ZZ}\right)^2h^{ZZ}-2 h(Z,Z) K_{Za}K^{Za}-h(Z,Z)K_{ab}K^{ab} 
\\
= 
& \frac{R_0^2}{R^2}+2K_{ZZ} (h^{ZZ})^{1/2} \frac{1}{\Rb} \, \Rbz-2 h(Z,Z) K_{Za}K^{Za}-h(Z,Z)K_{ab}K^{ab},
\endaligned
\ee
and, as before, we define $\alephz$ by 
$$
\habz= \frac{1}{\Rb} \, \Rbz  \, h_{ab}+\Rb \, \alephz, 
$$
so that the trace of $\alephz$ vanishes:
$\alephz h^{ab}=0$, 
which follows from the definition of $\Rbz$. One then has 
$$
\aligned
-h(Z,Z)K_{ab}K^{ab}
& = - {1 \over 2} \, \left(\frac{1}{\Rb} \, \Rbz\right)^2-\frac{\Rb^2}{4}\alephz \alephcdz h^{ad} h^{bd}
 = - {1 \over 2} \, \left(\frac{1}{\Rb} \, \Rbz^2\right)^2-2 \Pbz^2 -\frac{1}{2}\Abz^2 e^{4P}
\endaligned
$$
and, moreover, 
$$
2K_{ZZ}(h^{ZZ})^{1/2} \frac{1}{\Rb} \, \Rbz= 2 \frac{1}{\Rb} \, \Rbz \Bigg( \nubz-\Pbz-\frac{1}{2\Rb} \, \Rbz \Bigg).
$$

We now consider the last term on the right-hand side of \eqref{eq:contribKwH}. 
From the definition of $\Gbz$ and $\Hbz$, it follows that
$$
\aligned
K_{Za}K^{Za}
& = K_{Za}K_{Zb}h^{ab}=K_{ZX}^2 h^{XX}+K_{ZY}h^{YY}+2 K_{ZX}K_{ZY}h^{XY}
\\
& = \frac{1}{4} e^{-2\nub+4\Pb} R \left( \Gbz+\Ab \Hbz \right)^2+ \frac{1}{4} e^{-2 \nu} \Rb^2 \Hbz^2.
\endaligned
$$
Collecting all the terms computed above, we have established that the Hamiltonian constraint equation reads 
\be
\aligned \nonumber
& - 2 \left( \frac{\Rb_\theta}{\Rb}\right)_\theta+ 2\frac{ \Rb_\theta}{R} \left( \nub_\theta-\Pb_\theta \right) 
  - (5/2) \left(\frac{ \Rb_\theta}{R} \right)^2
  - \frac{1}{2 \Rb^2} \left( \Rbz \right)^2
  + {2 \over \Rb} \Rbz \, \left(\nubz-\Pbz \right) 
\\
& -2 \left( \Pb_\theta^2+ \Pbz^2 \right) -\frac{1}{2}\left(\Ab_\theta^2 +\Abz^2 \right)e^{4\Pb}
-\frac{1}{2} e^{-2\nub+4\Pb} \Rb \left( \Gbz+\Ab \Hbz \right)^2-\frac{1}{2} e^{-2\nub}\Rb^2 \Hbz^2 = 0,  
\endaligned
\ee
From a straightforward density argument it then follows that this equation is equivalent to \eqref{201}. 
On the other hand, the twist equations \eqref{203} are obtained easily by observing that the geometric formulation is equivalent to 
$
Z \big( h(Z,Z)^{1/2} \big( \Rb \big)^{-1} K_a^Z \big)=0$
and, then, using the decomposition of $K$.

We now consider the last momentum constraint equation \eqref{202}.
For this, we compute all the terms appearing in the first equation of \eqref{weak-momentum} one by one.
For the first term we have 
$$
\aligned
-Z( tr^2K)
 = -Z\left( e^{-\nub+\Ub}\Rbz \, \big( \Rb \big)^{-1} \right) 
& = -Z( e^{-\nub} \Rbz )e^U \big( \Rb \big)^{-1}+\Rbz \, \Rb^{-2}\Rb_\theta e^{-\nub+\Ub}
-\Rbz \, \big( \Rb \big)^{-1}(-\nub_\theta-\Ub_\theta)e^{-\nub+\Ub} 
\\
& = -Z\left( e^{-\nub} \Rbz \Rb^{-1/2}\right) e^{\Pb}-Z(\Pb) e^{-\nub+\Pb}\Rbz  \, \Rb^{-1/2}. 
\endaligned
$$
For the second term we find 
$$
\aligned
-h(Z,Z)^{1/2} tr \chi K^Z_Z
& = -\Rb_\theta \big( \Rb \big)^{-1} K^Z_Z
\\
& = \Rb_\theta R^{-1/2} e^{-\nub + \Pb} \left( \nubz - \Pbz-2 \Rbz (2\Rb)^{-1} \right)
\endaligned
$$
and, for the last term,  
$$
\aligned
-\Gamma^a_{Zb} K_a^b 
& = -\frac{1}{2} h^{ac} Z(h_{bc}) h^{bd} K_{ad} 
 = -\frac{1}{4}  e^{-\nub + \Ub} h^{ac}h^{bd} Z(h_{bc}) \hbdz. 
\endaligned
$$
Using the fact that the traces of $\alephz$ and $Z(\aleph)$ vanish, we obtain 
$$
\aligned
-\Gamma^a_{Zb} K_a^b 
= -\frac{1}{2} e^{-\nu+P} \big(\Rb\big)^{-3/2} \Rbz \, \Rb_\theta-\frac{1}{4}e^{-\nu+P}
 \, \Rb^{1/2} h^{bd} h^{ac} Z(\aleph_{bc}) \alephadz, 
\endaligned
$$
Finally, in view of  
$$
-\frac{1}{4}e^{-\nu+P}\Rb^{1/2} h^{bd} h^{ac} Z(\aleph_{bc}) \alephadz
= - R^{1/2} e^{-\nub+\Pb} \left( 2 \Pbz \, \Pb_\theta+ \frac{1}{2}\Abz \, \Ab_\theta e^{4\Pb}\right), 
$$
the last momentum constraint equation follows by collecting all the terms. 
\end{proof}

%---------------------------------------------------------------------------------------------

\subsection{Evolution equations in admissible coordinates}
 
In this section, we rely on the geometric formulation introduced earlier and 
derive the Einstein equations in admissible coordinates.  

\begin{proposition}[Weak version of the evolution equations in admissible coordinates]
\label{prop:efeac}
Let $(\Mcal,g)$ be a weakly regular $T^2$-symmetric spacetime with admissible coordinates $(t,x,y,\theta)$. 
Then, $(\Mcal,g)$ satisfies the weak formulation of the Einstein equations \eqref{weak-Einstein}
if and only if $(\Sigma_t,h(t))$ satisfies the constraint equations on each slice and the following equations are satisfied:
\begin{eqnarray} 
\label{ee:wn}
&& 0=T(\rho \nu_t )-Z(\rho^{-1}\nu_\theta)-\rho (P_t-\frac{R_t}{2R})^2+\rho^{-1} (P_\theta-\frac{R_\theta}{2R})^2
    -\frac{e^{4P}}{4} \left( \rho A_t^2-\rho^{-1} A_\theta^2 \right)  +\frac{3}{4R^4}\rho^{-1}e^{2\nu}K^2,  \qquad \qquad 
\\
\label{ee:pw}
&& 0=\left(\rho\left(P_t+\frac{R_t}{2R}\right)\right)_t-\left(\rho^{-1}\left(P_\theta+\frac{R_\theta}{2R}\right)\right)_\theta 
- \rho \frac{R_t U_t}{R}+ \rho^{-1} \frac{R_\theta U_\theta}{R}  
-\frac{\rho}{2}e^{4P}A_t^2 -\frac{\rho^{-1}}{2}e^{4P}A_\theta^2,
\\
&& 0=\left( \rho A_t \right)_t - \left( \rho^{-1} A_t \right)_t 
          -\rho \frac{R_t A_t}{R}-\rho^{-1} \frac{R_\theta A_\theta}{R} 
       -4 \left( \rho^{-1} A_\theta \left( P_\theta + \frac{R_\theta}{2R}\right)-\rho A_t \left(P_t +\frac{R_t}{2R}\right) \right),
 \label{ee:aw} 
\\
&& 0=\left( \rho R_t \right)_t - \left( \rho^{-1} R_\theta \right)_\theta - \frac{1}{2 R^3} \rho^{-1} e^{2\nu} K^2,
 \label{ee:rw}
\\
&&0 = \left( \rho R^2 e^{-2\nu}H_t\right)_t, \qquad \qquad \qquad 0 = \left( \rho R^2 e^{-2\nu+4P}(G_t+AH_t) \right)_t. 
\label{twisteqcoor1} 
\end{eqnarray}
\end{proposition}

\begin{proof}
The equations \eqref{twisteqcoor1} are easily obtained from $R_{Za}=0$, as in Proposition~\ref{constant-twist}.
We consider now the equations $R_{cd}=0$ which read
$$
\aligned
0=
& T\left( n g(Z,Z)^{1/2} \Gamma^T_{dc} \right)+Z\left( n g(Z,Z)^{1/2} \Gamma^Z_{dc}\right)
   + n g(Z,Z)^{1/2} \bigg( \Gamma^a_{at} \Gamma^T_{dc} +\Gamma^a_{aZ} \Gamma^Z_{dc} -\Gamma^T_{dZ} \Gamma^Z_{tc}\bigg)
\\
& + n g(Z,Z)^{1/2} \bigg(-\Gamma^Z_{dt} \Gamma^T_{Zc}-\Gamma^T_{da} \Gamma^a_{tc} 
    -\Gamma^a_{dt} \Gamma^T_{ac} -\Gamma^Z_{da} \Gamma^a_{Zc} -\Gamma^a_{dZ} \Gamma^Z_{ac} \bigg).
\endaligned
$$
First, we note the following identities:
$$
\aligned
& n g(Z,Z)^{1/2}= \rho n^2,  \qquad \qquad \qquad \Gamma^T_{dc}= \frac{1}{2n^2}g_{dc,t},
\qquad \qquad 
T \left( n g(Z,Z)^{1/2} \Gamma^T_{dc} \right) = 1/2 T(\rho g_{dc,t} ),
\\
& Z\left( n g(Z,Z)^{1/2} \Gamma^Z_{dc}\right)=-1/2 Z (\rho^{-1} g_{dc,\theta} ),
\qquad \qquad 
\Gamma^a_{at} \Gamma^T_{dc} =\frac{R_t}{R} \frac{1}{2n^2} g_{dc,t},
\\ 
& \Gamma^a_{aZ}\Gamma^Z_{dc}= -\frac{R_\theta}{R} \frac{1}{2}g^{Z,Z} g_{dc,\theta},
\qquad \qquad \qquad \qquad 
     -2 \Gamma^T_{dZ} \Gamma^Z_{tc} = -1/2\frac{K_d  K_c}{R^2},
\\
& \Gamma^T_{da} \Gamma^a_{tc}=\frac{1}{4n^2} g_{da,t} g_{bc,t}g^{ab},
\qquad \qquad \qquad \qquad 
\Gamma^Z_{da} \Gamma^a_{Zc}=-\frac{1}{4\rho^2 n^2} g_{da,\theta} g_{bc,\theta}g^{ab},
\endaligned
$$
where $K_d=K$ if $d=y$ and $0$ otherwise.
To investigate the last two expressions in more details, we set $g_{ab} =: R \aleph_{ab}$.  Then, we have 
$$
g_{da,t} g_{bc,t}g^{ab}= 2 R_t \aleph_{dc,t} +\frac{R_t^2}{R}\aleph_{dc}+ R \aleph_{da,t} \aleph_{bc,t} \aleph^{ab}
=2 \frac{R_t}{R} g_{dc,t} -\frac{R_t^2}{R^2}g_{dc}+ R \aleph_{da,t} \aleph_{bc,t} \aleph^{ab}=
$$
Now we compute, for $d=c=x$, 
$$
\aligned
\aleph_{ax,t} \aleph_{bx,t} \aleph^{ab}
= &\left(2 P_t e^{2P} \right)^2\left( e^{-2P} +A^2 e^{2P} \right) 
\\
   & + 2 (-A e^{2P})\left( 2P_te^{2P} \right)\left( 2P_t A e^{2P}+A_t e^{2P} \right)
   +\left( 2 P_t A e^{2P}+A_t e^{2P} \right)^2e^{2P}
\\
= & e^{2P} \left( 4P_t^2+A_t^2 e^{4P} \right),
\endaligned
$$
for $d=c=y$, 
$$
\aligned
\aleph_{ay,t} \aleph_{by,t} \aleph^{ab}
= &\left( A_t e^{2P}+2P_t A e^{2P} \right)^2 \left( e^{-2P} +A^2 e^{2P} \right)
\\
   &+2(-A e^{2P}) \left( -2P_t e^{-2P}+2 A A_t e^{2P}+A^2 2 P_t e^{2P}\right)\left(A_t e^{2P}+2P_t A e^{2P} \right)
\\
  &+\left( -2P_t e^{-2P}+2 A A_t e^{2P}+A^2 2 P_t e^{2P} \right)^2 e^{2P}
\\
= & \left( 4 P_t^2 +A_t^2 e^{4P} \right) \left( e^{-2P} +A^2 e^{2P} \right), 
\endaligned
$$
and for $d=x$ and $c=y$
$$
\aligned
\aleph_{ax,t}\aleph_{by,t}\aleph^{ab}
= &\left( 2P_t e^{2P}\right) \left( A_t e^{2P}+2P_t A e^{2P} \right)\left( e^{-2P}+A^2e^{2P} \right) 
\\
   &+(-A e^{2P} ) \left( 2P-t e^{2P}\right) \left( -2P_t e^{-2P}+2 A A_t e^{2P}+2P_t A^2 e^{2P} \right) 
\\
   & +(-A e^{2P}\left(A_t e^{2P} +2P_t A^2 e^{2P} \right)^2
 \\
   & + e^{2P}\left( A_t e^{2P}+2P_t A e^{2P} \right)\left(-2P_t e^{-2P}+2 A A_t e^{2P} +2P_t A^2 e^{2P} \right) 
\\
= & 4 P_t ^2 A e^{2P}+A A_t^2 e^{6P}=A e^{2P} \left( 4P_t^2 + A_t^2 e^{4P}\right). 
\endaligned
$$
Similar expressions are valid for $\aleph_{ac,\theta}\aleph_{bd,\theta}\aleph^{ab}$ by 
replacing the $t$-derivatives by $\theta$-derivatives.

Putting everything together, we obtain for $d=c=x$ 
$$
\aligned
0 = 
& (1/2) \, T( \rho g_{xx,t} ) - (1/2) Z( \rho^{-1} g_{xx,\theta})-\frac{\rho}{2} \frac{R_t}{R} g_{xx,t}
      +\frac{\rho^{-1}}{2} \frac{R_\theta}{R} g_{xx,\theta} 
\\ 
& -\frac{\rho}{2} \left( 2\frac{R_t}{R}g_{xx,t}-\frac{R_t^2}{R^2}g_{xx}+ \left(4P_t^2+A_t^2 e^{4P}\right)g_{xx}\right) 
\\
& +\frac{\rho^{-1}}{2} \left(2\frac{R_\theta}{R}g_{xx,\theta}-\frac{R_\theta^2}{R^2}g_{xx}+\left( 4P_\theta^2+A_\theta^2 e^{4P}\right)g_{xx}\right).
\endaligned
$$
Finally, substituting $g_{xx}=\left( P_t+ \frac{R_t}{2R} \right)$, one obtains easily the equation \eqref{ee:pw}.
The wave equation for $A$, \eqref{ee:aw} is derived similarly.

To derive equation \eqref{ee:rw}, we note that, for sufficiently regular solutions, 
$$
\aligned
& g^{cd} \left( \frac{1}{2}T\left(\rho g_{cd,t} \right)-\frac{1}{2}Z\left( \rho^{-1} g_{cd,\theta}\right) \right)
\\
& = T\left(\rho  \frac{R_t}{R}\right)-Z\left( \rho^{-1} \frac{R_\theta}{R} \right) 
       -\frac{1}{2} \rho g^{ad}g^{cd}g_{ab,t}g_{cd,t}+\frac{1}{2} \rho^{-1} g^{ad}g^{cd}g_{ab,\theta}g_{cd,\theta},
\endaligned
$$
from which \eqref{ee:rw} follows. Finally, a straightforward density argument then shows that \eqref{ee:rw} remains true 
under our regularity assumptions. 
We now consider the equation $R_{ZZ}=0$ and, in view of the definition, we have
$$
\aligned
R_{ZZ}
= & T(\Gamma^T_{ZZ})-Z(\Gamma^T_{tZ})-Z(\Gamma^a_{aZ})
      -\Gamma^a_{Zb} \Gamma^b_{az} +\Gamma^a_{aZ}\Gamma^Z_{ZZ}
      +\Gamma^a_{at}\Gamma^T_{ZZ}+2\Gamma^T_{Za} \Gamma^a_{Zt}
\\
&+\Gamma^T_{ZZ} \left( \Gamma^T_{TT}-\Gamma^Z_{Zt}\right)
+\Gamma^T_{tZ}\left( \Gamma^Z_{ZZ}-\Gamma^T_{Zt}\right). 
\endaligned
$$
We evaluate sucessively each of the terms above  and obtain 
$$
\aligned 
T(\Gamma^T_{ZZ})&=T \Big( -\frac{1}{n}K(Z,Z) \Big) = T\Big( \frac{1}{2n^2}g_{ZZ,t} \Big),
\qquad \quad 
-Z(\Gamma^T_{tZ})=-Z\left( 1/n Z(n) \right),
\\
-Z( \Gamma^a_{aZ})&=-Z\left( Z (\ln R)\right),
\qquad \qquad \qquad \qquad \qquad 
\Gamma^a_{at}\Gamma^T_{ZZ}= T(\ln R) \frac{1}{2 n^2} g_{ZZ,t}. 
\endaligned
$$

The algebraic expressions of the products 
$$
\Gamma^a_{Zb} \, \Gamma^b_{aZ}, \qquad 
\Gamma^a_{aZ} \, \Gamma^Z_{ZZ},
\qquad 
\Gamma^T_{Za} \, \Gamma^a_{Zt}
$$
have 
already been computed in terms of the metric functions (for the derivation of the constraint equations), i.e. 
$$
\aligned
-\Gamma^a_{Zb} \Gamma^b_{aZ}&= -\frac{1}{4}Z(h_{cb})Z(h_{ad})h^{ac}h^{bd}=-2 P_\theta^2-\frac{1}{2}A_\theta^2e^{4P},
 \\
\Gamma^a_{aZ}\Gamma^Z_{ZZ}&= Z(\ln R) \left( -\frac{R_\theta}{2R}+\nu_\theta-P_\theta\right),
\\
\Gamma^T_{Za}\Gamma^a_{ZT}&= K_Z^a K_{aZ}=\rho^2 \left( 1/4 e^{-2\nu+4P}R(G_t+A H_t )^2
  +\frac{1}{4}e^{-2\nu}R^2 H_t^2\right)=\frac{1}{4}R^2e^{-2\nu}K^2,
\endaligned
$$
where $K$ denotes the only non-vanishing twist constant.
The last two terms in the definition of $R_{ZZ}$ give
$$
\aligned
\Gamma^T_{ZZ} \left( \Gamma^T_{TT}-\Gamma^Z_{Zt}\right)
& = \frac{1}{4n^4}g_{ZZ,t}g^{ZZ}T(\rho^{-2}),
\\   
\Gamma^T_{tZ}\left( \Gamma^Z_{ZZ}-\Gamma^T_{Zt}\right)
& = \frac{n}{2} g^{ZZ}Z(n)Z(\rho^2).
\nonumber
\endaligned
$$
Adding all the terms together, we obtain the equation 
$$
\aligned
0 = \, 
& T \left( \rho^2 \left(\nu_t-P_t-\frac{R_t}{2R} \right) \right)-\rho \rho_t \left( \nu_t -P_t -\frac{R_t}{2R} \right)
-Z \left( \nu_\theta-P_\theta - \frac{R_\theta}{2R} \right)
\\
& + \frac{\rho_\theta}{\rho}\left( \nu_\theta-P_\theta-\frac{R_\theta}{2R}\right)
  - Z\left(\frac{R_\theta}{R}\right)+\frac{R_t}{R} \rho^2 \left( \nu_t-P_t-\frac{R_t}{2R} \right)
\\
& +\frac{R_\theta}{R}\left( \nu_   \theta-P_\theta-\frac{R_\theta}{2R} \right) 
- 2 P_\theta^2-\frac{1}{2}A_\theta^2 e^{4P} +\frac{1}{4}R^2e^{-2\nu}K^2. 
\endaligned
$$
Equation \eqref{ee:wn} then follows by using equation \eqref{ee:pw} as well as \eqref{201} to eliminate all 
second-order derivatives of $P, R$. 
\end{proof}

%----------------------------------------------------------------------------------

\subsection{Field equations in conformal coordinates}
 
Applying Proposition \ref{prop:efeac} to the special case of conformal coordinates, we obtain the following result. 

\begin{proposition}[Weak version of the field equations in conformal coordinates]
Let $(\Mcal,g)$ be a weakly-regular $T^2$-symmetric spacetimes and let $(\tau,\xi,x,y)$ be
a system of conformal admissible coordinates for $(\Mcal,g)$ in which the metric\footnote{The variable $P$ 
is now replaced by $U:=P-1/2 \ln R$, as this leads to certain some computational simplifications later on.}
 takes the following form:
\be
\label{metric:conformal} 
\aligned 
g = & e^{2(\nu - U)} \big( - d\tau^2 +  d\xi^2 \big) 
 +e^{2U} \big(dx + A \, dy + \big(G + AH\big) \, d\xi \big)^2
+ e^{-2U} R^2 \big( dy + H \, d\xi \big)^2.
\endaligned 
\ee
Then, the weak version of the Einstein equations \eqref{weak-Einstein} 
is equivalent to the following system of evolution and constraint equations:  
\begin{enumerate}
\item Four constraint equations: 
\begin{eqnarray}
&&0=U_\tau^2 + U_\xi^2+\frac{e^{4U}}{4R^2}(A_\tau^2 + A_\xi^2) + \frac{R_{\xi\xi}}{R}-\frac{\nu_\tau \, R_\tau}{R}
- \frac{\nu_\xi R_\xi}{R}
+ {e^{2\nu} \over 4R^4} \, K^2, \label{ee:confconst}
\\
&&0=2U_\tau U_\xi +\frac{e^{4U}}{2R^2}A_\tau A_\xi+\frac{R_{\xi \tau}}{R}-\frac{\nu_\xi R_\tau}{R}-\frac{\nu_\tau R_\xi}{R}, 
\\
&&K_\xi=0,   \qquad \quad K_\tau=0. 
\end{eqnarray}

\item Four evolution equations:
\begin{eqnarray}
&& U_{\tau\tau}-U_{\xi \xi} =\frac{R_\xi U_\xi}{R}-\frac{R_\tau U_\tau}{R}+\frac{e^{4U}}{2R^2}(A_\tau^2-A_\xi^2), 
\\
&& A_{\tau\tau}-A_{\xi \xi} =\frac{R_\tau A_\tau}{R}-\frac{R_\xi A_\xi}{R}+4(A_\xi U_\xi-A_\tau U_\tau), 
\\
&& R_{\tau\tau}-R_{\xi \xi} = \frac{e^{2\nu}}{2R^3}K^2, 
\\
&& \nu_{\tau\tau}-\nu_{\xi \xi} =U_\xi^2-U_\tau^2+\frac{e^{4U}}{4R^2}(A_\tau^2-A_\xi^2)-\frac{3e^{2\nu}}{4R^4}K^2. \label{ee:nuce}
\end{eqnarray}

\item Two auxiliary equations:
\be
\label{ee:confaux}
G_\tau+A H_\tau =0, 
\qquad 
G_\tau=\frac{e^{2\nu}}{R^3}K. 
\ee
\end{enumerate}
\end{proposition}

%-------------------------------------------------------------------------------------------------------------------

\subsection{Field equations in areal coordinates}
\label{areal}
  
Similarly, in the case of areal coordinates, we obtain the following equations. 

\begin{proposition}[Weak version of the field equations in areal coordinates]
Let $(\Mcal,g)$ be a weakly regular $T^2$-symmetric spacetime and let $(t,x,y,\theta)$ be areal admissible coordinates.
Then, the weak version of the Einstein equations \eqref{weak-Einstein} is equivalent to
the following evolution and constraint equations: 
\begin{enumerate}

\item Four evolution equations for the metric coefficients $U, A, \eta, a$:
\begin{align}
\label{weakform1}
& (R \, a^{-1} U_R)_R - (R \, a \, U_\theta)_\theta 
= 2 R \, \Omega^U, 
\\ 
\label{weakform2}
& (R^{-1} \, a^{-1} A_R)_R - (R^{-1}  \, a \, A_\theta)_\theta 
= e^{-2U} \Omega^A, 
\\
\label{weakform3} 
&( a^{-1} \eta_R)_R - ( a \, \eta_\theta)_\theta 
= \Omega^\eta 
- R^{- 3/2}\big( R^{3/2} \big( a^{-1} \big)_R\big)_R,
\\
\label{eq:lnalphar}
&( 2 \ln a )_R = -R^{-3} K^2 \, e^{2 \eta}, 
\end{align} 
where 
%% $\eta=\nu + \ln a$ and 
the right-hand sides are defined by 
$$
\aligned 
& \Omega^U := (2R)^{-2} e^{4U} \, \big( a^{-1} \, A_R^2 - a \, A_\theta^2 \big), 
\qquad \qquad 
\Omega^A := 4 R^{-1}  e^{2U} \, \big( - a^{-1} \, U_R A_R + a \, U_\theta A_\theta \big), 
\\
& \Omega^\eta := (- a^{-1} U_R^2 + a \, U_\theta^2) 
+ (2R)^{-2} e^{4U} ( a^{-1} A_R^2 - a \, A_\theta^2). 
\endaligned 
$$

\item Two constraint equations for the metric coefficient $\eta$: 
\begin{align}
\label{weakconstraintsr}
\eta_R +\frac{1}{4} \, R^{-3} \, e^{2\eta}K^2&= a \, RE, 
\qquad\qquad 
\eta_\theta  = R \, F, 
\end{align}
where 
$$
\aligned 
& E:= \big( a^{-1} \, U_R^2 + a \, U_\theta^2 \big) 
   + (2R)^{-2} e^{4U} \big( a^{-1} \, A_R^2 + a \, A_\theta^2 \big),  
\\      
& F := 2 U_R U_\theta + 2 R^{-2}  e^{2U} A_R A_\theta. 
\endaligned 
$$

\item Four auxiliary equations for the twists: 
\be
\label{twisteq} 
\aligned
&\left( R \, e^{4U-2\eta} a \left( G_R+A H_R \right) \right)_\theta =0, 
\qquad 
\left(R^3 \, e^{-2\eta} a \, H_R \right)_\theta = 0, 
\\ 
&\left( R \, e^{4U-2\eta} \, a \left( G_R+A H_R \right) \right)_R = 0, 
\qquad 
\left(R^3 \, e^{-2\eta} \, a  H_R\right )_R=0.
\endaligned
\ee

\item Two equations for the metric coefficients $\displaystyle G, H$: 
\be
\label{metricGH}
\aligned 
&  G_R = - A K e^{2 \eta} \, a^{-1} R^{-3}, 
\qquad H_R = K e^{2 \eta} \, a^{-1} R^{-3}.
\endaligned
\ee
\end{enumerate}  
\end{proposition}

%==============================================================================
 
\section{First properties of weakly regular $T^2$-symmetric manifolds} 
\label{sec45}

\subsection{Properties of the area function}

In this section, we collect some properties of weakly regular $T^2$-symmetric manifolds which will be useful 
for the analysis of the initial value problem of Sections~\ref{sec5} and \ref{sec56}.
First of all, we derive some properties of the area function which are immediate consequences of the field equations. 
The first one is an additional $L^1$ regularity for the second derivatives of $R$.

From the constraint equations \eqref{201}-\eqref{202} and the assumed regularity, we see that the second-order derivatives 
$R_{\theta \theta}$ and $R_{t \theta}$ may be written as a sum of functions that have $L^1$ regularity, at least.  
Moreover, in view of the evolution equation \eqref{ee:rw}, $R_{tt}$ also has $L^1$ regularity. 
The additional regularity \eqref{addi} will be crucial to prove local well-posedness of the system in Section~\ref{sec5}. 
Furthermore, 
for sufficiently regular $T^2$-symmetric spacetimes, it is known that $\nabla R$ is timelike unless the spacetime is flat
\cite{Chrusciel,Rendall2}. That this is still true at our level regularity is the subject of the second statement below.

\begin{proposition}[Properties of the area function] 
\label{prop:arr} 
Let $(\Mcal,g)$ be a vacuum $T^2$-symmetric Lorentzian manifold 
and let $(t,\theta,x,y)$ be admisssible coordinates. 

1. The area function $R=R(t,\theta)$ has the following {\rm additional regularity properties:}
\be
\label{addi}
R \in L^\infty_\loc(W^{2,1}(S^1)), 
\qquad  
R_t \in L^\infty_\loc(W^{1,1}(S^1)), 
\qquad
R_{tt} \in L^\infty_\loc(L^1(S^1)). 
\ee

2. Provided this manifold is non-flat, that is, $g$ does not coincide with 
a smooth metric on $\Mcal$ whose curvature tensor vanishes, then the {\rm gradient $\nabla R$ 
is timelike,} i.e.  
\be
g(\nabla R, \nabla R) < 0 \quad \text{ in } \Mcal.
\ee  
\end{proposition}

This, in particular, establishes the {\sl existence of an areal coordinate system} of class $C^1$
for any weakly regular $T^2$-symmetric spacetime. Note also that an alternative statement of the 
second item of Proposition~\ref{prop:arr} 
is as follows: for any weakly regular $T^2$-symmetric initial data set, one has either 
$$
\Big( \Rbz \Big)^2 - \Big( \Rb_\xi \Big)^2 > 0,
$$
or else the initial data is trivial, i.e. $\Rb,\Ab,\Ub$ are constants and $\Rbz,\Abz,\Ubz$ vanish identically.

\begin{proof} It remains to establish the second item. 
We follow here an argument due to Chrusciel~\cite{Chrusciel} and 
Rendall~\cite{Rendall2} for sufficiently regular spacetimes. In our weak regularity class, 
it follows that the norm $g(\nabla R, \nabla R)$ is a measurable and bounded function 
defined almost everywhere., at least. however, it follows from the first item of this proposition
that $R$ is actually of class $C^1$ in both variables $t,\theta$.
Define $\lambda^\pm := \rho R_t\pm R_\theta$ and $H := \nu_\theta-P_\theta+\nu_t -P_t$.
Taking the sum and the difference of the two constraint equations \eqref{201}-\eqref{202} lead to
$
Z(\lambda^\pm)=-\lambda^\pm H +N,
$
where $N$ can be checked to belong to $L^\infty_{loc}(L^2(S^1))$ and be non-positive almost everywhere. 
From the last two equations and the continuity of $\lambda^\pm$, it follows that 
either $\lambda^+=0$ or $\lambda^+$ never vanishes. A similar conclusion holds for $\lambda^-$. 
Moreover, periodicity of $R$ excludes the possibility that $\lambda^+ > 0$ and $\lambda^- <0$, 
as well as the possibility that $\lambda^+ < 0$ and $\lambda^- >0$. Thus, it follows that 
either $\lambda^+ \lambda^- > 0$ or else
$\lambda_\pm=0$ and $N=0$. In the latter case, $U$ and $A$ are constant functions and the spacetime is flat. 
\end{proof}

%-------------------------------------------------------------------------

\subsection{From conformal to areal coordinates}
\label{sec52}

To solve the initial value problem, we will need two different coordinate systems, 
one being better suited for the local-in-time analysis (i.e.~the conformal coordinate system) and 
the other being better suited for the long-time control of the growth of the initial norms (i.e.~the areal coordinate system). 
However, since the construction of these coordinates depends on the metric, the weak regularity of the metric 
imposes a restriction on  
the regularity of these coordinates as functions of the original coordinates. In this section, 
we prove that despite this difficulty, 
the weak regularity of the metric coefficients is invariant under such a transformation.
We begin with the following technical result which establishes additional regularity in time. 

\begin{lemma}[Additional regularity in time] 
\label{lem:contl2s}
Consider a $2$-dimensional Lorentzian manifold $(\mathcal{Q},\gt)$
\be
\label{500}
\mathcal{Q} := [t_0,t_1) \times S^1, \qquad \quad
\gt := -\rho \, dt^2+\rho^{-1} \, d\xi^2, 
\ee
where $\rho=\rho(\tau,\xi)$ is assumed to be class $C^1$.
Let $f \in L^\infty_\loc([t_0,t_1), H^1(S^1)) \cap W^{1,\infty}_\loc([t_0,t_1), L^2(S^1))$ be a weak solution to the wave equation
$$
\Box_\gt  f = q, 
$$
where the right-hand side satisfies $q \in L_\loc^2([t_0,t_1), L^2(S^1))$. 
Then, $f$ is actually more regular and belongs to  
$C^0 (H^1(S^1)) \cap C^1(L^2(S^1))$.
\end{lemma}

\begin{proof} 
Let a time interval $[t_0,t_2] \subset [t_0, t_1)$ be fixed, 
let $f_0^\eps$, $f_{t,0}^\eps$, and $q^\eps$ be smooth functions approximating $f(t_0,\cdot)$, $f_t(t_0,\cdot)$,
and $q$ in the topology of $H^1(S^1)$, $L^2(S^1)$, and $L^2([t_0,t_2]\times S^1)$, respectively. 
Let $f^\eps$ be the solution to the corresponding wave equation with source $q^\eps$ and initial data $(f_0^\eps,f_{t,0}^\eps)$. 
Observe that $f^\eps$ is of class $C^1$, at least, and set $\Delta f := f - f^\eps$,
 $\Delta q : = q-q^\eps$, etc. 
Then, a standard energy estimate implies for all $t \in [t_0,t_2]$ 
$$
\| \Delta f_t (t)||^2_{L^2} + || \Delta f_\theta (t) ||^2_{L^2}
\lesssim 
|| f_0-f_0^\eps ||^2_{H^1(S^1)}+||f_{t,0}-f^\eps_{t,0}||^2_{L^2(S^1)}+|| \Delta f_t \, \Delta q||_{L^1([t_0,t_2] \times S^1)},
$$
where the implied constant depends on the Lipschitz constant of $\rho$ and $t_0, t_1$. 
Applying Cauchy-Schwarz to the latter term above, we arrive at 
a Lipschitz continuity estimate which implies convergence of $f^\eps$ toward $f$. 
\end{proof}

Note that in conformal coordinates, $\rho=1$ in \eqref{500} and hence is indeed $C^1$, 
while for areal coordinates $\rho=a^{-1}$ for which we prove $W^{2,1}$ (thus $C^1$)
regularity in Section~\ref{sec56}. 
Moreover, we will prove later in Section~\ref{sec5} that the source terms in the wave equations for 
$R,U,A$ are indeed in $L^2_\loc$ so that the above lemma applies with $(\mathcal{Q},\gt )$ chosen to be 
the quotient space $\Mcal/T^2$ with its induced metric and differential structure 
given by either conformal or areal coordinates. 

These observations lead us to the following important result which, in particular, shows
that the regularity of the metric functions does not change under a change of coordinates 
from conformal to areal coordinates or vice versa.

\begin{proposition}[From conformal to areal coordinates and vice-versa]
\label{coordcomp}
Let $(\Mcal_{\mathcal{C}},g)$ be a weakly regular vacuum $T^2$-symmetric spacetime and assume that $\mathcal{C}=(\tau,\xi,x,y)$ 
are admissible conformal coordinates. It follows from Proposition~\ref{prop:arr} 
that there exists an areal coordinate system $\mathcal{A}=(R,\theta,x,y)$ 
(with $\nabla R$ timelike) that is $C^1$-compatible with $\mathcal{C}=(\tau,\xi,x,y)$. 
Let $\Mcal_{\mathcal{A}}$ be the topological manifold $\Mcal_{\mathcal{C}}$ endowed 
with the (unique) $C^\infty$-differential structure compatible with $(R,\theta,x,y)$. 
Then, $(R,\theta,x,y)$ are admissible coordinates for the manifold $(\Mcal_{\mathcal{A}},g)$
and, in particular, the Einstein field equations hold in areal coordinates. 

Similarly, let $(\Mcal_{\mathcal{A}},g)$ be a weakly regular vacuum $T^2$-symmetric spacetime and 
let $\mathcal{A}=(R,\theta,x,y)$ be admissible areal coordinates. 
It follows from Lemma \ref{lem:exiconf} and the improved regularity on the coefficient $a$ 
that there exists a conformal coordinate system $\mathcal{C}=(\tau,\xi,x,y)$
that is $C^1$-compatible with $\mathcal{A}=(\tau,\xi,x,y)$. 
Let $\Mcal_{\mathcal{C}}$ be the topological manifold $\Mcal_{\mathcal{A}}$ endowed 
with the (unique) $C^\infty$-differential structure compatible with $(\tau,\xi,x,y)$. 
Then, $(\tau,\xi,x,y)$ are admissible coordinates for the manifold $(\Mcal_{\mathcal{C}},g)$
and, in particular, the Einstein field equations hold in conformal coordinates. 
\end{proposition}

\begin{proof} We establish this result for the transformation from conformal to areal coordinates,
the proof of the second statement being similar. 
Note first that since the change of coordinates is of class $C^1$, 
the measures of volume associated with $(\tau, \xi)$ and $(R,\theta)$  
are equivalent, 
hence we may talk about $L^p$ functions unambigously.
Lemma \ref{936} ensures that the assumptions of Lemma \ref{lem:contl2s} are satisfied. 
Standard energy estimates and density argument then show that $U, A$, as functions of $(R,\theta)$, 
are of class $C_R(H^1_{\theta}) \cap C^1_R(L^2_\theta)$. By density, the weak version of the Einstein equations 
must hold in areal coordinates. It then follows from the constraint equations that $\nu$ and $a$ are 
in $C^0_R(W^{1,1}_\theta)\cap C^1_R(L^1_\theta)$ and $C^1_{R,\theta}$, respectively. 
Note finally that,  by construction, $R$ is $C^\infty$ in areal coordinates.
\end{proof}
 
%-------------------------------------------------------------------------

\subsection{Regularization of initial data sets with constant area of symmetry}   

We now establish that any given weakly regular $T^2$-symmetric initial data set with constant area $R=R_0$ 
can be uniformly approximated by smooth $T^2$-symmetric initial data set. 
In view of (\ref{weakconstraintsr}),  
the initial data for the functions $G, H$ do not enter the constraint equations, 
hence we may suppress here any reference to these functions. Therefore, we set 
$$
\Xb := \big( \Ubzero, \Abzero, \Ubone, \Abone, \abar, \etab_0, \etab_1 \big), 
$$
which represents an initial data set for the reduced equations (\ref{weakform1}).
We are interested in the existence of suitable regularizations of $\Xb$.

\begin{lemma}[Regularization of initial data sets in areal coordinates] 
\label{411} 
Let $\Xb$ be an initial data set for the reduced Einstein equations, in particular satisfying the constraint equations 
\eqref{weakconstraintsr} (with $U_0$ replaced by $\Ubzero$, etc.). 
Then, there exist smooth functions defined on $S^1$ 
$$
\Xb^\eps = \big( \Ubzero^\eps, \Abzero^\eps, \Ubone^\eps, \Abone^\eps, \abar^\eps, \etab^\eps_0, \etab^\eps_1 \big), 
\qquad \eps \in (0,1), 
$$
referred to as a {\bf regularized initial data set,} 
such that $\Xb^\eps$
satisfies the reduced Einstein constraint equations \eqref{weakconstraintsr} 
and converges almost everywhere (for the Lebesgue measure on $S^1$) with, moreover, 
$$
\aligned
& (\Ubzero^\eps, \Abzero^\eps, \Ubone^\eps, \Abone^\eps) \to (\Ubzero, \Abzero, \Ubone, \Abone) 
\quad \text{ in } L^2(S^1),
\\
& \abar^\eps \to \abar \quad \text{ in } W^{1,\infty}(S^1), 
\\
&
(\etabzero^\eps, \etabone^\eps) \to (\etabzero, \etabone) \quad \text{ in } L^1(S^1). 
\endaligned
$$
\end{lemma} 

Importantly, the method of proof of this lemma given now can also be applied to 
establish the existence of weakly regular $T^2$-symmetric initial data sets with constant $R$ 
whose regularity is precisely the one introduced in Definitions\ref{def21} and \ref{def:t2tri},
apart from the assumptions on $R$. 

\begin{proof} By convolution of the data $\Xb$ and relying
on the regularity assumed on the initial data set, 
one can define smooth functions 
$
\Ubzero^\eps$, $\Abzero^\eps$, $\Ubone^\eps$, $\Abone^\eps$, $\abar^\eps 
$
defined on $S^1$ such that, as $\eps \in (0,1)$ approaches $0$, the functions 
$\Ubzero^\eps$, $\Abzero^\eps$, $\Ubone^\eps$, $\Abone^\eps$ converges in $L^2(S^1)$ 
toward $\Ubzero$, $\Abzero$, $\Ubone$, $\Abone$, respectively,  
while $\abar^\eps$ converges to $\abar$ in $W^{1,\infty}(S^1)$. 

In order to obtain a complete set of regularized initial data, we also have to regularize  
the functions $\etabzero$ and $\etabone$ in such a way 
that the constraint equations \eqref{weakconstraintsr} hold for each $\eps$. 
To this end, to each regularized set 
$\Yb^\eps := (\Ubzero^\eps, \Abzero, \Ubone^\eps,\Abone^\eps, \abar^\eps)$, 
we associate the function and scalar
$$
\aligned 
\omega[\Yb^\eps]  
& := 2 R \, \big( \Ubzero^\eps \Ubone^\eps + R^{-2} e^{2\Ub^\eps} \Abzero^\eps \Abone^\eps \big),
\qquad \qquad 
\Omega[\Yb^\eps]  := \int_{S^1} \omega[\Yb^\eps] \, d\theta. 
\endaligned
$$ 
It follows 
% from the equations \eqref{weakconstraints} 
that the function $\omega[\Yb^\eps]$ converges in the space $L^1(S^1)$ toward $\etab_1$, 
and that the sequence $\Omega^\eps$ 
(is uniformly bounded and) converges to $0$. 

Assuming first that we have been able to choose the regularization $\Yb^\eps$ 
so that
$\Omega[\Yb^\eps] = 0$ for each $\eps$, and let us fix an arbitrary value $\theta_* \in S^1$. 
Then, by defining 
$$ 
\etab^\eps(\theta) := \eta(\theta_*) + \int_{\theta_*}^\theta \omega[\Yb^\eps]  \, d\theta',  
$$
we see that the functions $\etab^\eps$ converge in $W^{1,1}(S^1)$ toward the initial data $\etab$.  
We can also define the function $\etabzero^\eps$ by
\be
\label{eq:etabz}
\aligned
& (\abar^\eps)^{-1} \, \etabzero^\eps
+ (\abar^\eps)^{-1} \, \frac{e^{2\etab^\eps} \, K^2}{4R^3} 
= R \, E[\Yb^\eps]
\\
& E[\Yb^\eps] := \,  (\abar^\eps)^{-1} \, (\Ubzero^\eps)^2 + \abar^\eps (\Ubone^\eps)^2  
 + (2R)^{-2} \, e^{4\Ub^\eps} \, \Big( (\abar^\eps)^{-1} \, (\Abzero^\eps)^2
+ \abar^\eps (\Abone^\eps)^2 \Big). 
\endaligned
\ee 
Here, the constant $K$ is precisely the twist constant of the original initial data set.
The right-hand side of (\ref{eq:etabz}) converges in $L^1(S^1)$
to the right-hand side of (\ref{weakconstraintsr}). 
We also claim that $(\abar^\eps)^{-1} \, e^{2\etab^\eps} \, K^2 \, R^{-3}$
converges to $(\abar)^{-1} \, e^{2\etab} \, K^2 \, R^{-3}$ in $L^1(S^1)$.
Indeed, $\abar^\eps$ converges to $\abar$ in $W^{1,\infty}(S^1)$ and thus in $L^\infty(S^1)$,
and, moreover,  $e^{2\etab^\eps}$ converges to  $e^{2\etab}$ in $L^1(S^1)$, 
as follows from the convergence of $\etab^\eps$ in $W^{1,1}$ and, thus, in $L^\infty(S^1)$.
Therefore, we see from equation (\ref{weakconstraintsr}) that $\etabzero^\eps$ converges
in $L^1(S^1)$ to $\etabzero$, and passing to a subsequence if necessary, we may also assume
almost everywhere convergence. Thus, $\etab^\eps$ and $\etabzero^\eps$
satisfy the requirement of the lemma. 

It remains to determine a regularization such that $\Omega[\Yb^\eps]$ vanishes.  
We start from an arbitrary regularized set $\Yb^\eps$ that may not satisfy the constraints. 
Without loss of generality, we may assume that
$\int_{S^1} (\Ubone)^2$ or $\int_{S^1} (\Abone)^2 > 0$ (or both) are positive. 
For, if both of these terms vanish, $\Ub$ and $\Ab$ are almost everywhere constant,
say $\Ub=U_*$, $\Ab=A_*$,  
and choosing for regularization $\Rcal := (U_*, \Ubzero^\eps, A_*, \Abzero^\eps, \ab^\eps)$, 
we  obtain $\Omega^\eps_\Rcal = 0$.

Assume, for instance, that $\int_{S^1} (\Ubone)^2 =: c$ is positive,  
the case that $\int_{S^1} (\Ab_\theta)^2$ is positive being similar. 
For all $\eps>0$ sufficiently small, we have 
$\int_{S^1} (\Ubone^\eps)^2 > c/2$ and, by assumption, 
$\Omega^\eps_\Rcal$ goes to zero with $\eps$. 
Setting  
$$
\delta^\eps := -\frac{\Omega[\Yb^\eps]}{2R\int_{S^1}(\Ub^\eps_\theta)^2},
$$
we now claim that 
$$
\Yb' := \big( \Ub^\eps, \Ubzero^\eps + \delta^\eps \, \Ubone^\eps, \Ab^\eps, \Abzero^\eps, \abar^\eps \big)
$$ 
satisfies the constraints. 
Indeed, one can check that, by construction, $\Omega[\Yb']=0$ and the conclusion follows from the estimate 
$$
|\delta^\eps| \leq \frac{\big|\Omega[\Yb^\eps]\big|}{2c}, 
$$
where the right-hand side converges to $0$ as $\eps \to 0$.
\end{proof}

%---------------------------------------------------------------------------- 

\subsection{Regularization of generic initial data sets}

In passing, we now establish a stronger version of the previous regularization scheme 
which is of independent interest and applies to generic initial data sets. 
This result is not needed for our main result in this article, but is included for completeness. 

\begin{proposition}[Regularization of generic initial data sets]
\label{prop:reggid}
Let $(\Sigma,h,k)$ be a weakly regular $T^2$-symmetric Riemannian manifold
 satisfying the weak version of the vacuum constraint equations. 
Assume that either the area of the symmetry orbits $\Rb=const$ on $\Sigma$ 
or the following condition holds (using the notation of Lemmas~\ref{206} and \ref{lem:decompK}): 
\be
\label{3399}
\int_0^{2\pi} f(\xi) e^{-\int^{2\pi}_{\xi'} gd\xi''} d\xi' \neq 0,
\ee
where $f$ and $g$ are defined by 
$$
\aligned
f = &\frac{\Rb_{\xi}}{\Rbz^2-\Rb_{\xi}^2}\frac{1}{2\Rb},
\\
g = & 2 \, \left( \Rbz^2-\Rb_\xi^2 \right)^{-1} \Bigg( -\Rb \Rb_{\xi} 
\left( \Ubz^2+\Ub_{\xi}^2 + \frac{e^{4\Ub}}{4\Rb^2}\left(\Abz^2+\Ab_{\xi}^2\right)
    +\frac{\Rb_{\xi \xi}}{\Rb}\right)
\\
& \quad +\Rb \Rbz \left((2 \Ubz \Ub_{\xi} 
+ \frac{e^{4\Ub}}{2\Rb^2}\Abz \Ab_{\xi} + \frac{\Rbz_{\xi}}{\Rb} \right) \Bigg),
\endaligned
$$
$\Ub=\Pb+\frac{1}{2}$, and $\Ubz=\Pbz+\Rbz/(2 \Rb)$.
Then, there exists a smooth family of $T^2$-symmetric metrics $h^\eps$ (parameterized by $\eps \in (0,1)$ )
invariant by the same $T^2$ action, together with  
a smooth family of $T^2$-symmetric, symmetric $2$-tensors $k^\eps$ invariant by the same $T^2$ action
such that the triple $(\Sigma, h^\eps,k^\eps)$ 
satisfies the constraints in the same conformal system of coordinates 
and $(h^\eps,k^\eps)$ converges to $(h,k)$ as $\eps$ goes to $0$ in the following topology: 
$$
\aligned
& \Ub^\eps,\Ab^\eps \rightarrow \Ub,\Ab \quad \text{ in } H^1(S^1), 
\qquad
& \Ubz^\eps,\Abz^\eps \rightarrow \Ubz,\Abz \quad \text{ in } L^2(S^1), 
\qquad 
\nub^\eps \rightarrow \nub \quad \text{ in } W^{1,1}(S^1)
\\
& \nubz^\eps \rightarrow \nubz \quad \text{ in } L^1(S^1), 
\qquad 
& \Rb^\eps \rightarrow \Rb \quad \text{ in } W^{2,1}(S^1), 
\qquad \qquad 
\Rbz^\eps \rightarrow \Rbz \quad \text{ in } W^{1,1}(S^1), 
\endaligned
$$
and the twist coefficients associated with $(h,k,X,Y)$ converge to the twists coefficients associated 
with $(h^\eps,k^\eps,X,Y)$.
\end{proposition}

\begin{proof}
For simplicity in the notation, we drop the bars 
and write $R_\tau$ for $\Rbz$. Without loss of generality, we may also assume that the initial data are not trivial 
and, in particular, that $A,U$ are not contants and $A_\tau,U_\tau$ do not vanish identically.
Since $R_\tau, R_{\xi}$ are of class $C^1$ at least 
and $R_{\tau}^2-R_{\xi}^2 > 0$ (cf.~Proposition~\ref{prop:arr}), 
by a continuity argument we obtain the lower bound 
$
R_{\tau}^2-R_{\xi}^2 \geq c > 0
$
for some contant $c$. It follows that the constraint equations are equivalent to 
$$
\aligned
0 & = R R_{\xi} \left( U_\tau^2+U_{\xi}^2 + \frac{e^{4U}}{4R^2}\left(A_\tau^2+A_{\xi}^2\right)+\frac{R_{\xi \xi}}{R} 
+\frac{e^{2\nu}K^2}{4R^2}\right) 
      - \nu_\xi R_{\xi}^2-\nu_{\tau} R_\tau R_\xi,
\\
0 & = R R_\tau \left( 2 U_\tau U_{\xi} + \frac{e^{4U}}{2R^2}A_\tau A_{\xi} + \frac{R_{\xi\tau}}{R} \right)
-\nu_\xi R_{\tau}^2-\nu_{\tau} R_\tau R_\xi, 
\endaligned
$$
where $K$ denotes the twist constant associated with $Y$, the other twist constant being set to $0$ 
(without loss of generality in view of Proposition~\ref{constant-twist}). 
Taking the difference of the last two equations, we obtain 
$$
\aligned 
\nu_{\xi} +\frac{R R_{\xi}}{R_\tau^2-R_{\xi}^2}\frac{K^2}{4R^2}e^{2\nu}
= & 
\left( R_\tau^2-R_\xi^2 \right)^{-1} \Bigg( -R R_{\xi} \left( U_\tau^2+U_{\xi}^2 
   + \frac{e^{4U}}{4R^2}\left(A_\tau^2+A_{\xi}^2\right)+\frac{R_{\xi \xi}}{R}\right) 
\\
& +R R_\tau \left( 2 U_\tau U_{\xi} + \frac{e^{4U}}{2R^2}A_\tau A_{\xi} 
+ \frac{R_{\xi\tau}}{R} \right) \Bigg).
\endaligned
$$ 

Let us rewrite this equations as 
$\left( e^{2\nu} \right)_{\xi}e^{-2\nu} + K^2 f e^{2\nu} =g$
with 
\begin{eqnarray}
f&=&\frac{R R_{\xi}}{R_\tau^2-R_{\xi}^2}\frac{1}{2R^2} \nonumber \\
g&=&2\left( R_\tau^2-R_\xi^2 \right)^{-1} \Bigg( -R R_{\xi} \left( U_\tau^2+U_{\xi}^2 
+ \frac{e^{4U}}{4R^2}\left(A_\tau^2+A_{\xi}^2\right)+\frac{R_{\xi \xi}}{R}\right) \nonumber\\
&&\hbox{}+R R_\tau \left( 2 U_\tau U_{\xi} 
+ \frac{e^{4U}}{2R^2}A_\tau A_{\xi} 
+ \frac{R_{\xi\tau}}{R} \right) \Bigg). \nonumber
\end{eqnarray}
Setting $\phi=e^{-2\nu}$, then we have
$\phi'+ \phi g =K^2f$ which may be solved as 
\begin{eqnarray} 
\label{nuformde}
e^{-2\nu}(\xi)=e^{-2\nu}(0) e^{-\int^{\xi}_0g(\xi') d\xi'} 
+ K^2 \int_0^{\xi} f(\xi) e^{-\int^{\xi}_{\xi'} g d\xi''} d\xi'.
\end{eqnarray}
We now use the above formula to define the regularized $\nu$, as follows. 
Regularize $R$, $R_\tau$, $U$, $U_\tau$, $A$ and $A_\tau$ by convolution.   

If $R_\xi=0$ uniformly, then $R=const$ initially, and hence, 
we may apply the regularization scheme developed in areal coordinates in the previous section. 
Thus, let us assume that $R_\xi \neq 0$ 
and that the technical assumption of the lemma holds.

Define $K^\eps$ by 
$$
(K^\eps)^{2} := e^{-2\nu(0)}\left( 1- e^{-\int^{2\pi}_0 g^\eps(\xi')d\xi'}\right) 
\left( \int_0^{2\pi} f^\eps(\xi) e^{-\int^{2\pi}_{\xi'} g^\eps d\xi''} d\xi'. 
\right)^{-1},
$$
and observe that from the strong convergence of $f^\eps$ to $f$ 
and $g^\eps$ to $g$ it follows that $(K^\eps)^2$ is well-defined 
and converges to $K^2$ (as $\eps$ goes to $0$). 
Define now $\nu^\eps$ as 
\begin{eqnarray} \label{nuform}
e^{-2\hat{\nu}^\eps}(\xi)=e^{-2\nu}(0) e^{-\int^{\xi}_0g^\eps(\xi') d\xi'} 
+ (K^\eps)^2 \int_0^{\xi} f^\eps(\xi) e^{-\int^{\xi}_{\xi'} g^\eps d\xi''} d\xi'.
\end{eqnarray}
It follows from the definition of $K^\eps$ 
that $\nu^\eps$ is periodic with period $2\pi$ 
(and hence can be identified with a smooth function on $S^1$) 
and converges to $\nu$ in $W^{1,1}$ as $\eps$ goes to $0$. 
Finally, we define $\nu_\tau^\eps$ 
so that the remaining constraint equation holds, i.e. by 
$$
\aligned
\nu_\tau^\eps
= 
& -\frac{1}{(R^\eps_\tau)^2-(R_{\xi}^\eps)^2}
\Bigg( R^\eps R^\eps_\xi\left( 2 U^\eps_\tau U^\eps_\xi+\frac{e^{4U^\eps}}{2(R^\eps)^2}A^\eps_\tau A_\xi^\eps+R^\eps_{\xi \tau}\right)
\\
& + R^\eps R_\tau^\eps\left( (U_\tau^\eps)^2
  +(U^\eps_\xi)^2+\frac{e^{4U^\eps}}{4R^\eps}\left( (A_\tau^\eps)^2
  +(A_\xi^\eps)^2\right)+ \frac{R^\eps_{\xi \xi}}{R^\eps} + \frac{e^{2\nu^\eps}(K^\eps)^2}{4(R^\eps)^2}\right)
\Bigg).
\endaligned
$$
The convergence of the right-hand side and the original validity of the constraints then imply 
that $\nu^\eps_\tau$ converges in $L^1$ to $\nu_\tau$.
\end{proof}

%===================================================================================
 
\section{Local geometry of weakly regular $T^2$--symmetric spacetimes} 
\label{sec5}

\subsection{Strategy of proof}

For the existence of weak solutions to the initial value problem associated with the Einstein equations 
under the assumed symmetry, we proceed as follows. 
\begin{enumerate}

\item[Step 1.] {\sl Local existence in conformal coordinates and blow-up criterion.}
First, we prove a compactness property for solutions to the conformal equations. 
This yields us, for any weakly regular initial data set, the existence of a local-in-time solution,
defined on a sufficiently small interval of (conformal) time 
$[\tau_1,\tau_1+\eps)$, where $\eps$ only depends on natural (energy-like) norms corresponding to the assumed (weak) regularity 
of the initial data. Together with this local existence result, we obtain a continuation criterion. 
This result is stated precisely in Theorem~\ref{prop:leccc}, below, and the rest of this section is devoted to its proof.

\item[Step 2.] {\sl Local existence in areal coordinates.} 
We can always arrange that the condition $\tau=\tau_1$ coincides with $R=R_1$ 
(cf.~the construction of conformal coordinates in Lemma~\ref{lem:exiconf}), 
and
since $R$ is strictly increasing with $\tau$ and weak solution to the conformal equations can be transformed 
to weak solutions to the areal equations (i.e.~the equations derived in Proposition~\ref{coordcomp}), 
we obtain a local solution to the
 equations in areal coordinates, defined on a small interval of areal time $[R_1,R_1+\eps)$. 
Moreover, we also obtain a continuation criterion in areal coordinates which states the solution ceases to exist
only if the natural energy-like norms are blowing up. 

\item[Step 3.] {\sl Global existence in areal coordinates.} 
Finally, performing a further analysis of the Einstein system in areal coordinates, we obtain 
a global-in-time control of the natural norms which will lead us to the desired global existence result. 
This step will be presented in Section~\ref{sec56}, below. 
\end{enumerate}
The above strategy is motivated by the following observations. Due to the quasilinear structure of the equations in areal coordinates, 
one cannot directly estimate the difference of solutions. While we do obtain a priori estimates 
for solutions in Step $3$, these estimates do not provide sufficiently strong compactness properties. 
A possible strategy (for general quasilinear systems) in order  to 
cope with this difficulty would be to prove compactness in a weaker functional space. 
However, under our weak regularity assumptions, 
the natural functions spaces for $U, A$ which one may think of 
would be $L^2$ (instead of $H^1$); 
however, one cannot control the behavior of the remaining metric coefficients $a, \nu$ by the $L^2$ norm of $U, A$. 
This is the reason why we propose here to rely on conformal coordinates (in which the equations become semi-linear)
in order to prove local-well-posedness. However, in conformal coordinates, the natural energy
associated with $U,A$ fails to be a-priori bounded, and this is why only local-in-time existence is obtained 
in conformal coordinates, one must introduce areal coordinates to get a  global-in-time result.  
In the rest of this section, we discuss the issue of local existence in conformal coordinates. 

%-----------------------------------------------------------------------------------------------------

\subsection{Local existence}

As explained above, the aim of this section is to prove the following result.

\begin{theorem}[Local existence in conformal coordinates]
\label{prop:leccc}
Let $(\Sigma,h,K)$ be a weakly regular $T^2$-symmetric initial data set. 
Assume that $(\Sigma,h,K)$ admits a regularization $(\Sigma^\eps,h^\eps,K^\eps)$ as described in Lemma~\ref{prop:reggid},
which, for instance, applies if the associated area function $R$ is constant on $\Sigma$. 
Let $(\xi',x',y')$ be admissible coordinates and $\Rbz$ be defined in as in Lemma~\ref{lem:decompK}. Assume finally 
that
$$
M_0 := \inf_{\Sigma} |\Rbz-\Rb_{\xi'}|\, \inf_{\Sigma} |\Rbz+\Rb_{\xi'}| 
$$
is non-vanishing (which holds for non-trivial data in view of Proposition~\ref{prop:arr}).
Then, there exists a weakly regular $T^2$-symmetric Lorentzian manifold $(\Mcal,g)$ 
endowed with admissible conformal coordinates $(\tau,\xi,x,y)$ such that the following conditions hold: 
\begin{enumerate}
\item $\Mcal=[\tau_0,\tau_1) \times \Sigma$ for some $\tau_0 < \tau_1$, 
 and the metric $g$ takes the conformal form \eqref{eq:metric}. 

\item $R$ is strictly increasing with $\tau$. 

\item$|\tau_1 - \tau_0| > 0$ depends only 
the initial norm $N_0$ of the initial data set, defined by  
\be
\label{def:normconf}
\aligned
N_0 := & ||\Ub,\Ab||_{H^1(S^1)}+||\Ubz,\Abz||_{L^2(S^1)}+||\nub||_{W^{1,1}(S^1)}+||\nubz||_{L^1}
\\
     & +||\Rb||_{W^{2,1}(S^1)}+||\Rbz||_{W^{1,1}(S^1)}+||\big( \Rb \big)^{-1}||_{L^\infty(S^1)}
+ {1 \over M_0}. 
\endaligned
\ee 

\item The metric coefficients have the following regularity:
$$
\aligned
& U,A \in C^0_\tau(H^1_{\xi}(S^1))\cap C^1_\tau(L^2_{\xi}(S^1)),
\qquad  
\nu\in C^0_\tau(W_{\xi}^{1,1}(S^1))\cap C^1_\tau(L^1_{\xi}(S^1)),
\\
& R \in C^0_\tau(W^{2,1}_\xi(S^1))\cap C^1_\tau(W^{1,1}_{\xi}(S^1)).
\endaligned
$$

\item Denoting by $\psi$ the embedding:
$\psi: \Sigma \rightarrow \Mcal$, $(\xi',x',y') \mapsto (\tau,\xi,x,y)$, 
one has 
$$
\Big(U,U_\tau, A,A_\tau, \nu, \nu_\tau, R, R_\tau, G,G_\tau,H,H_\tau\Big)(\tau_0)
=\left(\Ub,\Ubz, \Ab,\Abz, \nub, \nubz, \Rb, \Rbz, \Gb,\Gbz, \Hb,\Hbz\right) \circ \psi.
$$

\item Let $(\Mcal',g')$ be another weakly regular $T^2$-symmetric manifold with admissible 
conformal coordinates satisfying all of the conditions   
above but with another embedding $\psi':\Sigma \rightarrow \Mcal'$. 
Then, there exists a neighborhood $\mathcal{U}  \subset \Mcal$ of $\psi(\Sigma)$, 
a neighborhood $\mathcal{U}'  \subset \Mcal'$ of $\psi'(\Sigma)$
and 
$C^\infty$ diffeomorphism $\phi : \mathcal{U}' \rightarrow \mathcal{U}$ such that ${g'}|_{\mathcal{U}'} = \psi^* g|_{\mathcal{U}}$ and 
$
\psi \circ \phi \circ \psi' =id_{\Sigma}. 
$ 
\end{enumerate}
\end{theorem}
 
To establish this result, we are going first  to derive a~priori estimates for any given smooth solution 
and, next,  a~priori estimates about the difference of two solutions. Compactness of the set of all 
solutions arising from a regularization of the initial data follows easily from these estimates. 
Interestingly, our estimate for the difference of two solutions  requires
a  property of higher-order integrability on curved spacetimes with weakly regular geometry, 
inspired from Zhou \cite{Zhou} who treated a system of $(1+1)$-wave maps 
on the (flat) $(1+1)$-Minkowski space. 
The uniqueness statement in the above theorem also follows from our estimates on the difference of two solutions, 
 once a system of conformal coordinates has been fixed, which is equivalent to fixing a system of admissible coordinates on $\Sigma$.

The derivation of a~priori estimates for smooth solutions given now relies on a bootstrap argument.  
To establish energy estimates for the wave equations for $U,A$, we need an upper bound on the sup norm 
of the first derivatives of $R$ as well as on the sup norm of $\nu$. 
Thus, we first prove energy estimates depending on these bounds  
and, next, use these energy estimates to improve 
the upper bounds, on sufficiently small time intervals at least.
 
%---------------------------------------------------------------------------------------------------------------

\subsection{A priori estimates for smooth solutions} 

We consider a smooth solution $(U,A,R,\nu)$ of the Einstein equations in conformal coordinates defined on 
some interval $[\tau_0,\tau_1)$ with $\tau_1 > \tau_0$. Moreover, we assume that the solution is non-trivial 
(i.e. does not lead to a flat spacetime) and that the time orientation has been chosen so that $R_\tau > 0$.
 
\begin{lemma}[Monotonicity of the area function]
Both functions $R_\tau \pm R_\xi$ are strictly increasing along the integral curves of $\tau \mp \xi =const$, 
as functions of $\tau \mp \xi$, respectively.  
Moreover, $R$ is a strictly increasing function of $\tau$ and in particular, for all $\tau \ge \tau_0$,
$$
R(\tau,\xi) \ge \min_{\tau' =\tau_0}R.
$$
\end{lemma}

\begin{proof} 
Introducing the notation  $ \del_u= \del_\tau- \del_\xi$, $ \del_v= \del_\tau+ \del_\xi$,
we observe that 
$$
R_{uv} \ge 0,
$$ 
which, in view of our assumptions on the initial data, leads to the desired claims.
\end{proof}

From now on, we set 
\be
\label{3995}
R_0 := \min_{\tau' =\tau_0}R,
\ee 
and we work with the energy-like functional  
$$
\Ecal_{\text{conf}} (\tau)
: = 
 \int_{S^1} \Bigg(
R\left(  U_\tau^2+U_\xi^2\right) +\frac{e^{4U}}{4R}\left( A_{\tau}^2+A_{\xi}^2\right)+ \frac{e^{2\nu}K^2}{4R^3}
\Bigg). 
$$

\begin{lemma}[Energy estimate] For all $\tau \ge \tau_0$, one has 
$$
\Ecal_{\text{conf}}(\tau) \le \Ecal_{\text{conf}}(\tau_0) \, e^{C(R_0)\left(||R||_{C^1(\tau,\xi)}+1\right)(\tau-\tau_0)}, 
$$
where $C(R_0) > 0$ depends only $R_0$.
\end{lemma}

\begin{proof}
From the constraint equations, it follows that
$$
\aligned
\Ecal_{\text{conf}}
& =\int_{S^1} -R_{\xi\xi}-\nu_{\tau}R_{\tau}-\nu_{\xi}R_{\xi}
   = \int_{S^1}-\nu_{\tau}R_{\tau}-\nu_{\xi}R_{\xi}  
\endaligned
$$
and, after several integration by parts, 
$$
\frac{d}{d\tau}\Ecal_{\text{conf}} 
= \int_{S^1}-\nu_\tau(R_{\tau\tau}-R_{\xi\xi})-R_\tau(\nu_{\tau\tau}-\nu_{\xi\xi}).
$$
Using the wave equations for $\nu$ and $R$, we then obtain 
$$
\frac{d}{d\tau}\Ecal_{\text{conf}}
=C(R_0)||R||_{C^1}\Ecal_{\text{conf}} - \int_{S^1} \nu_\tau e^{2\nu}\frac{K^2}{2R^3}.
$$
The desired result then follows by integration in time, using and Gronwall's lemma and after 
integating by parts to control the second term above.
\end{proof}

\begin{lemma}[First-order estimates on the area function]
By defining 
\be
\label{def:nnr}
M(R)(\tau) : = \inf_{\xi \in S^1} R_u(\tau,.) \,  \inf_{\xi \in S^1}R_v(\tau,.), 
\ee
the area function satisfies 
$$
\aligned
||R||_{C^1}(\tau) 
& \le C \, \left( (\tau-\tau_0) ||e^{2\nu}||_{L^\infty[\tau_0,\tau]\times S^1}
+||R||_{C^1}(\tau_0)\right), 
\\
\left(R_\tau^2-R_\xi^2\right)(\tau) 
& \ge M(R)(\tau_0),
\endaligned
$$
where the constant $C=C(R_0,K)>0$ only depends on $R_0$ and the twist constant $K$.
\end{lemma}

\begin{proof} Both estimates are straightforward consequences of the wave equation satisfied by the function $R$. 
The second uses the fact that $R_\tau^2-R_\xi^2=R_u R_v$ and that both $R_u$ and $R_v$ 
are monotonically increasing in respectively $v$ and $u$.
\end{proof}

A direct consequence of the constraint equations is now stated. 

\begin{lemma}[First-order estimate on $\nu$] The metric coefficient $\nu$ satisfies 
\begin{eqnarray} \label{es:nul1}
||\nu_\xi||_{L^1}(\tau)+|| \nu_\tau ||_{L^1} (\tau)
\le 
C  \, 
||R||_{C^1\left( [\tau_0,\tau] \times S^1 \right)} 
\, \Big( \Ecal_{\text{conf}}(\tau) 
+||R_{\xi \xi}||_{L^1}(\tau)+||R_{\xi \tau}||_{L^1}(\tau) \Big), 
\end{eqnarray}
where $C= C\left( R_0,M(R)(\tau_0) \right) > 0$ is a constant. 
\end{lemma}

Finally, we have the following additional estimate on $R$.

\begin{lemma}[Higher-order estimates on the area function]
The area function satisfies the following second-order estimates: 
$$
\aligned
|| R_{\xi \xi}||_{L^1(S^1)}(\tau)
& \le C(R_0)\int^\tau_{\tau_0} \left( || \nu_{\xi}||_{L^1}(\tau')
   +||R||_{C^1\left( [\tau_0,\tau'] \times S^1\right)}\right) ||e^{2\nu}||_{L^\infty}d\tau' 
     +||R_{\xi \xi}||_{L^1(S^1)}(\tau_0), 
\\
|| R_{\xi \tau}||_{L^1(S^1)}(\tau)
& \le C(R_0)\int^\tau_{\tau_0}\left( || \nu_{\tau}||_{L^1}(\tau')
    +||R||_{C^1\left( [\tau_0,\tau'] \times S^1 \right)}\right)||e^{2\nu}||_{L^\infty}d\tau'
    +||R_{\xi \tau}||_{L^1(S^1)}(\tau_0). 
\endaligned
$$
\end{lemma}

\begin{proof} This is a simple commutation argument for the wave equation of $R$. Recall that $R$ satisfies an equation of the form
$
R_{uv}=\Omega_R,
$
hence we have
$$
R_{\xi u}(\xi,v) = \int_v  \del_\xi \Omega_R+R_{\xi u}(\xi,v_0).
$$
Similar expressions holds for $R_{\xi v}$, $R_{\tau u}$ and $R_{\tau v}$.
The result follows since $\Omega_R= e^{2\nu}K^2 / (2R^3)$.
\end{proof}

To close the argument and arrive at the desired uniform estimate, 
we consider the following bootstrap assumptions:
\be
 \label{btassum1}
|| \nu||_{L^\infty}(\tau) 
< 5C_1 \, \left(||\Rb||_{C^1(S^1)}+||\Rbz||_{C^0(S^1)}\right)\left( \Ecal_{\text{conf}}(\tau_0)
   +||\Rb_{\xi\xi}||_{L^1}(\tau_0)+||\left( \Rbz \right)_{\xi}||_{L^1}(\tau_0)+ 1 \right)
+\frac{1}{\pi}|| \nub ||_{L^1}, 
\ee
\be
|| R ||_{C^1}\left( [\tau_0,\tau] \times S^1 \right) < 2 \, \left(||\Rb||_{C^1(S^1)}+||\Rbz||_{C^0(S^1)}\right), 
\label{btassum2}
\ee
where $C_1 = C_1(R_0,M_0) > 0$ is the constant arising in \eqref{es:nul1}.
Let $\delta > 0$ be fixed,
and $\Bcal \subset [\tau_0,\tau_0+\delta]$ be the largest spacetime region which is 
 included in $[\tau_0,\tau_0+\delta]$ and in which \eqref{btassum1}-\eqref{btassum2} hold. 
Then, $\Bcal$ is clearly non-empty and open. 
We  show that for all sufficiently small $\delta$ 
(in terms of the initial norm of the data \eqref{def:normconf}, only)  
we can improve \eqref{btassum1}-\eqref{btassum2}, i.e. that the following holds. 
 
\begin{lemma}
If $\delta > 0$ is sufficiently small (depending only on the initial norm \eqref{def:normconf}) then $\Bcal$ is closed.
\end{lemma}

\begin{proof} 
It follows from the previous estimates and the bootstrap assumptions that if $\delta$ is sufficiently small, 
depending only on the initial norm of the data, we have 
$$
|| \nu_{\xi}||_{L^1(S^1)}(\tau)+|| \nu_{\tau}||_{L^1(S^1)}(\tau) 
\leq
4C_1 \, \left(||\Rb||_{C^1(S^1)} 
  +||\Rbz||_{C^0(S^1)}\right) 
\, 
\Big( \Ecal_{\text{conf}}(\tau_0)+||R_{\xi\xi}||_{L^1}(\tau_0)+||R_{\xi\tau}||_{L^1}(\tau_0)+ 1/2 \Big). 
$$ 
Since 
$$
|| \nu||_{L^\infty}(\tau) \le \frac{1}{2\pi}|| \nu||_{L^1}(\tau_0)+|| \nu_{\xi}||_{L^1}(\tau)+\frac{1}{2\pi}(\tau-\tau_0)|| \nu_\tau||_{L^1}(\tau),
$$
we have improved \eqref{btassum1}, and then
 \eqref{btassum2} is easily improved using the wave equation for $R$.  
\end{proof} 

Hence, we have established the following result. 

\begin{proposition}[A priori estimates in conformal coordinates] 
There exists a real $\delta > 0$ depending only on the initial norm of the data \eqref{def:normconf}
such that, on $[\tau_0,\tau+\delta]$,
\be
\aligned
N(\tau)
: &=  ||U,A||_{H^1(S^1)}(\tau)+||U_\tau,A_\tau||_{L^2(S^1)}+||\nu||_{W^{1,1}(S^1)}+||\nu_\tau||_{L^1}
\\
& \quad +||R ||_{W^{2,1}(S^1)}+||R_{\tau}||_{W^{1,1}(S^1)}+||R^{-1}||_{L^\infty(S^1)}+N(\nabla R)^{-1}(\tau)
\\
& \le C 
\endaligned
\ee
where $C:= C\big( N(\tau_0), M(R)(\tau_0)\big)$ is a constant. 
\end{proposition}
 
%------------------------------------------------------------------------------

\subsection{Higher-integrability in spacetime}

In order to prove compactness of sequences of solutions, we need a better control over the source terms 
arising in the equations satisfied by the metric coefficients $U,A$. 
To this end, we establish now a higher integrability property in spacetime for these source terms.  
Our method is inspired from Zhou \cite{Zhou} who treated $1+1$ dimensional wave map systems.
The following result is actually stated in a more general form than needed for the proof of local well-posedness in conformal coordinates, 
but the full statement is relevant for the analysis in areal coordinates (cf.~Section~\ref{sec56}, below). 

\begin{lemma}[Spacetime higher-integrability estimate] 
\label{936}
Let $w_-, w_+: [R_0, R^\star] \times \RR \to \RR$ be weak solutions in $L^\infty_t L^2_\theta$
to the equations $\del_R w_\pm \pm \del_\theta (a \, w_\pm) = h_\pm$, respectively, 
where the coefficient $a:  [R_0, R^\star] \times \RR \to \RR$ belongs to $L^\infty$ 
and satisfies 
$0 < a_0 \leq a \leq a_1$ and $h_\pm:  [R_0, R^\star] \times \RR \to \RR$ in $L^\infty_t L^1_\theta$ 
are given functions.  
Then, for each $L>a_1 R$ one has 
$$
{d \over dR} N^I + 2 a_0 \, N^{II} \leq N^{III}, 
$$
with 
$$
\aligned
N^I(R)  :=& \int_{-L + a_1 R}^ {L - a_1 R} \int_{\theta_+}^{L - a_1 R} |w_+(R,\theta_+)| \,  |w_-(R,\theta_-)| \, d\theta_+ d\theta_-, 
\\
N^{II}(R)  :=& \int_{-L + a_1 R}^{L- a_1 R} |w_+(R, \cdot)| \,  |w_-|(R, \cdot) \, d\theta, 
\\
N^{III}(R)  :=& 
\sum_\pm \int_{-L + a_1 R}^ {L - a_1 R} |h_\pm(R, \cdot)|\, d\theta \, \int_{-L + a_1 R}^{L - a_1 R} |w_\mp(R, \cdot)| \, d\theta. 
\endaligned 
$$ 
\end{lemma}

\begin{proof} It is not difficult to check that
$$
\del_R |w_\pm| \pm \del_\theta (a \, |w_\pm|) \leq |h_\pm|.  
$$
On the other hand, from the definitions, we obtain 
$$
\aligned
{d \over dR} N^I(R) 
\leq
& \int_{-L + a_1 R}^ {L - a_1 R} \int_{\theta_+}^{L - a_1 R} 
\big( - \del_\theta (a \, |w_+|) + |h_+| \big) (R,\theta_+) \,  |w_-(R,\theta_-)| \, d\theta_- d\theta_+
\\
& 
+ \int_{-L + a_1 R}^ {L - a_1 R} \int_{\theta_+}^{L - a_1 R} 
|w_+(R,\theta_+)| \, \big( \del_\theta (a \, |w_-|) + |h_-| \big) (R,\theta_-) \, d\theta_- d\theta_+
\\
& 
- a_1 \int_{-L + a_1 R}^{L - a_1 R} | w_+(R, -L + a_1 R)| \, |w_-(R,\theta)| \, d\theta  
- a_1 \int_{-L + a_1 R}^{L - a_1 R} | w_+(R, \theta)| \, |w_-(R,L - a_1 R)| \, d\theta
\endaligned
$$
and, therefore, 
$$
\aligned
{d \over dR} N^I(R) 
\leq 
&
- 2 \int_{-L + a_1 R}^ {L - a_1 R} a(R, \theta) \,|w_+(R,\theta)| \, |w_-|(R,\theta) \, d\theta
\\
& 
+\int_{-L + a_1 R}^ {L - a_1 R} \int_{\theta_+}^{L - a_1 R}
\Big( |h_+| (R,\theta_+) \,  |w_-(R,\theta_-)| + |w_+| (R,\theta_+) \,  |h_-(R,\theta_-)| \Big) \, d\theta_- d\theta_+
\\ 
& 
- \int_{-L + a_1 R}^{L - a_1 R} \big(a_1-a(R,\theta)\big)| w_+(R, -L + a_1 R)| \, |w_-(R,\theta)| \, d\theta 
\\
& 
- \int_{-L + a_1 R}^{L - a_1 R} \big(a_1-a(R,\theta)\big)| w_+(R, \theta)| \, |w_-(R,L - a_1 R)| \, d\theta. 
\endaligned
$$
Using the lower and upper bound of the function $a$, we obtain the desired estimate. 
\end{proof}

%---------------------------------------------------------------------------------------

\subsection{Well-posedness theory for weak solutions}

We are now in a position to complete the proof of Theorem~\ref{prop:leccc}
concerning local existence of solutions for the system \eqref{ee:confconst}-\eqref{ee:nuce} 
by establishing estimates for the difference of two solutions. 
Let $(U^{\eps_1}, A^{\eps_1},\nu^{\eps_1},R^{\eps_1},K^{\eps_1})$ 
and $(U^{\eps_2},A^{\eps_2},\nu^{\eps_2},R^{\eps_2},K^{\eps_2})$
be two $C^\infty$ solutions to the system \eqref{ee:confconst}-\eqref{ee:nuce}, with respectively twist constant $K^{\eps_1}$ and $K^{\eps_2}$, 
defined on a cylinder $[\tau_0,\tau_1] \times S^1$, where $\tau_1=\tau_0+\delta$, with $\delta$ small enough 
so that the uniform estimates of the previous section hold for both solutions. 
Denote by $N^i(\tau)$ ($i=1,2$) the norms of the solutions at time $\tau$, i.e. 
\begin{eqnarray}
N^i(\tau)&=&||U^{\eps_i},A^{\eps_i}||_{H^1(S^1)}(\tau)+||U^{\eps_i}_\tau,A^{\eps_i}_\tau||_{L^2(S^1)}
 +||\nu^{\eps_i}||_{W^{1,1}(S^1)}+||\nu_\tau^{\eps_i}||_{L^1}\nonumber \\
&&\hbox{}+||R^{\eps_i} ||_{W^{2,1}(S^1)}+||R_{\tau}^{\eps_i}||_{W^{1,1}(S^1)}
 +||\left( R^{\eps_i}\right)^{-1}||_{L^\infty(S^1)}
+ {1 \over M(R^{\eps_i})(\tau)}.
\end{eqnarray}
From the uniform estimates established above, it follows that for $i=1,2$, there exists 
a positive constant $C^i$, depending only on $N^i(\tau_0)$, such that
$$
N^i(\tau) \le C^i.
$$
We define $\Delta U :=U^{\eps_2}-U^{\eps_1}$, $\Delta A: =A^{\eps_2}-A^{\eps_1}$, ... 
and we set 
\begin{eqnarray}
N^\Delta(\tau)&:=&||\Delta U,\Delta A||_{H^1(S^1)}(\tau)
+||\Delta \nu, \Delta R_\xi, \Delta R_\tau||_{W^{1,1}}(\tau)
+|| \Delta U_\tau,\Delta A_\tau||_{L^2(S^1)}(\tau) \nonumber \\
&&\hbox{}+||\nu_\tau||_{L^1}(\tau) 
+||\Delta R,\Delta (R^{-1})||_{C^1(S^1)}(\tau)
+||\Delta R_\tau||_{C^0(S^1)}(\tau).\nonumber
\end{eqnarray}
Then, $\Delta U$, $\Delta A$, etc. satisfy the equations 
$$
\aligned
& \Delta U_{\tau \tau}-\Delta U_{\xi \xi} = \Omega^{\Delta U}, 
\qquad \quad 
\Delta A_{\tau \tau}-\Delta A_{\xi \xi} = \Omega^{\Delta A},
\\
& \Delta \nu_{\tau \tau}-\Delta \nu_{\xi \xi} = \Omega^{\Delta \nu}, 
\qquad \quad \quad 
\Delta R_{\tau \tau}-\Delta R_{\xi \xi} = \Omega^{\Delta R}, 
\endaligned
$$
with error terms given by 
$$
\aligned 
\Omega^{\Delta U} =  
& -\frac{R_\tau^{\eps_2}}{R^{\eps_2}} U_\tau^{\eps_2}
+\frac{R_\tau^{\eps_1}}{R^{\eps_1}} U_\tau^{\eps_1} 
+\frac{R_\xi^{\eps_2}}{R^{\eps_2}} U_\xi^{\eps_2}
-\frac{R_\xi^{\eps_1}}{R^{\eps_1}} U_\xi^{\eps_1} 
\\
& + \frac{e^{4U^{\eps_2}}}{2(R^{\eps_2})^2}\left( (A^{\eps_2}_\tau)^2-(A^{\eps_2}_\xi)^2\right)
  - \frac{e^{4U^{\eps_1}}}{2(R^{\eps_1})^2}\left( (A^{\eps_1}_\tau)^2-(A^{\eps_1}_\xi)^2\right), 
\\
\Omega^{\Delta A} =  
& \frac{R_\tau^{\eps_2}}{R^{\eps_2}} A_\tau^{\eps_2}
-\frac{R_\tau^{\eps_1}}{R^{\eps_1}} A_\tau^{\eps_1} 
-\frac{R_\xi^{\eps_2}}{R^{\eps_2}} A_\xi^{\eps_2}
+\frac{R_\xi^{\eps_1}}{R^{\eps_1}} A_\xi^{\eps_1} 
\\
& +4 \left(A^{\eps_2}_\xi U^{\eps_2}_\xi-A^{\eps_2}_\tau U^{\eps_2}_\tau \right)
-4 \left(A^{\eps_1}_\xi U^{\eps_1}_\xi-A^{\eps_1}_\tau U^{\eps_1}_\tau \right),
\endaligned
$$
and
$$
\aligned 
\Omega^{\Delta \nu}  = 
&{(U_\xi^{\eps_2})}^2-(U_\tau^{\eps_2})^2-(U_\xi^{\eps_1})^2+(U_\tau^{\eps_1})^2 
 \\
&    + \frac{e^{4U^{\eps_2}}}{4(R^{\eps_2})^2}\left( (A^{\eps_2}_\tau)^2-(A^{\eps_2}_\xi)^2\right)
       - \frac{e^{4U^{\eps_1}}}{4(R^{\eps_1})^2}\left( (A^{\eps_1}_\tau)^2-(A^{\eps_1}_\xi)^2\right) 
\    - \frac{3(K^{\eps_2})^2}{4 (R^{\eps_2})^4}e^{2\nu^{\eps_2}}+\frac{3(K^{\eps_1})^2}{4 (R^{\eps_1})^4}e^{2\nu^{\eps_1}},
\\ 
\Omega^{\Delta R} =
&-\frac{(K^{\eps_2})^2}{2 (R^{\eps_2})^3}e^{2\nu^{\eps_2}}+\frac{(K^{\eps_2})^2}{2 (R^{\eps_1})^2}e^{2\nu^{\eps_1}}.
\endaligned
$$

Moreover, we also have from the constraint equations 
$$ 
\Delta \nu_\tau = \Omega^{\Delta \nu_\tau},
\qquad \quad 
 \Delta \nu_\xi = \Omega^{\Delta \nu_\xi},
$$
where $\Omega^{\Delta \nu_\tau}$ and $\Omega^{\Delta \nu_\xi}$ are obtained from the equations
$$
\aligned
\nu^{\eps_i}_{\tau} 
= &-\frac{1}{(R^{\eps_i}_\tau)^2-(R_{\xi}^{\eps_i})^2}
   \Bigg( R^{\eps_i} R^{\eps_i}_\xi\left( 2 U^{\eps_i}_\tau U^{\eps_i}_\xi+\frac{e^{4U^{\eps_i}}}{2(R^{\eps_i})^2}A^{\eps_i}_\tau A_\xi^{\eps_i}+R^{\eps_i}_{\xi \tau}\right)
\\
  & +R^{\eps_i}R_\tau^{\eps_i}\left( (U_\tau^{\eps_i})^2+(U^{\eps_i}_\xi)^2+\frac{e^{4U^{\eps_i}}}{4R^{\eps_i}}\left( (A_\tau^{\eps_i})^2+(A_\xi^{\eps_i})^2\right)+ \frac{R^{\eps_i}_{\xi \xi}}{R^{\eps_i}} + \frac{e^{2\nu^{\eps_i}}(K^{\eps_i})^2}{4(R^{\eps_i})^2}\right)
\Bigg)
\endaligned
$$
$$
\aligned
\nu^{\eps_i}_{\xi}
= &-\frac{R^{\eps_i} R^{\eps_i}_{\xi}}{(R^{\eps_i})_\tau^2 
  -(R^{\eps_i})_{\xi}^2}\frac{(K^{\eps_i})^2}{4(R^{\eps_i})^2}e^{2\nu^{\eps_i}} 
\\
& +\left( (R^{\eps_i})_\tau^2-(R^{\eps_i})_\xi^2 \right)^{-1} \Bigg( -R^{\eps_i} R^{\eps_i}_{\xi} \left( (U^{\eps_i})_\tau^2+(U^{\eps_i})_{\xi}^2 + \frac{e^{4U^{\eps_i}}}{4(R^{\eps_i})^2}\left((A^{\eps_i})_\tau^2+(A^{\eps_i})_{\xi}^2\right)+\frac{(R^{\eps_i})_{\xi \xi}}{(R^{\eps_i})}\right)
\\
& +R^{\eps_i} R^{\eps_i}_\tau \left( 2 U^{\eps_i}_\tau U^{\eps_i}_{\xi} + \frac{e^{4U^{\eps_i}}}{2(R^{\eps_i})^2}A^{\eps_i}_\tau A^{\eps_i}_{\xi} 
+ \frac{R^{\eps_i}_{\xi\tau}}{R^{\eps_i}} \right) \Bigg). 
\endaligned
$$

We now arrive at one of our key estimates, i.e.~a Lipschitz continuity property for solutions to the Einstein equations 
in terms of their initial data. 
Note that the small-time restriction below is made for convenience for the application of Lemma~\ref{936}. 
The following statement completes the proof of Theorem~\ref{prop:leccc}.

\begin{proposition}[Continuous dependence property]
Provided that $\tau_1-\tau_0 \le \pi$, one has 
$$
N^\Delta(\tau_1) \le C \, N^\Delta(\tau_0),
$$
where $C > 0$ only depends on the constants $C^i$.
\end{proposition}

\begin{proof}
We apply Lemma~\ref{936} first with $w_+=(\Delta A_+)^2=(\Delta A_\tau + \Delta U_\xi)^2$ and $w_-=A_-^2=(A_\tau-A_\xi)^2$, 
where $A$ stands for any of the components $A^{\eps_i}$. This leads us to
$$
|| (\Delta A_+)A_-||^2_{L^2([\tau_0,\tau]\times S^1)} 
\le 4 \sum_{\pm}\int^{\tau}_{\tau_0}\int_{S^1} |h_\pm| d\xi\int_{S^1} |w_{\mp}|d\xi
$$
for any $\tau \in [\tau_0,\tau_1]$, with 
$$
\aligned
h_+&= 2 (\Delta A_+) \Omega^{\Delta A},
\qquad \qquad
h_-= 2A_-\Omega^A,
\\
|w_+| &\le 2 (\Delta A_\tau)^2+2(\Delta A_{\xi})^2,
 \quad 
|w_-| \le 2 A_\tau^2 + 2 A_\xi^2.
\endaligned
$$
Thus, we have 
\be
\label{es:daa}
|| (\Delta A_+)A_-||^2_{L^2([\tau_0,\tau]\times S^1)} 
\le C \, N^2 \int_{\tau_0}^{\tau}\int_{S^1} \Delta A_+ |\Omega^{\Delta A}|d\xi d\tau
   + \int_{\tau_0}^{\tau}(N^{\Delta})^2(\tau)\int_{S^1} A_- |\Omega^{A}|d\xi d\tau, 
\ee
where $N$ is the maximum of $N^1(\tau_0)$ and $N^2(\tau_0)$ 
and where $C>0$ is a numerical constant.

For the second term on the right-hand side, recall also the estimate
$$
\aligned
\int_{\tau_0}^{\tau}\int_{S^1} A_- \Omega^A 
\leq C \, N \, \bigg( &||A_- R_+||_{L^2(\tau,\xi)}||A_- ||_{L^2(\tau,\xi)} + ||R_- A_+||_{L^2(\tau,\xi)}||A_- ||_{L^2(\tau,\xi)} 
\\
& + ||A_- U_+||_{L^2(\tau,\xi)}||A_- ||_{L^2(\tau,\xi)} + ||U_- A_+||_{L^2(\tau,\xi)}||A_- ||_{L^2(\tau,\xi)}\bigg),
\endaligned
$$
where $A$ stands for any of the $A^{\eps_i}$ and 
where $||.||_{L^2(\tau,\xi)}$ stands for $||.||_{L^2([\tau_0,\tau_1]\times S^2)}$.
Together, with the a-priori estimate of the previous section,
we then obtain 
$$
\int_{\tau_0}^{\tau_1}\int_{S^1} A_- \Omega^A \le CN^3.
$$

For the first term on the right-hand side of \eqref{es:daa}, 
we note that 
$$
\aligned
|\Omega^{\Delta A}| 
\leq \sum_{\pm}\bigg( 
& 1/2 {R_0^{-1}}\left| \Delta R_\pm\right| |A_\mp^{\eps_2}| +1/2 R_0^{-1}\left| \Delta A_\pm\right| |R_\mp^{\eps_1}|+1/2|\Delta R^{-1}||A_\pm R_\mp| 
\\
& + 2 \, |\Delta A_\pm ||U^{\eps_2}_{\mp}|+2|A^{\eps_1}_\pm ||\Delta U_\mp| \bigg), 
\endaligned
$$
where $R_0>0$ is the minimum of $(R^\eps_i)^{-1}$, 
for $i=1,2$ on the initial data.  Then, we have 
$$
|\Omega^{\Delta A}| \le C N \sum_{\pm} \left| \Delta R_\pm\right| |A_\mp^{\eps_2}| +\left| \Delta A_\pm\right| |R_\mp^{\eps_1}|+|\Delta R^{-1}||A_\pm R_\mp|+2|\Delta A_\pm ||U^{\eps_2}_{\mp}|+2|A^{\eps_1}_\pm ||\Delta U_\mp| \nonumber \\
$$
for some numerical constant $C>0$. Thus, using Cauchy-Schwarz inequality, we find 
\begin{eqnarray}
\int_{\tau_0}^{\tau}\int_{S^1} \Delta A_+ |\Omega^{\Delta A}|d\xi\tau &\le
& C \, N \, \bigg( 
||\Delta A_+ A_-^{\eps_2}||_{L^2(\tau,\xi)}||\Delta R_+ ||_{L^2(\tau,\xi} 
+ ||\Delta A_+ A_-^{\eps_2}||_{L^2(\tau,\xi)}||\Delta R^{-1} R_+ ||_{L^2(\tau,\xi)} \nonumber \\
&& \qquad \hbox{}+||\Delta A_+ U_-^{\eps_2}||_{L^2(\tau,\xi)}||\Delta A_+ ||_{L^2(\tau,\xi)}  
         +||\Delta A_- U_+^{\eps_2}||_{L^2(\tau,\xi)}||\Delta A_+ ||_{L^2(\tau,\xi)} \nonumber \\
&& \qquad \hbox{}+||\Delta A_+ A_-^{\eps_2}||_{L^2(\tau,\xi)}||\Delta U_+ ||_{L^2(\tau,\xi)} 
           +||\Delta U_- A_+^{\eps_2}||_{L^2)\tau,\xi)}||\Delta A_+ ||_{L^2(\tau,\xi)} \bigg) \nonumber \\
&&\hbox{}+\int_{\tau_0}^\tau ||A_+^{\eps_2} \Delta R_-||_{L^2(\xi)}||\Delta A_+ ||_{L^2(\xi)} \nonumber \\
&&\hbox{}+||\Delta A_+ R_-^{\eps_2}||_{L^2(\xi)}||\Delta A^+ ||_{L^2(\xi)}
+||\Delta A_- R_+^{\eps_2}||_{L^2(|\xi)}||\Delta A_+ ||_{L^2(\xi)} \nonumber \\
&&\hbox{}+||\Delta R^{-1}||_{L^\infty(\xi)} ||\Delta A_+ R_-^{\eps_2}||_{L^2(\xi)}||A_+||_{L^2(\tau,\xi)}.
\end{eqnarray}

We estimate all products 
$$
||A_\pm \Delta R_\mp|| \, ||\Delta A_\pm||,
\qquad \quad
||\Delta A_{\mp}R_\pm|| \, ||\Delta A_\pm||, 
\qquad \quad
||\Delta R^{-1}||_{L^\infty(\xi)} .||\Delta A_+ R_-^{\eps_2}||.||A_+ ||,
$$
and obtain 
$$
||A_\pm^{\eps_2} \Delta R_\mp||_{L^2(\tau)}||\Delta A_\pm||_{L^2(\tau)}
\le C \, ||\Delta R_{\pm}||_{C^0} N ||\Delta A_\pm||_{L^2(\tau)} 
\le C \, N \, ||N^{\Delta}||^2 
$$
and, similarly, 
\begin{eqnarray}
||\Delta A_{\mp}R_\pm||_{L^2(\tau)} ||\Delta A_\pm||_{L^2(\tau)} \le C N ||N^{\Delta}||^2, \nonumber  \\
||\Delta R^{-1}||_{C^0} ||\Delta A_+ R_-^{\eps_2}||_{L^2(\tau)}||A_+ ||_{L^2(\tau)} \le CN^2 ||N^{\Delta}||^2.
\end{eqnarray}
Similar estimates hold by replacing $+$ by $-$,  $A$ by $U$, and $\Delta A$ by $\Delta U$, and so we have 
\begin{eqnarray}
\sum_{i,j}\sum_{\pm}|| u^i_\pm \Delta u^j_\mp||^2_{L^2(\tau,\xi)} \le 
C(N^3+N^2)\int_{\tau_0}^\tau||N^{\Delta}||^2 d\tau'.
\end{eqnarray}
where $(u_1,u_2)=(U,A)$ stands for either $(U^{\eps_1},A^{\eps_1})$ or $(U^{\eps_2},A^{\eps_2})$ 
and $\Delta u^j=(\Delta U, \Delta A)$. In view of 
$$
\frac{d}{d\tau}\int_{S^1} \Delta A_\tau^2+\Delta A_\xi^2
=\int_{S^1} 2 \Delta A_\tau \Omega^{\Delta A},
$$
we have proved that
\begin{eqnarray}
\left(\int_{S^1}\left( \Delta A_\tau^2+ \Delta U_\tau^2 +\Delta \nu_\tau^2+\Delta A_\xi+ \Delta U_\xi^2 +\Delta \nu_\xi^2 \right)d\xi \right)(\tau) \le \nonumber \\
\left(\int_{S^1}\left( \Delta A_\tau^2+ \Delta U_\tau^2 +\Delta \nu_\tau^2
+\Delta A_\xi^2+ \Delta U_\xi^2 +\Delta \nu_\xi^2 \right)d\xi \right)(\tau_0) + C_N\int^\tau_{\tau_0} (N^\Delta)^2(\tau) d\tau', 
\end{eqnarray}
where $C_N> 0$ only depends on $N$.

For $R$, we proceed as before, by integration along null lines, to check that 
$$
\aligned
||\Delta R_{\pm}||_{C^0} 
\leq 
& C_N \int^\tau_{\tau_0}  \left( ||\Delta R^{-1}||_{L^\infty(\xi)} +||\Delta \nu||_{L^\infty(\xi)}\right)d\tau' 
\\
\leq 
& C_N  \, \int^\tau_{\tau_0} \left( ||\Delta R^{-1}||_{L^\infty(\xi)} 
+||\Delta \nu||_{W^{1,1}+||\Delta \nu_\tau||_{L^1}}\right)d\tau'. 
\endaligned
$$
Similar estimates for higher derivatives hold in $L^1$  
after commuting the equations as in the previous section. Using 
$$
\left|\frac{1}{x_1^2-y_1^2}-\frac{1}{x_2^2-y_2^2}\right| \le 
\frac{|x_1^2-x_2^2|+|y_1^2-y_2^2|}{|x_1^2-y_1^2||x_2^2-y_2^2|}
$$
to estimate the differences for the terms containing
$1/ \Big( (R^{\eps_i}_\tau)^2-(R^{\eps_i}_\xi)^2 \Big)$, 
we also obtain easily the necessary estimates for $\Omega^{\Delta \nu_\xi}$ 
and $\Omega^{\Delta \nu_\tau}$.

Finally, we have trivially the following estimates 
for $\Delta A,\Delta U$ in $L^2$ (and not derivatives thereof):
$$
\frac{d}{d\tau}||U,A||_{L^2}^2 \le (N^\Delta)^2(\tau)
$$
and similalry, we have estimates on $\Delta \nu$ 
and $\Delta R$ simply from the definition of $N^{\Delta}$.
Thus, putting everything together, we have the following estimate 
from which the result follows: 
\[
N^\Delta(\tau)^2 \le N^\Delta(\tau_0)^2 
+ C_N \int^\tau_{\tau_0} (N^\Delta)^2(\tau') d\tau'.
\qedhere 
\]
\end{proof}

%=============================================================================

\section{Global geometry of weakly regular $T^2$--symmetric spacetimes}  
\label{sec56}

\subsection{Continuation criterion}

We are now in a position to complete the proof of Theorem \ref{theorem62} in this section. 
Combining Theorem~\ref{prop:leccc} and Proposition~\ref{coordcomp}, we obtain that, 
for any weakly regular $T^2$-symmetric initial data with contant $R=R_0$, there exists 
a weakly regular $T^2$-symmetric Lorenztian manifold arising from this data with admissible areal coordinates. 
Consider one such development and let $R_1$ denote the final time of existence of this solution. 
Note that, in conformal coordinates, we have the following lower bound:
\be
\label{lowerb}
R_\tau \ge \frac{1}{2}\left( \inf_{\tau=\tau_0} R_u + \inf_{\tau=\tau_0} R_v \right),
\ee 
where we have used the notation of the previous section. Since, the conformal time of existence given 
by Theorem~\ref{prop:leccc} only depends on the initial norm \eqref{def:normconf}, it follows that the areal time 
of existence of the solution is bounded below by a constant depending only \ref{def:normconf}. 
Hence, we have the following continuation criterion. 

\begin{lemma}[Continuation criterion]
Let $(U,A,\eta,a)$ be a solution to the equations \eqref{weakform1}--\eqref{weakconstraintsr} with 
the regularity 
$U,A \in C^0_R(H^1_{\theta}(S^1))\cap C^1_R(L^2_{\theta}(S^1))$, 
$\eta \in C^0_R(W_{\theta}^{1,1}(S^1))\cap C^1_R(L^1_{\theta}(S^1))$, 
$a,a^{-1} \in C^0_R(W_{\theta}^{2,1}(S^1))\cap C^1_R(W^{1,1}_{\theta}(S^1))$,  
and defined on an interval of time $R \in [R_0,R_1)$. Assume that $R_1 < \infty$ and that the norm 
$$
N : = ||U,A||_{H^1}(R)+||U_R,A_R||_{L^2}(R)+||\eta,a_R,a_\theta ||_{W^{1,1}}(R)+||\eta_R,a_{RR},a_{R\theta},a_{\theta \theta}||_{L^1}(R)+||a,a^{-1}||_{L^\infty}(R)
$$
is uniformly bounded on the interval $[R_0,R_1)$. Then, the solution can be extended beyond $R_1$ with the same regularity.
\end{lemma}

As a consequence, we can prove the existence of global solutions in areal coordinates provided 
we derive uniform estimates on the above norm, as we do now in the rest of this section. 
Moreover, since one can approximate (locally in time, at least) weakly regular solutions by smooth solutions,
we consider, in the rest of this section, a smooth solution $(U,A,\eta,a)$ to the equations \eqref{weakform1}--\eqref{weakconstraintsr}.
defined on some interval of time $[R_0,R^\star)$ for some $R^\star> R_0$.  
We search for now bounds that are uniform on this interval $[R_0, R^\star)$. 
Constants that depend on the (natural norms of the) 
initial data, only, are denoted by $C$, while constants that also depend on $R_\star$ are denoted by $C^\star$.  

%--------------------------------------------------------------------------------------

\subsection{Uniform energy estimates in areal coordinates} 

Both energy-like functionals 
$$
\aligned 
\Ecal(R) :=& \int_{S^1} E(R,\theta) \, d\theta,
\qquad  \quad 
E :=  a^{-1}  (U_R)^2 + a  \, (U_\theta)^2
+ {e^{4U} \over 4R^2} \left( a^{-1}  (A_R)^2 + a  \, (A_\theta)^2 \right)
\endaligned 
$$
and 
$$
\aligned 
\Ecal_K(R) :=& \int_{S^1} E_K(R,\theta) \, d\theta,
\qquad \quad 
E_K :=E + {K^2  \over 4 R^4} \, e^{2 \eta} \, a^{-1} 
\endaligned 
$$ 
are non-increasing in time, since 
$$
\aligned
&{d \over dR} \Ecal(R) 
= - {K^2 \over 2 R^3} \int_{S^1} E \, e^{2 \eta} \, d\theta 
- {2 \over R} \int_{S^1} \left(
a^{-1}  \, (U_R)^2  + {1 \over 4R^2} \, e^{4U}  \, a  \, (A_\theta)^2\right) \, d\theta,
\\
& {d \over dR} \Ecal_K(R) 
= - {K^2 \over  R^5} \int_{S^1} \, a^{-1}e^{2 \eta} \, d\theta 
- {2 \over R} \int_{S^1} \left(
a^{-1}  \, (U_R)^2  + {e^{4U} \over 4R^2} a \, (A_\theta)^2\right) \, d\theta.
\endaligned
$$
These functionals yields a uniform control for all times $R \geq R_0$.

\begin{lemma}[Energy estimates] 
\label{energ}
The following energy bounds hold
$$ 
\aligned 
&\sup_{R \in [R_0,R^\star)} \Ecal(R) \leq \Ecal(R_0), 
\qquad 
\qquad 
\sup_{R \in [R_0,R^\star)} \Ecal_K(R) \leq \Ecal_K(R_0), 
\endaligned 
$$
as well as the spacetime bounds  
$$
\int_{R_0}^\infty \int_{S^1} \Big( 
c_1   \, (U_R)^2 \, a^{-1} 
+ 
c_2  \, (U_\theta)^2 \, a
+ 
c_3 \,  (A_R)^2 \, a^{-1}  
+ 
c_4 \, (A_\theta)^2 \, a \Big) \, dRd\theta
\leq \Ecal(R_0)
$$
with 
$$
\aligned
& c_1 = {2 \over R}   + {K^2 \over 2R^3} \, e^{2 \eta}, 
\qquad 
c_2 := {K^2 \over 2R^3} \, e^{2 \eta},
\\
& c_3 = {K^2 \over 8R^5} \, e^{4U + 2\eta}, 
\qquad 
c_4 := {1 \over 2R^3} \, e^{4U} 
 + {K^2 \over 8R^5} \, e^{4U+2\eta}, 
\endaligned
$$
and     
$$
\int_{R_0}^\infty \int_{S^1} {K^2 \over R^5} \, e^{2 \eta} \, a^{-1} \, dR d\theta \leq \Ecal_K(R_0).
$$
\end{lemma}

Moreover, since the function $a$ is bounded above and below on the initial slice $R=R_0$, 
the initial energy $\Ecal(R_0)$ is comparable with the $H^1$ norm of the data $\Ub, \Ab$, that is, 
$$
C_1 \, \Ecal(R_0) \leq
\| (\Ub, \Ubz, \Ab, \Abz)\|_{L^2(S^1)} \leq C_2 \, \Ecal(R_0)  
$$ 
for constants $C_1, C_2>0$ depending on the {\sl sup norm} of the data
at time $R=R_0$, only. 
To have similar inequalities at arbitrary times $R$ requires a sup-norm bound on the other metric coefficients, 
which we derive below. 

We now derive direct consequences of the energy estimate in Lemma ~\ref{energ}.  

\begin{lemma}[Upper bound for the function $a$] 
\label{lem:consener}
The function $a$ satisfies the upper bound
$$
\sup_{[R_0, R^\star) \times S^1} a \leq \sup_{S^1} \ab,
$$
as well as
$$
\frac{1}{2R} \, \int_{S^1} |(1/a)_R| \, d\theta \leq \Ecal_K(R_0).
$$
\end{lemma}

\begin{proof} From \eqref{eq:lnalphar} we see that $a$ decreases when $R$ increases, 
which implies the desired sup-norm bound for $a$. 
The other estimate follows immediately from the 
equations \eqref{eq:lnalphar} and \eqref{weakconstraintsr}, 
since 
\[ 
0 \leq - 2 \, a_R \, a^{-1} \le \frac{K^2}{R^3} \, e^{2\eta} \, a^{-1} = 4 \, R \, (E_K - E) \leq 4R \, E_K.  
\qedhere
\]
\end{proof}

\begin{lemma}[Estimates for the function $\eta$] 
\label{lem:bnu}
The function $\eta$ satisfies the integral estimates
$$ 
\aligned
& {1 \over R} \, \int_{S^1}  |\eta_R| \, a^{-1} \, d\theta \leq \Ecal_K(R_0), 
\qquad\qquad
 {1 \over R} \, \int_{S^1}| \eta_\theta|  \, d\theta \leq \Ecal(R) \leq \Ecal(R_0)
\endaligned
$$ 
and the pointwise estimate 
$$
\aligned
|\eta(R, \theta)|  
\leq R \, \Ecal(R_0) + \Big| \int_{S^1} \etab \, d\theta' \Big|
+ \big(\sup_{S^1} \ab \big) \, \frac{R^2- R_0^2}{2} \, \Ecal_K(R_0). 
\endaligned
$$
\end{lemma}

\begin{proof} We have 
$$
\aligned
 |\eta_\theta| \le R \, E,
\qquad
\qquad
 |\eta_R| \, a^{-1} \leq R \, E + \frac{a^{-1}}{4R^3}e^{2\eta}K^2 = R \, E_K. 
\endaligned
$$
On the other hand, in view of Lemma~\ref{lem:consener}, for any $\theta, \theta' \in S^1$
we have 
$$
|\eta(R,\theta) - \eta(R,\theta')| \leq R \, \Ecal(R).
$$
Thus, by integrating in $\theta'$, we find 
$$
\int_{S^1}\eta(R,\theta') \, d\theta' - 2 \pi \, R \, \Ecal(R) 
\leq 2\pi \, \eta(R,\theta) 
\leq 2\pi \, R \, \Ecal(R) + \int_{S^1} \eta(R,\theta') \, d\theta'.
$$
On the other hand, we have 
$$
\left|\int_{S^1} \eta(R,\theta')d\theta'\right| \le \bigg|\int_{S^1} \int_{R_0}^R \eta_R(R,\theta')d\theta'\bigg|
+ \bigg| \int_{S^1}\etab \, d\theta' \bigg|, 
$$
and we can  evaluate the second term on the right-hand side above from Lemma~\ref{lem:consener}, as follows: 
$$
\bigg|\int_{S^1} \int_{R_0}^R \eta_R(R,\theta')d\theta'\bigg| \le \left(\sup_{S^1} a(R,.) \right) 
\, \frac{R^2- R_0^2}{2} \, \Ecal_{K}(R). 
$$
The desired conclusion then follows from the energy estimates in Lemma~\ref{energ}
and the upper bound on $a$ in Lemma~\ref{lem:consener}.  
\end{proof}

%--------------------------------------------------------------------------------------------------------

\subsection{Conclusion of the proof of Theorem~\ref{theorem62}}

We now conclude by deriving 
We already know that $a$ is non-increasing and, so, bounded above. Deriving a lower bound is more delicate. 

\begin{lemma}[Lower bound for the function $a$] 
 \label{lem:alpha-1}
The function $a$ satisfies 
$$
a^{-1} \leq C^\star.  
$$ 
\end{lemma}

\begin{proof}
Using Lemma \ref{lem:bnu}, we find 
$(a^{-2})_R \le C \, R^{-3} \, e^{C \, R^2}$
and, by integration,
$$
\aligned 
a(R,\theta)^{-2} - \ab(\theta)^{-2}  
& \leq \int^R_{R_0}  C \, \frac{e^{C \, R'^2}}{R'^3} \, dR' 
\\
& \leq  \int^R_{R_0} C \, \frac{2R' e^{C \, R'^2}}{2R'^4} \, dR' \leq  C_1 \,\left( e^{C \, R^2} - e^C \right). 
\endaligned
$$
By estimating $\ab(\theta)^{-2}$, this concludes the proof.
\end{proof}

\begin{lemma}[Estimates of the functions $U,A$]
The functions $U,A$ satisfy the integral estimate 
$$
\int_{S^1} \big( U_t^2 + A_t^2 + U_\theta^2 + A_\theta^2 \big) \, d\theta
\leq C^\star, 
$$ 
and the pointwise estimate 
$$ 
\sup_{[R_0, R^\star] \times S^1} \big( |U|+|A| \big) 
\leq C^\star. 
$$ 
\end{lemma}

\begin{proof}
It follows immediately from the energy estimates and the estimates for $a$ and $a^{-1}$ that 
\[ 
\int_{S^1} \big( U_\theta^2 + e^{4U} \, A_\theta^2 \big) \, d\theta
\leq C^\star, 
\qquad
\qquad
\int_{S^1} \big( U_t^2 + e^{4U} \, A_t^2 \big) \, d\theta \leq C.    \qedhere
\]
\end{proof}

\begin{lemma}[Additional estimate for the function $a$]
The mixed derivative of the metric coefficient $a$ is controled by the energy density 
$$
\left| (\ln a)_{R \theta}   \right| 
\leq
{K^2 \over 2 \, R^2} \, e^{2\eta} E,  
$$ 
and, therefore, its $\theta$-derivative satisfies the pointwise estimate
$$
|a_\theta| \leq C^\star. 
$$
\end{lemma}

\begin{proof} 
Taking the $\theta$ derivative of
$
(\ln a)_R = - e^{2 \eta} \, K^2/ (2R^3), 
$
we obtain 
$$
\aligned
\big| (\ln a)_{R \theta} \big| 
& = \big| \frac{K^2}{4 \, R^3}e^{2\eta}2\eta_\theta \Big| 
   \leq \frac{K^2}{2 \, R^3}e^{2\eta} \, R \, E, 
\endaligned
$$
since  $|\eta_\theta| \le R \, E$.  From the identity 
\be 
\label{eq:idnualpha}  
(a^{-1} e^{2\eta} )_R = 2 R e^{2\eta} E, 
\ee
we obtain  
$$ 
\left| (\ln a)_{R \theta}  \right| 
\leq
\frac{K^2}{4 \, R^3} (a^{-1} e^{2\eta} )_R.
$$
The second statement follows immediately by integration and using Lemmas~\ref{lem:alpha-1} and \ref{lem:bnu}.
\end{proof}

Finally, we obtain further control on the metric coefficient $a$.

\begin{lemma}[Higher-order estimates on $a$]
The following uniform estimates hold: 
$$
|| a_{R\theta},a_{RR},a_{\theta \theta}||_{L^1}(R) \leq C^\star.
$$
\end{lemma}

\begin{proof}
For the mixed derivative $a_{R\theta}$, this follows from the pointwise estimate derived the previous lemma 
and the energy bounds. For the derivative $a_{RR}$, this follows from the $L^1$ uniform estimate on $\eta_R$
 by commuting the evolution equation for $a$.
For $a_{\theta \theta}$, we proceed as follows. Note first that
$$
\left( a^{-2} (e^{2\eta})_R\right)_{R} - \left( e^{2\eta}\right)_{\theta \theta}
= 4 e^{2\eta} \left( a^{-2}\eta_R^2- \eta_\theta^2 \right)+ 2 e^{2 \eta }\left( \left( a^{-2}\eta_R \right)_R - \eta_{\theta\theta} \right).
$$
The second term on the right-hand side is known to be uniformly bounded in $L^1$, using the wave equation for $\eta$. 
For the first term, we note that it involves the product $(a^{-1}\eta_R + \eta_\theta)(a^{-1}\eta_R-\eta_\theta)$. 
This is a null product which rewritten in terms of $U$ and $A$ and up to uniformly bounded factors is the sum of 
uniformly bounded functions and the null products 
$(a^{-1} U_R+a U_\theta)^2(a^{-1} U_R-a U_\theta)^2$, $(a^{-1} A_R+a A_\theta)^2(a^{-1} A_R-a A_\theta)^2$. 
However, these are bounded in spacetime $L^1$ as an application of Lemma \ref{936}. 
On the other hand, we have 
$$
\aligned
\label{ee:le}
\left( \ln a\right) _{R\theta \theta} 
= & \frac{K^2}{2R^3}\left(- e^{2\eta} \right)_{\theta \theta} 
  = \frac{K^2}{2R^3}\left( \left( a^{-2} (e^{2\eta})_R\right)_{R}+F \right), 
\endaligned
$$
where $F$ is a function bounded uniformly in $L^1([R_0,R*] \times S^1 )$.
The result then follows by integration of the previous equation, using an integration by parts and the $L^1$ estimate on 
$\eta_R$ to control the term arising from $\left( a^{-2} (e^{2\eta})_R\right)_{R}$. 
\end{proof}

This completes the derivation of global-in-time uniform estimates and, hence, the proof of Theorem~\ref{theorem62}.  
We can now reformulate our existence result in coordinates, as follows.

\begin{theorem}[Global existence in areal coordinates] 
For any weakly regular initial data set with constant area $R=R_0 >0$, 
the system of partial differential equations describing $T^2$--symmetric spacetimes in areal coordinates
admits a weak solution $U, A, \nu, a, G, H$, satisfying 
the regularity conditions \eqref{regl}, defined on the whole interval $[R_0,\infty)$ and which is unique among the set of functions 
satisfying $\eqref{regl}$. The solution then constructed has the following regularity: 
$$
\aligned
& U, A \in C^0_R(H^1_{\theta}(S^1))\cap C^1_R(L^2_{\theta}(S^1)), 
 \qquad 
\eta \in C^0_R(W_{\theta}^{1,1}(S^1))\cap C^1_R(L^1_{\theta}(S^1)), 
\\
& a,a^{-1} \in C^0_R(W_{\theta}^{2,1}(S^1))\cap C^1_R(W^{1,1}_{\theta}(S^1)), 
\\
& G,H \in C^0_R(L^\infty(S^1)), \qquad  G_R,H_R \in C^0_R(W_{\theta}^{1,1}(S^1))\cap C^1_R(L^1_{\theta}(S^1)). 
\endaligned
$$
\end{theorem}

We emphasize that {\sl additional regularity} of the metric  is 
established here, 
which was not required to express Einstein's field 
equations in the weak sense,
but was deduced from the structure of the Einstein equations under the assumed symmetry. 

%============================================================================

\section*{Acknowledgments}

The first author (PLF) was partially supported by the Centre National de la Recherche Scientifique (CNRS) 
and the Agence Nationale de la Recherche (ANR) through the grant 06-2-134423 entitled
``Mathematical Methods in General Relativity'' (Math-GR). 
The second author (JS) thanks the 
Max Planck Institute for Gravitational Physics (Albert Einstein Institute) for financial support. 
This work was done when the second author (JS) was visiting the Laboratoire Jacques-Louis Lions. 
 
%============================================================================


\small 

\begin{thebibliography}{10} 

\bibitem{BLSS} \auth{Barnes A.P., LeFloch P.G., Schmidt B.G., and Stewart J.M.,}
The Glimm scheme for perfect fluids on plane-symmetric Gowdy spacetimes, 
Class. Quantum Grav. 21 (2004), 5043--5074.

\bibitem{BergerChruscielMoncrief} \auth{Berger B.K., Chru\'sciel P., and Moncrief V.,}
On asymptotically flat spacetimes with $G_2$-invariant Cauchy surfaces, 
Ann. Phys. 237 (1995), 322--354. 

\bibitem{BergerChruscielIsenbergMoncrief} 
\auth{Berger B.K., Chru\'sciel P., Isenberg J., and Moncrief V.,}
Global foliations of vacuum spacetimes with $T^2$ isometry, 
Ann. Phys. 260 (1997), 117--148.   

\bibitem{Choquet-Bruhat2} \auth{Choquet-Bruhat Y.,}
{\em General relativity and the Einstein equations,}
Oxford Math. Monographs, 2009.

\bibitem{ChoquetBruhatGeroch} \auth{Choquet-Bruhat Y. and Geroch R.,} 
Global aspects of the Cauchy problem in general relativity, 
Comm. Math Phys. 14 (1969), 329--335.

\bibitem{Christodoulou} \auth{Christodoulou D.,} 
Bounded variation solutions of the spherically symmetric Einstein-scalar field equations, 
Comm. Pure Appl. Math. 46 (1992), 1131--1220.  

\bibitem{Chrusciel} \auth{Chru\'sciel P.,} 
On spacetimes with $U(1) \times U(1)$ symmetric compact Cauchy surfaces,
Ann. Phys. 202 (1990), 100--150.

\bibitem{ChruscielIsenbergMoncrief} \auth{Chru\'sciel P., Isenberg J., and Moncrief V.,} 
Strong cosmic censorship in polarized Gowdy spacetimes, 
Class. Quantum Grav. 7 (1990), 1671--1680. 

\bibitem{EardleyMoncrief} \auth{Eardley D. and Moncrief V.,}
The global existence problem and cosmic censorship in general relativity, 
Gen. Relat. Grav. 13 (1981), 887--892.  

\bibitem{ChoquetBruhat} \auth{Foures-Bruhat Y.,}
Th\'eor\`emes d'existence pour certains syst\`emes d'\'equations aux d\'eriv\'ees partielles non-lin\'eaires,
Acta Matematica 88 (1952), 141--225.

\bibitem{Gowdy} \auth{Gowdy R.,} 
Vacuum spacetimes with two-parameter spacelike isometry groups and compact invariant hypersurfaces: 
topologies and boundary conditions, 
Ann. Phys. 83 (1974), 203--241.  

\bibitem{HughesKatoMarsden} \auth{Hughes T.J.R., Kato T., and Marsden J.E.,}
Well-posed quasi-linear second-order hyperbolic systems with applications to nonlinear elastodynamics and general relativity,
Arch. Rational Mech. Anal. 63 (1976), 273--294.

\bibitem{IsenbergMoncrief} \auth{Isenberg J. and Moncrief V.,}
Asymptotic behavior of the gravitational field and the nature of singularities in Gowdy spacetimes, 
Ann. Phys. 99 (1990), 84--122. 

\bibitem{IsenbergWeaver} \auth{Isenberg J. and Weaver M.,}
On the area of the symmetry orbits in $T^2$--symmetric spacetimes, 
Class. Quantum Grav. 20 (2003), 3783--3796.  

\bibitem{KlainermanRodnianski} \auth{Klainerman S. and Rodnianski I.,}
Rough solutions of the Einstein vacuum equations,
Ann. Math. 161 (2005), 1143--1193.

\bibitem{LeFlochMardare} \auth{LeFloch P.G. and Mardare C.,}
Definition and weak stability of spacetimes with distributional curvature, 
Port. Math. 64 (2007), 535--573.

\bibitem{LeFlochRendall} \auth{LeFloch P.G. and Rendall A.,}
A global foliation of Einstein-Euler spacetimes with Gowdy symmetry on $T^3$,  
Arch. Rational Mech. Anal. (2011). 
% See also ArXiv:1004.0427v1. 

\bibitem{LeFlochSmulevici0} \auth{LeFloch P.G. and Smulevici J.,}
Global geometry of $T^2$--symmetric spacetimes with weak regularity, 
Note C.R. Acad. Sc. 348 (2010), 1231--1233. 
% See also ArXiv:1006.2933.

\bibitem{LeFlochSmulevici2} \auth{LeFloch P.G. and Smulevici J.,}
Weakly regular $T^2$--symmetric spacetimes. Future geodesic completeness, 
in preparation. 

\bibitem{LeFlochSmulevici3} \auth{LeFloch P.G. and Smulevici J.,}
Weakly regular $T^2$--symmetric spacetimes, in preparation. 

\bibitem{LeFlochStewart} \auth{LeFloch P.G. and Stewart J.M.,}
Shock waves and gravitational waves in matter spacetimes with Gowdy symmetry, 
Port. Math. 62 (2005), 349--370. 

\bibitem{LeFlochStewart2} \auth{LeFloch P.G. and Stewart J.M.,}
The characteristic initial value problem for 
plane symmetric spacetimes with weak regularity, 
Class. Quantum Grav. (2011). 
% Preprint ArXiv:1004.2343v1.  

\bibitem{Moncrief} \auth{Moncrief V.,}
Global properties of Gowdy spacetimes with $T^3 \times \RR$ topology, 
Ann. Phys. 132 (1981), 87--107. 

\bibitem{Rendall-crush} \auth{Rendall A.D.,} 
Crushing singularities in spacetimes with spherical, plane, and hyperbolic symmetry, 
Class. Quantum Grav. 12 (1995), 1517--1533. 

\bibitem{Rendall2} \auth{Rendall A.D.,} 
Existence of constant mean curvature foliations in spacetimes with two-dimensional local symmetry,
Commun. Math. Phys. 189 (1997), 145--164.

\bibitem{Ringstrom} \auth{Ringstr\"om H.,}
On a wave map equation arising in general relativity,
Comm. Pure Appl. Math. 57  (2004), 657--703.

\bibitem{Ringstrom1}\auth{Ringstr\"om H.,} 
Curvature blow-up on a dense subset of the singularity in $T^3$-Gowdy, 
Jour. Hyper. Diff. Equa. 2 (2005), 547--564. 

\bibitem{Ringstrom2}\auth{Ringstr\"om H.,} 
Strong cosmic censorship in $T^3$-Gowdy spacetimes, 
Ann. Math. 170 (2009), 1181--1240. 

\bibitem{SmithTataru} \auth{Smith H.F. and Tataru D.,}
Sharp local well-posedness results for the nonlinear wave equation,
Ann. of Math. 162 (2005), 291--366. 

\bibitem{Smulevici1} \auth{Smulevici J.,}
Strong cosmic censorship for $T^2$--symmetric spacetimes with positive cosmological constant and matter, 
Ann. Henri Poincar\'e 9 (2009), 1425--1453.

\bibitem{Smulevici2} \auth{Smulevici J.,}
On the area of the symmetry orbits in spacetimes with toroidal or hyperbolic symmetry, 
Analysis \& PDE, to appear.  

\bibitem{Zhou} \auth{Zhou, Y.,}
Uniqueness of weak solutions of $1+1$ dimensional wave maps,
Math. Z.  232  (1999),  707--719. 
\end{thebibliography}
\end{document}